\documentclass{aa} 

\usepackage{graphicx}
\usepackage{txfonts}

\usepackage[T1]{fontenc}
\usepackage{ae,aecompl}
\usepackage[utf8]{inputenc}

\usepackage{graphicx}	
\usepackage{amsmath}	
\usepackage{amssymb}	
\usepackage{lscape}
\usepackage{tikz}   
\usepackage{gensymb}
\usepackage{tikz-3dplot} 
\usepackage{import}
\usepackage{lmodern}
\RequirePackage{fix-cm}

\usepackage{graphicx}	
\usepackage{amsmath}	
\usepackage{amssymb}	
\usepackage{grffile}

\DeclareRobustCommand{\rchi}{{\mathpalette\irchi\relax}}
\newcommand{\irchi}[2]{\raisebox{\depth}{$#1\chi$}} 

\newcommand{\rlight}{r_{\rm L}}

\begin{document} 
	
\title{Young radio-loud gamma-ray pulsar light-curve fitting}
	
\author{
J. P\'etri\inst{1}
\and
D. Mitra\inst{2,3}
}
\institute{Universit\'e de Strasbourg, CNRS, Observatoire astronomique de Strasbourg, UMR 7550, F-67000 Strasbourg, France.\\
\email{jerome.petri@astro.unistra.fr}  
\and 
National Centre for Radio Astrophysics, Tata Institute for Fundamental Research, Post Bag 3, Ganeshkhind, Pune 411007, India
\and
Janusz Gil Institute of Astronomy, University of Zielona G\'ora, ul. Szafrana 2, 65-516 Zielona G\'ora, Poland     
}
	
\date{Received ; accepted }

\abstract
{Since the launch of the Fermi gamma-ray telescope, several hundreds of radio-loud gamma-ray pulsars have been detected, many belonging to millisecond pulsars but also some belonging to the young pulsar population with spin periods larger than 30~ms.}
{Observing simultaneously pulsed radio and gamma-ray emission from these stars helps to constrain the geometry and radiation mechanisms within their magnetosphere and to localize the multiple photon production sites. In this paper, we fit the time-aligned gamma-ray light-curves of young radio-loud gamma-ray pulsars. We assume a dipole force-free magnetosphere where radio photons emanate from high altitude above the polar caps and gamma-rays originate from outside the light-cylinder, within the striped wind current sheet.}
{We compute a full atlas of radio and gamma-ray pulse profiles depending on the magnetic axis obliquity and line of sight inclination with respect to the neutron star rotation axis. By applying a $\rchi^2$ fitting technique, we are able to pin down accurately the magnetosphere geometry. Further constrains are obtained from radio polarization measurement following the rotating vector model, including aberration and retardation effects.}
{We found a good agreement between our model and the time-aligned single or double peaked gamma-ray pulsar observations. We deduce the magnetic inclination angle and the observer line of sight with respect to the rotation axis within a small error bar. The distinction between radio-loud or radio-quiet gamma-ray pulsars or only radio pulsars can entirely be related to the geometry of the associated emitting regions.}
{The high altitude polar cap model combined with the striped wind represents a minimalistic approach able to reproduce a wealth of gamma-ray pulse profiles for young radio pulsars. Based on self-consistent force-free simulations, it gives a full geometrical picture of the emission properties without resorting to detailed knowledge of the individual particle dynamics and energetics.}



\maketitle



\section{Introduction}

The observed pulsed emission properties of pulsars in the radio and high-energy bands, like their light-curves and spectra are very sensitive to their global geometry defined by their electromagnetic field topology and the angles on one hand between the rotation axis and the magnetic dipole axis and on the other hand between the rotation axis and the line of sight. A good first guess about the knowledge on the electromagnetic field is given by the now comprehensive simulations of dipole force-free magnetospheres for aligned rotators \citep{contopoulos_axisymmetric_1999, komissarov_simulations_2006, parfrey_introducing_2012, cao_spectral_2016} and oblique rotators \citep{spitkovsky_time-dependent_2006, petri_pulsar_2012, kalapotharakos_extended_2012}. More detailed models include some dissipation through resistivity like done in \cite{li_resistive_2012, kalapotharakos_fermi_2017, cao_oblique_2016}. Even kinetic simulations are available \citep{cerutti_particle_2015}. However a force-free fluid approach already suffices to construct realistic radio and gamma-ray pulse profiles, furnishing severe constrains on the underlying geometry.

Several works in the past indeed showed that a simultaneous radio and gamma-ray light-curve fitting is valuable to pin down the geometry. For instance \cite{petri_unified_2011} showed that in the framework of a force-free split-monopole solution, simple analytical expressions for the radio time lag and the gamma-ray peak separation can be derived. Meanwhile \cite{seyffert_geometric_2011} used an emission model for gamma-rays (relying on outer gaps or two pole caustics) and the constraints from radio polarization to deduce the geometry of several pulsars, soon after the publication of the first Fermi gamma-ray pulsar catalogue \citep{abdo_first_2010}. \cite{pierbattista_light-curve_2015} performed a comprehensive analysis of light-curve modelling of young gamma-ray pulsars assuming different geometries like polar cap, slot gap, outer gap and one pole caustic but did not include the striped wind. They also pointed out the importance of joint radio/$\gamma$-ray fit to constrain the geometry. Some refinements of this approach are due to \cite{pierbattista_young_2016}. A complete atlas of gamma-ray pulse profiles for several magnetospheric models, summarizing the pulse properties and merit of each of them can be found in \cite{watters_atlas_2009}.

Other useful constraints on the emission sites come from detailed radio polarization observations. However, these polarization data if coming from millisecond pulsars, are difficult to interpret because of the presence of strong non dipolar fields at the photon production sites. Nevertheless \cite{benli_constraining_2021} were able to put constrains on some of these millisecond pulsars by fitting the time aligned gamma-ray light-curves without resorting to accurate radio pulse profile modelling. Their model is based on accurate dipole force-free magnetosphere simulations. Fortunately, the situation is drastically better for young radio-loud gamma-ray pulsars. Indeed, thank to radio polarization measurements according to the rotating vector model \citep{radhakrishnan_magnetic_1969}, aberration/retardation effects \citep{blaskiewicz_relativistic_1991} help to localize the altitude of radio emission which is about 5\% of the light-cylinder radius for the sample studied in \cite{mitra_nature_2017}. In this paper, we apply the fitting procedure used by \cite{benli_constraining_2021} to young pulsars for which the radio emission height is better constrained and the polarization data reasonably follow the rotating vector model thus relying on a pure dipole field with high confidence.

In this paper, we re-explore the work done by \cite{petri_unified_2011} by using a realistic dipole force-free magnetosphere solution extracted from our numerical simulations. The paper is organized as follow. Sec.~\ref{sec:radio} summarizes the radio observations using polarization data to constrain emission heights and the geometry. Sec.~\ref{sec:Magnetosphere} summarizes the emission properties of the current sheet within the force-free split monopole framework of \cite{michel_rotating_1973} and \cite{bogovalov_physics_1999}. Time-aligned radio and gamma-ray light curves are computed for a bunch of geometric configurations and summarized in several sky maps. Then it generalizes this approach to the more realistic dipole field, smoothly joining the stellar surface to the striped wind and referred as the dipole force-free magnetosphere. Sec.~\ref{sec:Results} shows the results of our fitting procedure for a good sample of young pulsars, constraining their geometry. Conclusions are drawn in Sec.~\ref{sec:Conclusion}.

\section{Radio observations}
\label{sec:radio}

Pulsars are broadband emitters and various frequencies emanating from different parts of the pulsar magnetosphere. The location of the broadband emission is best constrained for the pulsed radio emission which is thought to arise from regions near the neutron star polar cap and the $\gamma$-rays which are thought to arise near the light cylinder. Magnetospheric simulations of pulsars assume a star centred dipolar magnetic field configuration, and are restricted to fast rotating pulsars (roughly 10 msec) due to limitation in numerical resolution and computation time. Thus to compare simulation results with observations, ideally it is best suited to use millisecond pulsars (MSP) and identify emission zones in pulsars where the magnetic field is dipolar. Such studies have been done earlier, see for instance \cite{benli_constraining_2021}, however in the case of MSPs it is difficult to constrain both the location and magnetic field structure in the radio emission region, and as we discuss below the young pulsar population can be used to get significantly better constraints. 

Radio polarization observations are particularly useful in this regard, since the polarization properties can be used to find both the location of the emission site and its magnetic field geometry. The polarization position angle (PPA) of the pulsar linear polarization shows a characteristic S-shape across the pulse profile. The PPA traverse can be interpreted in terms of the rotating vector model (RVM, \cite{radhakrishnan_magnetic_1969}), which states that the PPA traverse reflects the change in the diverging dipolar magnetic field line planes as the pulse profile sweeps past the observer. According to the RVM, the PPA as a function of the pulse phase $\phi$ is given by, 
\begin{equation}
\Psi (\phi) 
= \Psi_{\circ} + \arctan \left( \frac{\sin{\alpha}\sin{(\phi-\phi_{\circ}})} {\sin{\zeta}\cos{\alpha} - \sin{\alpha}\cos{\zeta}\cos{(\phi-\phi_{\circ})}}\right)
\label{eq1}
\end{equation}
where $\alpha$ is the angle between the rotation axis and the dipole magnetic axis, $\beta$ is the angle between the magnetic axis and the observer line of sight and $\zeta=\alpha+\beta$ the angle between the rotation axis and the observer line of sight. $\Psi_{\circ}$ corresponds to the steepest gradient point of the RVM which occurs at the longitude $\phi_{\circ}$, such that 
\begin{equation}
\left. \frac{d\Psi}{d\phi}\right|_{\phi_{\circ}} = \frac{\sin\alpha}{\sin\beta} .
\end{equation}
While in principle fitting eq.~\eqref{eq1} to the PPA traverse in pulsars can constrain the magnetic field geometry $\alpha$ and $\beta$, in practice these parameters are highly correlated, so they remain unconstrained (see e.g. \cite{everett_emission_2001}). Nonetheless a good fit of the observed PPA to the RVM indicates that the pulsar radio emission arises from regions of dipolar magnetic field regions.

In this study we focus on 31 young pulsars with periods longer that 50~msec (see Table~\ref{tab:sample}). Young pulsars are generally known to be highly polarized and their PPA are often consistent with the RVM. Out of the 31 pulsars in our sample, we could obtain archival polarization data for 21 pulsars at 1.4~GHz from \cite{johnston_polarimetry_2018} and \cite{theureau_psrs_2011}. Further we could reliably fit the RVM given by eq.~\eqref{eq1} for 17~pulsars. 
The reduced $\rchi^2$ values for the fits corresponding to the $\alpha$ and $\zeta = \alpha+\beta$ values given in Table~\ref{tab:meilleur_fit} are given in the seventh column in Table~\ref{tab:sample}, and in most cases they indicate that the RVM are good fit to the PPA traverse. The large reduced $\rchi^2$ value for PSR J0908--4913 and J0835--4510 mostly occurs due to certain abrupt changes in the PPA traverse towards the edge of the profiles. Such changes in pulsar average PPA traverse can occur due to orthogonal polarization moding or emission across the profile arising due to a range of heights (see \cite{mitra_effect_2004, mitra_absolute_2007}). However for both these pulsars the overall PPA traverse is consistent with the RVM when these kinky regions are excluded. 

For some pulsars mentioned as `No RVM',  it was not possible to constrain the RVM either due to scattering (PSR~J0248+6021, PSR~J1019--5749, J1730--3350) or due to low polarization (PSR~J1509--5830) or due to extremely flat PPA traverse (PSR~J1016--5857, J1028--5819). 
For several pulsars in our sample, RVM fits and $\rchi^{2}$ contours have been reported by \cite{rookyard_investigating_2015,weltevrede_mapping_2009,kramer_high-precision_2008}, and our
results are in good agreement with these earlier studies. For the cases where RVM fits was possible, it can be concluded that the radio emission arises from regions of dipolar magnetic field lines. While it is desirable to model the polarization for all the pulsars in our sample, in the absence of such data currently however we assume that this conclusion is applicable for our whole sample of young pulsars.

Next we turn our attention to finding the location of the radio emission region. It has been suggested by \cite{blaskiewicz_relativistic_1991} that due to rotation of the pulsar a delay ($\Delta\phi$) is introduced between the center of the pulse profile and the steepest gradient point of the PPA, as a result of aberration/retardation (A/R) effect. This delay, in the linear approximation, is related to the radio emission height~$h$ from the stellar surface and the pulsar period $P$ as $\Delta \phi = 1440 \, h / c \, P$ (deg), where $c$ is the velocity of light. The radio emission heights using the A/R delay has been estimated in a large sample of pulsars by several studies (e.g. \cite{blaskiewicz_relativistic_1991,mitra_comparing_2004, weltevrede_profile_2008}). 
The emission height as a function of pulsar period calculated using the A/R method is found to originate about 500~kilometers above the neutron star surface (see  \cite{mitra_nature_2017}). As the emission height appears to be constant over a wide period range, the value of $\Delta \phi$ is expected to be much larger in younger pulsars than older pulsars. As a result more robust radio emission heights can be obtained in the younger pulsar population.

However, there are certain limitations in applying the A/R delay method for emission height estimates. In order to estimate $\Delta \phi$ the longitude at the leading ($\phi_l$) and trailing ($\phi_t$) edge of the profile is obtained as well as the longitude corresponding to the steepest gradient point ($\phi_{\circ}$) of the PPA traverse from RVM fits. The A/R effect predicts a positive $\Delta \phi$, where $\Delta_\phi = \phi_\circ - (\phi_t - \phi_l)/2$. It has been noted by several studies, like \cite{mitra_comparing_2004, weltevrede_profile_2008} that there are pulsars for which $\Delta \phi$ is negative, and hence do not reflect the A/R effect. The limitations stem from the fact that the A/R method requires $\phi_l$ and $\phi_t$ to correspond to last open magnetic field line which is symmetrically placed with respect to the magnetic axis. However, $\phi_l$ and $\phi_t$ are measured at a few times above the noise level at the edge of the profile, and this can lead to errors due to weak emission near the profile edges. \cite{mitra_toward_2011} showed that single pulse studies can be beneficial to estimate profile edges, as some single pulses can significantly stronger than the average profile with the edge emission being brighter and prominent. In addition, the $\phi_{\circ}$ measurements can also be affected due to the presence of orthogonal polarization modes, and single pulses can be used to model the RVM significantly better by disentangling the orthogonal polarization moding effects.

Our sample pulsars however has only average profile data, and in Table~\ref{tab:sample} we give the $\phi_l$ and $\phi_t$ measured at 5 times the rms level measured in the off pulse region. The fiducial point $\phi_{\circ}$ obtained by the RVM fits are also given in the Table. Using these values we estimated the emission heights~$h$ as shown in Table~\ref{tab:sample}. Reliable $h$ could be estimated for 8 cases, and was found to lie below 10\% of the light cylinder radius, which we assume to be a good estimate for our pulsar sample.

\begin{table*}
	\begin{tabular}{lccccccccc}
		\hline
		PSR	& P	& $\delta$ & $\Delta$ & $\phi_l$ & $\phi_t$ & $\chi^2$ & $\phi_{\circ}$ & $\Delta \phi$ & Height ($h$)\\	
		\hline
		(J2000) &(ms) & &	&  deg        &  deg        &                &  deg        &  deg     &(km) \\
		\hline\hline
		J0248+6021  & 217 & 0.336 $\pm$ 0.017 & --- & -7.7$\pm$0.2&65.4$\pm$0.2              &&NO RVM            & &   \\
		J0631+1036  & 288 & 0.497 $\pm$ 0.022 & --- &-7.7$\pm$0.3 & 14.4$\pm$0.3 &1.16 & 14$\pm$2  &11$\pm$2  & 660$\pm$120  \\
		J0659+1414  & 385 & 0.224 $\pm$ 0.010 & --- &-18.9$\pm$0.4& 15.5$\pm$0.4 &2.05 & 22$\pm$7  &23$\pm$7 & 1884$\pm$561  \\
		J0742--2822 & 167 & 0.627 $\pm$ 0.005 & --- &-12.6$\pm$0.3& 17.5$\pm$0.3 &42.7 & 8$\pm$2   &6$\pm$2  & 208$\pm$69  \\
		J0835--4510  & 89  & 0.129 $\pm$ 0.001 & 0.433 $\pm$ 0.001 &-38.6$\pm$0.2& 37.6$\pm$0.2 &3245 & 4.3$\pm$0.5&5$\pm$1 & 93$\pm$20  \\
		J0908--4913$^{\dagger}$ & 107 & 0.102 $\pm$ 0.005 & 0.501 $\pm$ 0.006 &-15.8$\pm$0.3& 12.3$\pm$0.3 &1466& 3.4$\pm$1  &5$\pm$1 & 111$\pm$22  \\
		J1016--5857 & 107 & 0.143 $\pm$ 0.003 & 0.423 $\pm$ 0.004&-19.3$\pm$0.3& 10.2$\pm$0.3 && NO RVM     &         &   \\
		J1019--5749 & 162 & 0.482 $\pm$ 0.010 & --- &-10.5$\pm$0.3& 52.7$\pm$0.3 && NO RVM     &         &   \\
		J1028--5819 & 91  &  0.195 $\pm$ 0.001 & 0.475 $\pm$ 0.001& -1.8$\pm$0.4&  0.4$\pm$0.3 && NO RVM     & &   \\
		J1048--5832 & 124 & 0.125 $\pm$ 0.001 & 0.426 $\pm$ 0.001&-16.2$\pm$0.3& 20.4$\pm$0.3 &6.6& 3.5$\pm$1  &1$\pm$1  &   \\
		J1057--5226$^{\dagger}$ & 197 & 0.304 $\pm$ 0.003 & 0.307 $\pm$ 0.004&-13.7$\pm$0.3& 23.2$\pm$0.3 &11.09& 0.0$\pm$18&-4$\pm$18&   \\
		J1119--6127 & 408 & 0.285 $\pm$ 0.015 & 0.204 $\pm$ 0.020&-15.1$\pm$0.3& 15.8$\pm$0.3 &1.5  & 26.0$\pm$10&25$\pm$10&2125$\pm$850   \\
		J1357--6429 & 166 & 0.359 $\pm$ 0.028 & --- &-15.8$\pm$0.3& 16.2$\pm$0.3 &1.33& 2.5$\pm$100&2$\pm$100&   \\
		J1420--6048 & 68  & 0.196 $\pm$ 0.011 & 0.312 $\pm$ 0.015 &-39.7$\pm$0.3& 12.3$\pm$0.3 &1.9& -10.7$\pm$3&3$\pm$3  &   \\
		J1509--5850 & 89  & 0.271 $\pm$ 0.011 & 0.264 $\pm$ 0.013&-7.1 $\pm$0.3&  8.1$\pm$0.3 && NO RVM     & &   \\
		J1648--4611 & 165 & 0.261 $\pm$ 0.010 & 0.298 $\pm$ 0.082 &-14.7$\pm$0.3&  9.1$\pm$0.3 &1.6& -9.6$\pm$10&-6$\pm$10&   \\
		J1702--4128 & 182 & 0.397 $\pm$ 0.038 & --- &-11.9$\pm$0.3& 16.5$\pm$0.3 &1.4& -17.9$\pm$30&-20$\pm$30&             \\
		J1709--4429 & 102 & 0.239 $\pm$ 0.001 & 0.244 $\pm$ 0.002&-30.9$\pm$0.3& 29.5$\pm$0.3 &1.07& 10.3$\pm$3 &11$\pm$3 &233$\pm$64   \\
		J1718--3825 & 75  & 0.397 $\pm$ 0.009 & --- &-10.9$\pm$0.3& 38.6$\pm$0.3 &1.26& 21.4$\pm$5 & 9$\pm$5 &140$\pm$78   \\
		J1730--3350 & 139 & 0.128 $\pm$ 0.007 & 0.419 $\pm$ 0.007 &-11.6$\pm$0.3& 50.8$\pm$0.3 && NO RVM     & &   \\
		J1747--2958 & 99  & 0.181 $\pm$ 0.003 & 0.392 $\pm$ 0.005&NRP          &              &&            & &   \\
		J1801--2451 & 125 & 0.060 $\pm$ 0.005 & 0.496 $\pm$ 0.020 &-10.4$\pm$0.3& 10.9$\pm$0.3 &0.87& -8.6$\pm$6 &-9$\pm$6&   \\
		J1835--1106 & 166 & 0.139 $\pm$ 0.006 & 0.421 $\pm$ 0.011 &-11.2$\pm$0.3& 11.2$\pm$0.3 &1.37& 4.2$\pm$5  &4$\pm$5 &   \\
		J1907+0602 & 107 & 0.209 $\pm$ 0.003 & 0.389 $\pm$ 0.004&NRP          &              &&            & &   \\
		J1952+3252  & 39  & 0.161 $\pm$ 0.002 & 0.478 $\pm$ 0.003&NRP          &              &&            & &   \\
		J2021+3651  & 104 & 0.132 $\pm$ 0.001 & 0.478 $\pm$ 0.001&NRP          &              &&            & &   \\
		J2030+3641  & 200 & 0.269 $\pm$ 0.010 & 0.309 $\pm$ 0.014 &NRP          &              &&            & &   \\ 
		J2032+4127  & 143 & 0.099 $\pm$ 0.001 & 0.516 $\pm$ 0.001&NRP          &              &&            & &   \\
		J2043+2740  & 96  & 0.132 $\pm$ 0.007 & 0.432 $\pm$ 0.010&NRP          &              &&            & &   \\
		J2229+6114  & 52   & 0.187 $\pm$ 0.007 & 0.299 $\pm$ 0.008&NRP          &              &&            & &   \\
		J2240+5832  & 140 & 0.118 $\pm$ 0.014 & 0.476 $\pm$ 0.014&-8.6$\pm$0.2  &7.4$\pm$0.2   &1.93&3.7$\pm$10 & 4$\pm$10 &    \\
		\hline
	\end{tabular}
	\caption{Radio profile of Young radio-loud gamma-ray pulsars extracted for the Fermi second 
		pulsar catalogue. The data for PSR J2240+5832 and PSR J0248+6021 has been obtained from \cite{theureau_psrs_2011}. The rest of the data has been obtained from \cite{johnston_polarimetry_2018}, and the abbreviation 
		`NRP' above stands for cases where no radio profile was available for analysis. The abbreviation
		`NO RVM' correspond to cases where the RVM fit was not possible for the data. The pulsar with superscript $\dagger$ are
                 interpulsars, where the $\phi_{\circ}$, $\Delta\phi$ and $h$ in the table
                 is estimated for the region below the main pulse.
		\label{tab:sample}}
\end{table*}

\section{Split monopole versus dipole magnetosphere}
\label{sec:Magnetosphere}

Computing multi-wavelength light-curves relies on some magnetosphere models. In this paper, we exclusively consider force-free models based on either a split-monopole or a dipole magnetic field. Before showing the results of the dipole magnetosphere fitting the observations, it is instructive to compare the split-monopole expectations to the more realistic dipole field.

\subsection{Split monopole}

The split monopole is a simple but elegant exact 3D force-free solution of a neutron star magnetosphere at large distances~$r \gg \rlight$, well outside the light-cylinder, in the wind zone. Note however that the magnetic field strength decreases only like a monopole that is with $B \propto r^{-2}$ and not like $B \propto r^{-3}$ as for a dipole field. This is of no concern in the present study because we focus essentially on geometrical properties and not on the energetics related to the electromagnetic field strength or dissipation and the associated particle dynamics. The most important feature of the split monopole is the position of its equatorial current sheet. This infinitely thin sheet is actually accurately localized by a simple expression given by a two-dimensional surface expressed in spherical polar coordinates $(r,\vartheta,\varphi)$ as
\begin{equation}
\label{eq:Rs}
r_{\rm s}(\vartheta,\varphi,t) = \beta_{\rm v} \, \rlight \, \left[ 
\pm \arccos ( - \cot\vartheta \, \cot\chi) + \frac{c\,t}{\rlight} - 
\varphi + 2\,\ell\,\pi \right]
\end{equation}
where $\Omega$ is the stellar rotation rate, $c$ the speed of light, $\alpha$ the inclination of the split monopole, $\beta_{\rm v} = V/c$ the wind speed, $\rlight=c/\Omega$ the radius of the light cylinder, $t$ the time as measured by a distant observer at rest, and $\ell$ an integer. The current sheet is connected to the stellar surface by monopolar magnetic field lines. To a very good approximation, we assume that the wind moves radially at exactly the speed of light, $V=c$.

The simultaneous time-aligned radio and gamma-ray pulse profile evolution with the geometric configuration has been extensively computed by \cite{petri_unified_2011}. The main features of this emission was a radio time lag~$\delta$ connected to the gamma-ray peak separation~$\Delta$ (if both gamma peaks are visible) expressed by
\begin{equation}
\label{eq:delai_radio}
\delta \approx \frac{1-\Delta}{2} .
\end{equation}
Moreover, the gamma-ray peak separation~$\Delta$ depends only on $\alpha$ and the inclination of the line of sight~$\zeta=\alpha+\beta$. These parameter were found to be related by
\begin{equation}
\label{eq:SeparationPic}
\cos(\pi\,\Delta) = |\cot \zeta \, \cot \alpha|.
\end{equation}
These expressions have been derived analytically with some approximations detailed in \cite{petri_unified_2011}.
According to eq.~\eqref{eq:delai_radio}, the gamma-ray peak separation~$\Delta$ is not independent of the radio time lag~$\delta$. This formula is actually a simple consequence of the geometrical behaviour and symmetries of the striped wind emission properties related to the polar cap radio emission. It assumes that the radio emission emanates from deep within the light-cylinder. In reality, as we will show, we must add an additional delay due to the variable radio emission height from pulsar to pulsar. Eq.~\eqref{eq:SeparationPic} correlates $\alpha$ and $\zeta$ depending on the peak separation~$\Delta$ independently of the radio properties. Here again, this formula is derived from pure geometrical considerations related to the current sheet structure as given by eq.~\eqref{eq:Rs}.

\begin{figure}
\centering
\includegraphics[width=\columnwidth]{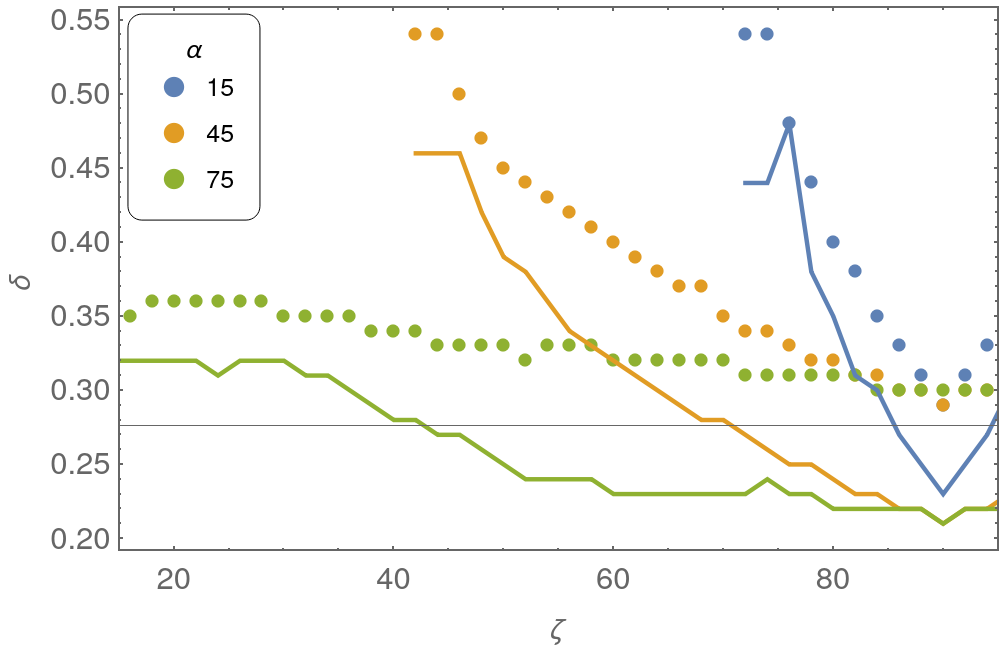}	
\caption{Time lag~$\delta$ between the radio and the closest gamma-ray peak for the split monopole model in dotted points and for the dipole model in solid curves, for $\alpha = \{15\degr,45\degr,75\degr$\}.\label{fig:delai_radio}}
\end{figure}

\begin{figure}
\centering
\includegraphics[width=\columnwidth]{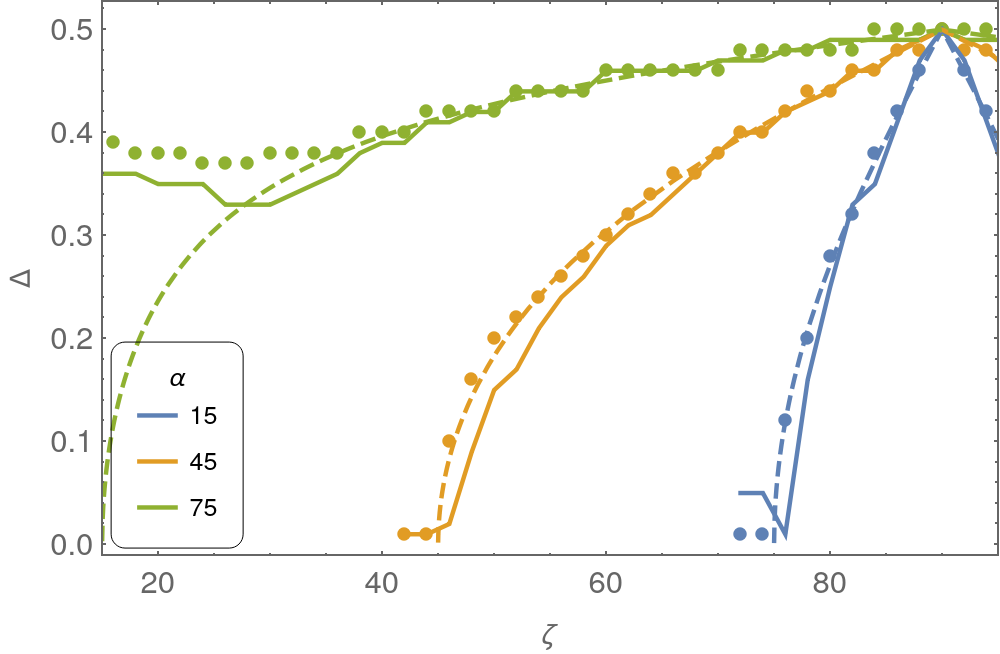}
\caption{Gamma-ray peak separation~$\Delta$ for the split monopole model in dotted points and for the dipole model in solid curves for $\alpha = \{15\degr,45\degr,75\degr\}$. The dashed lines represent the expectations from eq.~\eqref{eq:SeparationPic}. \label{fig:gamma_pic_separation}}
\end{figure}

In the present work, we compute numerically the radio and gamma-ray light curves assuming a Gaussian beam shape around the polar cap for radio emission up to the actual emission height at approximately~$0.05\,\rlight$ and a thin current layer around the current sheet depicted by eq.~\eqref{eq:Rs} for several inclination angles~$\alpha$. The results for the radio time lag~$\delta$ are shown in coloured dotted points in Fig.~\ref{fig:delai_radio} for $\alpha = \{15\degr,45\degr,75\degr$\}. The associated gamma-ray peak separation~$\Delta$ is shown in coloured dotted points in Fig.~\ref{fig:gamma_pic_separation}. The latter figure shows the good agreement between our simulations and the analytical expectation in eq.~\eqref{eq:SeparationPic}, also shown in dashed coloured lines. The solid lines correspond to the dipole model, see below.

\begin{figure}
\includegraphics[width=\columnwidth]{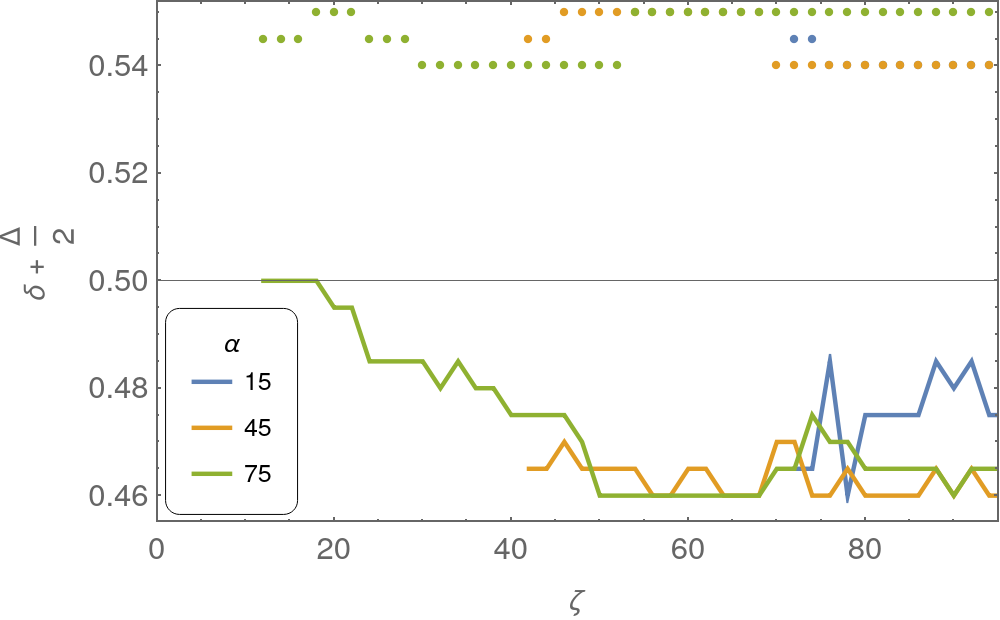}
\caption{The sum $\delta+\Delta/2$ for the split monopole model in dotted point and the dipole model in solid curves for $\alpha = \{15\degr,45\degr,75\degr\}$. The solid line shows the expect theoretical value of $1/2$.\label{fig:fit_delai}}
\end{figure}

As a check of the accuracy of relation eq.~\eqref{eq:delai_radio}, we plotted the sum $\delta+\Delta/2$ in coloured dotted points in Fig.~\ref{fig:fit_delai}. It always lies around $0.55$ and therefore remaining close to the expected value of $0.5$ whatever the geometry of the magnetosphere (arbitrary $\zeta$ and $\alpha$).

\subsection{Dipole magnetosphere}
\label{sec:FFE_Magnetosphere}

The split monopole gives a good first guess to the structure of the striped wind. However, it does not connect properly the stationary region in the vicinity of the stellar surface to the wave zone outside the light-cylinder. A more realistic case must take into account the dipole nature of the field inside the light-cylinder and not a split monopole. Therefore, in order to estimate the discrepancy between the split monopole results shown in the previous section and the physical situation of a true magnetosphere, we use the solutions from force-free numerical simulations of pulsar magnetospheres, treating self-consistently the electromagnetic field, and relying on our previous publications in \cite{petri_pulsar_2012}.

However, we emphasize that young radio pulsars with periods above several tenths of milliseconds are difficult to model numerically because the ratio between the neutron star radius~$R$ and the light-cylinder radius~$\rlight$ is very small $a = R/\rlight \ll 1$. Because the simulation box must resolve all scales from the neutron star size to the light-cylinder length, numerical simulations require very high resolutions in 3D rendering it impossible to reckon the electromagnetic field with a decent computational time. Therefore in all simulations the ratio $R/\rlight$ is artificially increased to millisecond periods in order to get tractable runs. However such large ratios will not significantly impact on the global magnetospheric geometry because variations are expected to scale as $(R/\rlight)^2$. For instance in the Deutsch solution \citep{deutsch_electromagnetic_1955}, the spindown corrections introduces a factor $(1-a^2)$, meanwhile the polar cap size decreases as $\sqrt{a}$ without significant changes in their shape (homothetic transformations), see for instance \cite{petri_general-relativistic_2018}. Therefore, without loss of precision, we can use a ratio $a=0.2$ as done in our force-free runs to compute young pulsars emission properties to good accuracy. An additional time lag can be added if necessary due to time of flight propagation effects.

We constructed a set of pulsar dipole magnetospheres with $a=0.2$ and obliquities $\alpha$ ranging from $0\degr$ to $90\degr$ in steps of $5\degr$. Then we computed the polar cap shapes, localizing the last open field lines as well as the current sheet outside the light-cylinder. The observer line of sight $\zeta$ ranges from $0\degr$ to $180\degr$ in steps of $2\degr$.

\begin{figure}
	\centering
	\begin{tabular}{cc}
		\includegraphics[width=0.5\linewidth]{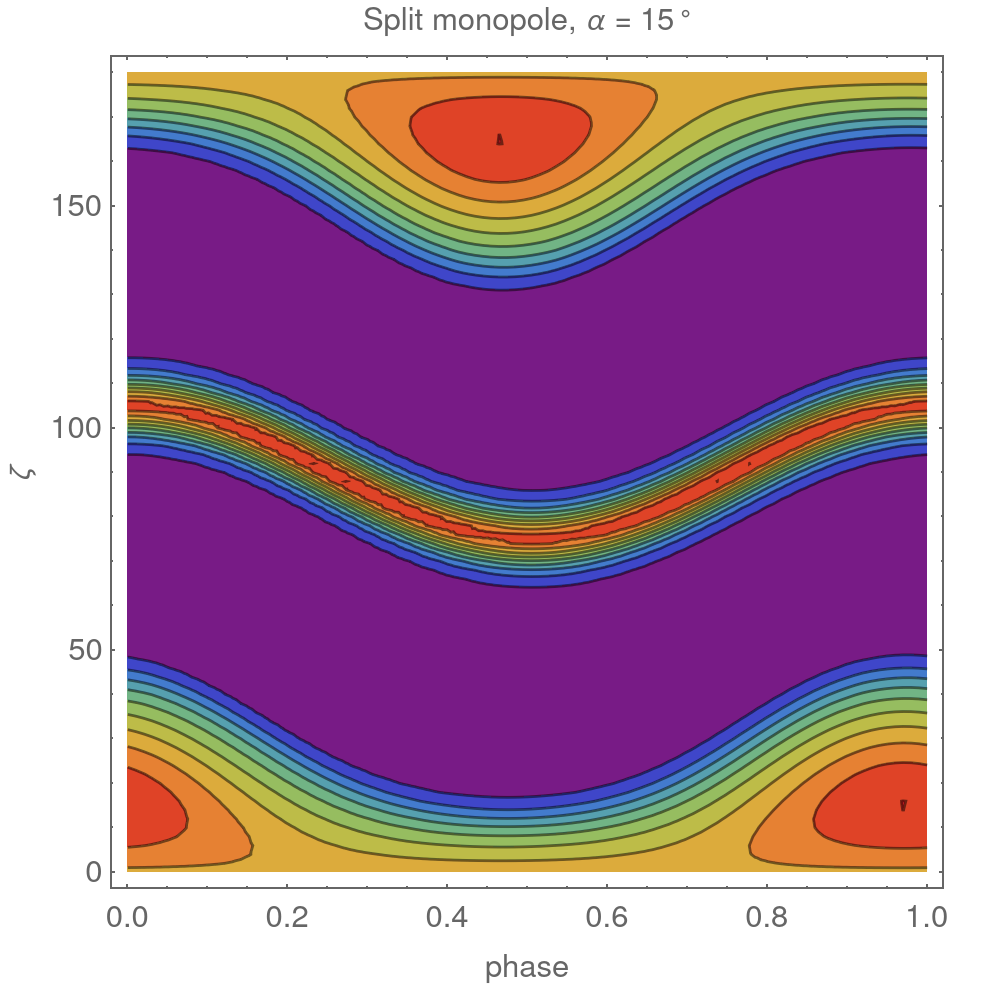} &
		\includegraphics[width=0.5\linewidth]{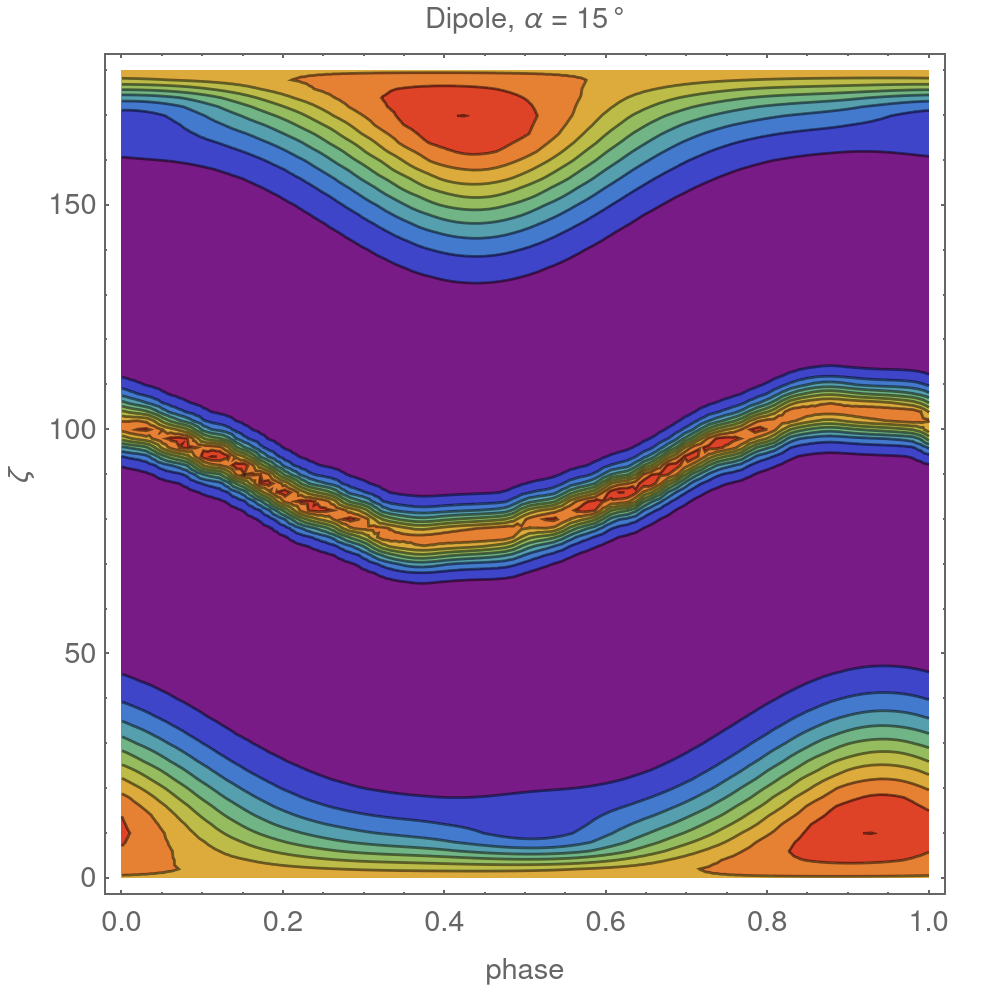} \\
		\includegraphics[width=0.5\linewidth]{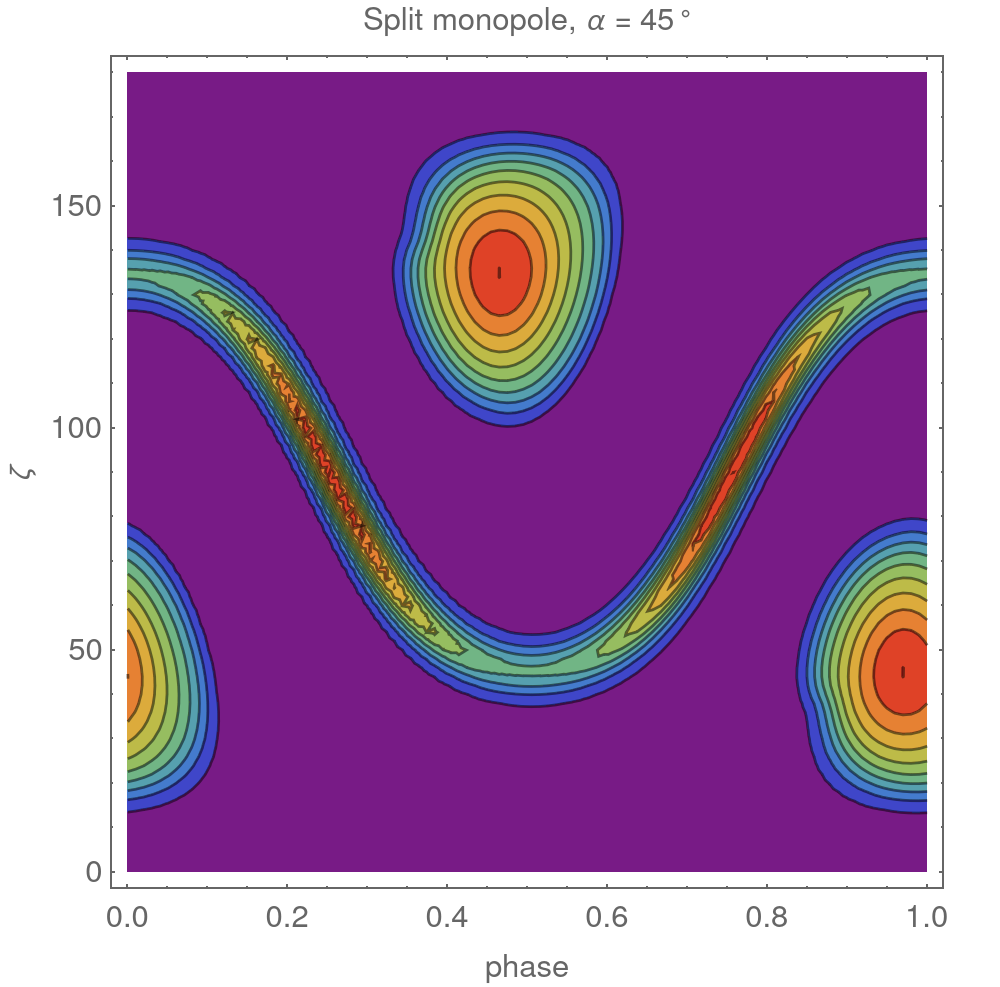} &
		\includegraphics[width=0.5\linewidth]{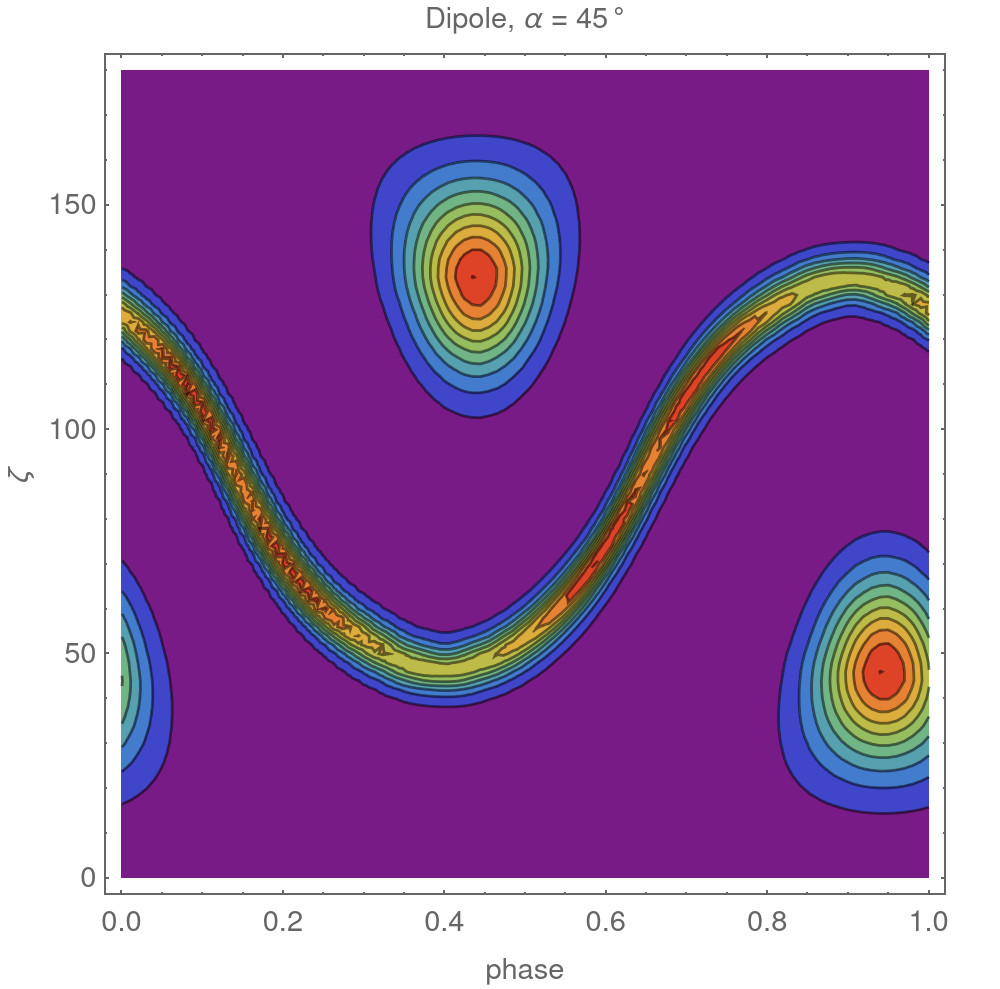} \\
		\includegraphics[width=0.5\linewidth]{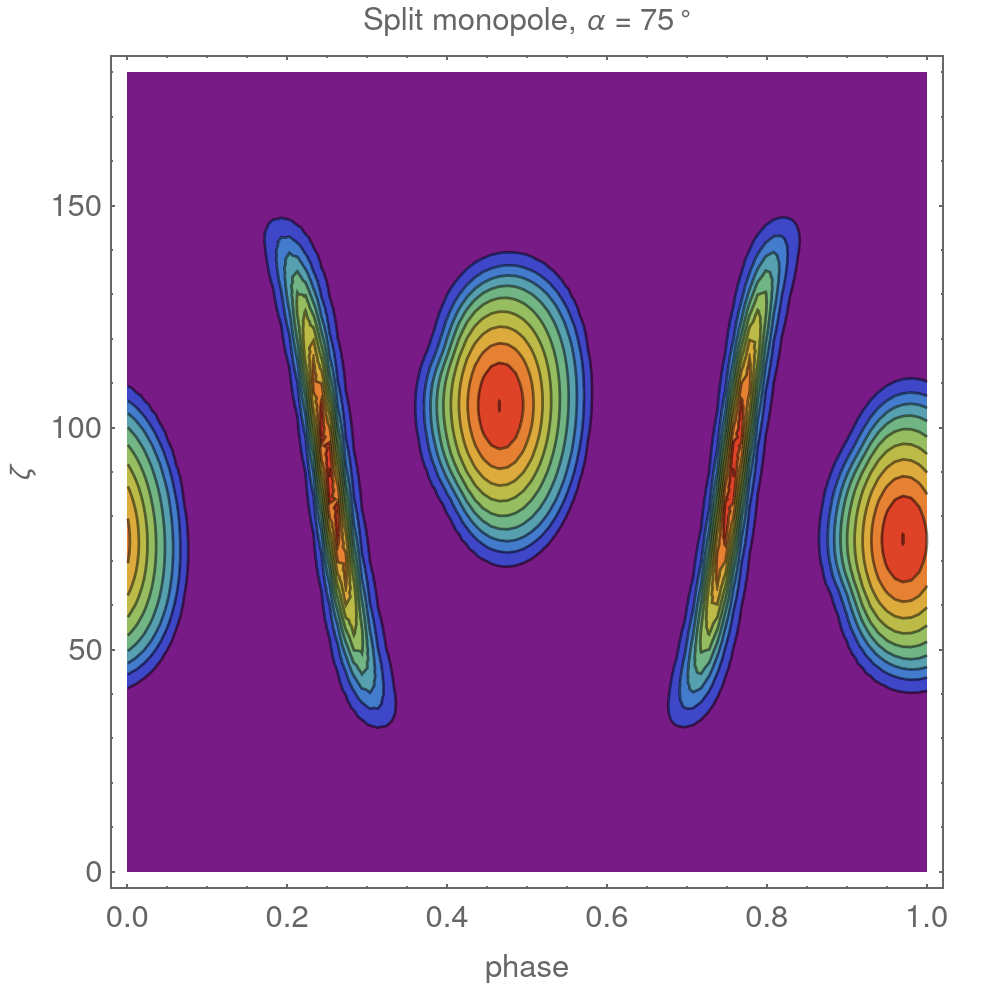} &
		\includegraphics[width=0.5\linewidth]{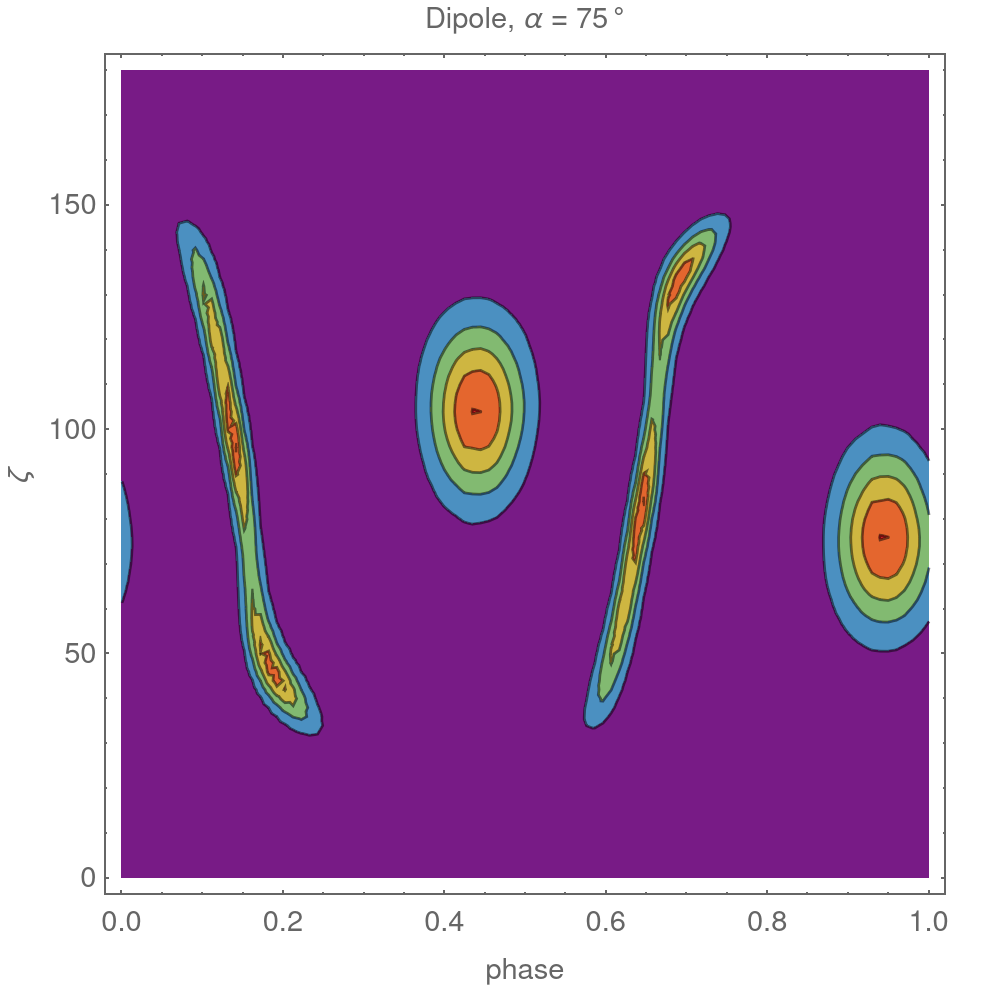}
	\end{tabular}
	\caption{Gamma-ray sky maps for the split monopole, left column, and the dipole magnetosphere, right column for $\alpha=15\degr,45\degr,75\degr$.}
	\label{fig:FFE_vs_Monopole_skymap}
\end{figure}

Some relevant sky maps for split monopole and dipole magnetospheres are shown in Fig.\ref{fig:FFE_vs_Monopole_skymap}.

As done in the previous section for the split monopole, we computed the radio time lag~$\delta$ as shown in coloured solid lines in Fig.~\ref{fig:delai_radio}. The associated gamma-ray peak separation~$\Delta$ is plotted in coloured solid lines in Fig.~\ref{fig:gamma_pic_separation}. Here again, we found a good agreement between the dipole model and the analytical expectations in eq.~\eqref{eq:delai_radio} and eq.~\eqref{eq:SeparationPic}.
We finally also checked the deviation for the simple law eq.~\eqref{eq:delai_radio} by computing $\delta+\Delta/2$ for all configurations. Remarkably we found only a small deviation with a value between 0.46 and 0.5 instead of the theoretical value of~0.5, see the coloured solid lines in Fig.~\ref{fig:fit_delai}.

Consequently, we have a simple tool to quickly guess the geometry of any radio loud gamma-ray pulsar by measuring its radio lag~$\delta$ and gamma-ray peak separation~$\Delta$. However, the angles $\zeta$ and $\alpha$ remain degenerate because a continuum of couples~$(\zeta,\alpha)$ give the same results. In order to leave the degeneracy, we must scrutinize individually each pulsar by fitting its gamma-ray light curve. Then as an a posteriori check, we verify its compatibility with measurements of the radio polarization position angle. This helps to drastically narrow down uncertainties in the geometrical configuration.

Moreover the emission height although situated at about 5\% of $\rlight$ is not firmly constrained. There exist still a slight freedom to shift the radio time lag to the leading or trailing direction depending on the exact location with respect to our fiducial point. Indeed, in our simulations, we assumed a radio beam radiated in the radial direction at a distance~$h_0$ from the stellar centre. This height has been numerically fixed to $h_0/\rlight=0.2$ for the dipole simulations. If the radio emission emanates from a distance~$h_1$ from the stellar centre, the time of flight delay compared to the fiducial altitude~$h_0$ is
\begin{equation}\label{eq:delta_t}
\Delta t = \frac{h_0-h_1}{c}
\end{equation}
corresponding to a phase shift in the light curve amounting to
\begin{equation}\label{eq:phi_r}
\phi_r = \frac{h_0-h_1}{2\,\pi\,\rlight}.
\end{equation}
Because the radio pulse profile is taking as phase zero for synchronisation purposes, the gamma-ray light-curves move in the opposite direction, to earlier phases with respect to the radio pulse profile, therefore $\phi_s=-\phi_r$. Consequently, allowing emission deeper within the magnetosphere $h_1<h_0$ shifts the gamma-ray pulse profile to earlier phases with a negative additional delay $\phi_s<0$ compared to our simulated time-aligned gamma-ray profiles. In the opposite case of higher emission altitudes $h_1>h_0$, the gamma-ray pulse profile shifts to later phases with a positive delay $\phi_s>0$. Therefore, in all of our fits, we added an offset phase~$\phi_s$ in order to take such uncertainty into account, as well as a possible miss in the data of the middle of the radio pulse profile. See also the discussion in \cite{benli_constraining_2021}. Knowing that emission heights are about $h_1\approx 0.05\,\rlight$, this offset is expected to be around $\phi_s \approx -0.15/2\pi \approx -0.02$. As can be computed from eq.~\eqref{eq:phi_r}, the phase shift induced by uncertainties in the radio emission height is weak, at most 2\% of the period. Including aberration and/or altitude dependent magnetic field sweep back \citep{phillips_radio_1992} will only at most double or triple this value. The good news is that we do not need an accurate location of the radio emission site. The bad news is that larger shifts, as we will found in our fittings requires another ingredients to justify 10\% or 15\% shift in the period. One possibility is to move the emission from the striped wind to larger distances, not starting right at the light-cylinder but at twice or three times $\rlight$. Indeed, shifting from $1\,\rlight$ to $2\,\rlight$ introduces a time lag (actually an advance in time corresponding to a shift to earlier phases) of approximately $1/2\,\pi\approx0.16=16$\% of the period.

\begin{figure}
	\includegraphics[width=\columnwidth]{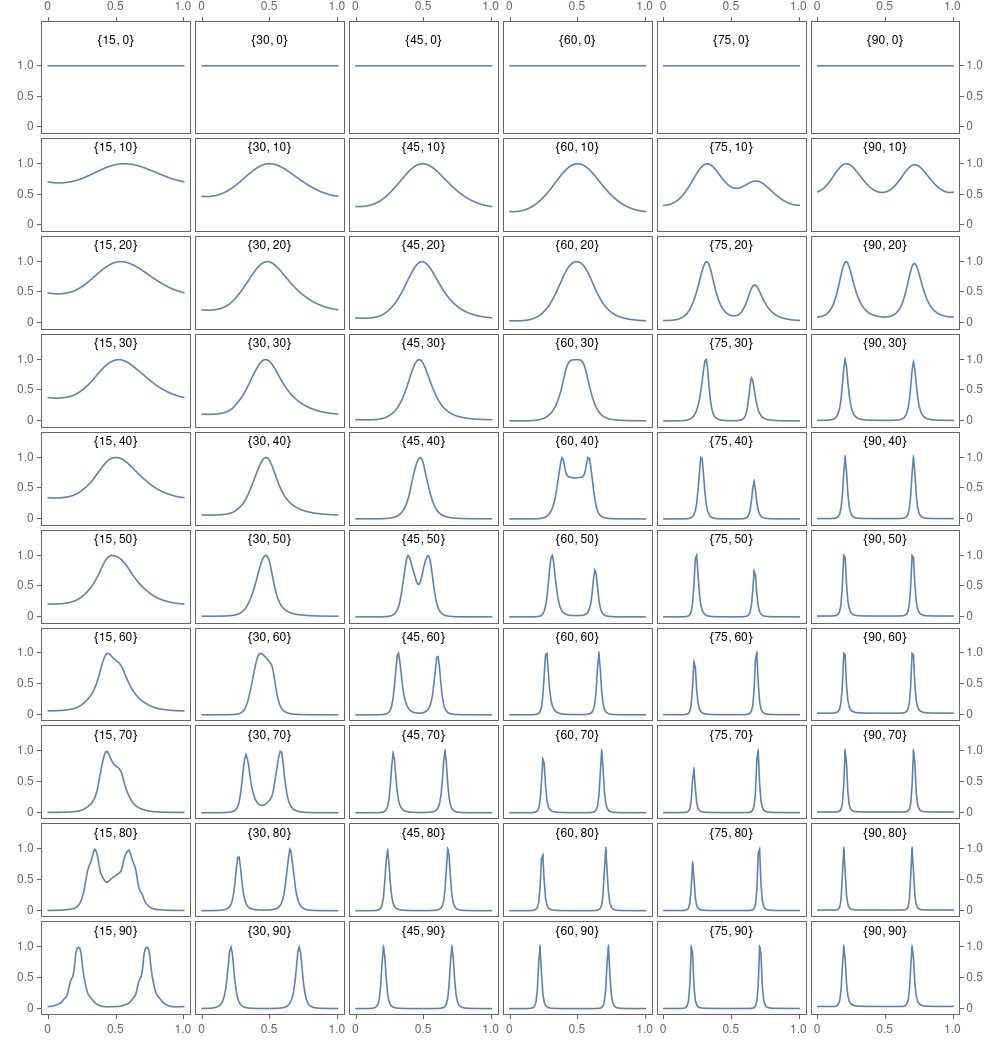}
	\caption{Atlas of gamma-ray light curves for $\alpha=\{15\degr,30\degr,45\degr,60\degr,75\degr,90\degr\}$ from left to right column and $\zeta=\{0\degr,...,90\degr\}$ from top to bottom line with a step of $10\degr$ in the format $\{\alpha,\zeta\}$.\label{fig:atlas}}
\end{figure}

In order to summarize all possible gamma-ray light curves, single peaked or double peaked, an atlas is shown in Fig.~\ref{fig:atlas} with the full range of obliquities~$\alpha$ and line of sight~$\zeta$. Note that all intensities are normalized to unity but in reality, we expect much fainter radiation when the observer line of sight does not cross or only grazes the current sheet in the wind. Due to the symmetry of the dipole, we do not show the south hemisphere light-curve atlas with either $\alpha>90\degr$ or $\zeta>90\degr$. Indeed, the radio and gamma-ray sky maps highlight a north south symmetry meaning that the configuration $(\alpha,\zeta)$ produces exactly the same light-curves as the symmetrical configuration $(\pi-\alpha,\pi-\zeta)$. Another important symmetry connects $(\alpha,\zeta)$ to $(\alpha,\pi-\zeta)$, the latter showing a light curve identical to the former except for a shift in phase of half a period. Therefore from the knowledge of the sky maps for the obliquity ranges $(\alpha,\zeta) \in [0\degr, 90\degr]^2$ only, we are able to produce any light-curve whatever $(\alpha,\zeta) \in [0\degr, 180\degr]^2$. We stress that this symmetry is broken when fitting the radio PPA. Very high quality polarization data are able to left the degeneracy from the gamma-ray sky maps, pinning down the angles to small uncertainties. These conclusions reveal primordial when studying pulsars for which we expect $\alpha>90\degr$ such as J0742-2822, J0908-4913 and J1702-4128. We will however use $\alpha \leq 90\degr$ to adjust gamma-ray light-curves even if $\alpha$ is constrained to be larger than $90\degr$ from RVM thanks to this symmetry.

Several kind of profiles are produced. Asymmetric single peaks are obtained for weak inclination angles $\alpha$ and $\zeta$, upper left part of the atlas. When moving downwards to the right, an unresolved double peak structure appears with two overlapping peaks showing a kind of bridge emission. For the largest angles $\alpha$ and $\zeta$, lower right part of the atlas, the two peaks are well separated. We note also that the dominant peak is either the first or the second depending on the observer line of sight. For instance, the case $\alpha=75\degr$, fifth column, starts with a dominant first peak becoming weaker when the observer looks through the equator, for $\zeta>50\degr$.

\begin{figure}
	\centering
	\includegraphics[width=\linewidth]{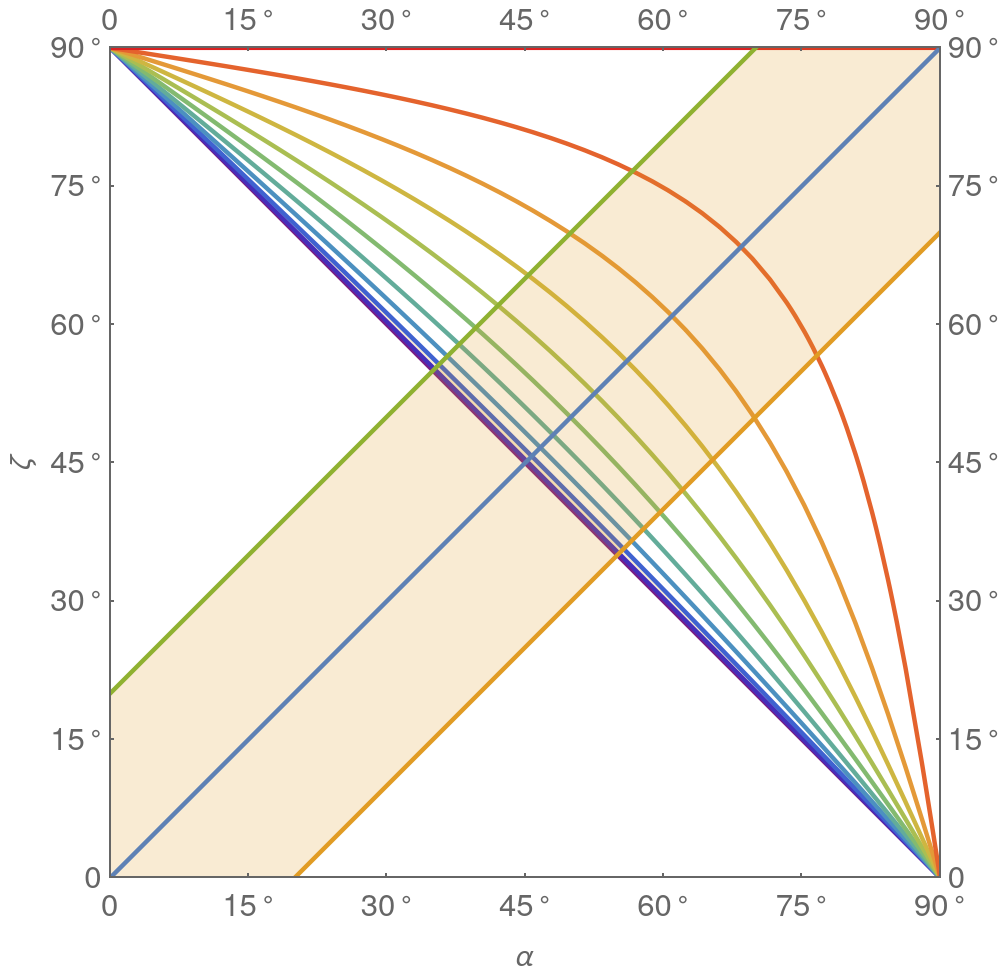}
	\caption{Isocurves of constant gamma-ray peak separation~$\Delta$ depending on $\alpha$ and $\zeta$. The orange shaded area corresponds to observable radio emission with beam half opening angle $\theta_{\rm em}=20\degr$.}
	\label{fig:separationgammaalphazeta}
\end{figure}

\begin{figure}
	\centering
	\includegraphics[width=\linewidth]{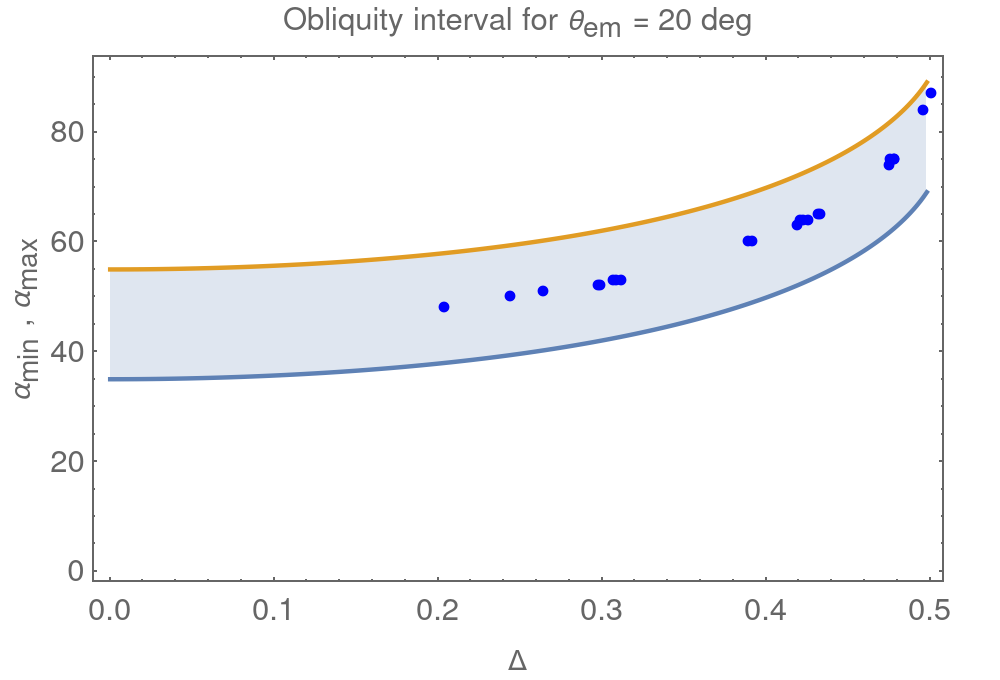}
	\caption{Constrain on the obliquity for a radio-loud gamma-ray pulsar with two peaks separated by~$\Delta$. The blue points correspond to the results of the fits performed on the pulsar sample chosen in this work, see section~\ref{sec:Results}.}
	\label{fig:alpha_min_max}
\end{figure}

The simultaneous observation of radio and gamma-ray pulses is conditioned to the line of sight crossing the radio emission cone. Assuming the formula for a static dipole and setting the emission height at a distance $r$ from the stellar centre, the half opening angle of this cone is
\begin{equation}\label{eq:emission_cone_half_angle}
\theta_{\rm em} = \frac{3}{2} \, \theta_{\rm pc} \approx \frac{3}{2} \, \sqrt{\frac{R}{\rlight}} \approx 1.3\degree \, \left( \frac{P}{1~s}\right)^{-1/2} .
\end{equation}
Actually, the radio emission escapes not from the polar caps for young pulsars, but at a substantial height above the stellar surface, around $r\approx 0.05 \, \rlight$ \citep{mitra_nature_2017}. The half opening angle is therefore insensitive to the period and equals 
\begin{equation}
\theta_{\rm em} = \frac{3}{2} \, \sqrt{\frac{r}{\rlight}} \approx 20\degree  .
\end{equation}
This means in other words that the line of sight must not deviate more than $\theta_{\rm em}$ from the magnetic moment axis, $\zeta \in [\alpha-\theta_{\rm em}, \alpha + \theta_{\rm em}]$ or for the angle~$\beta \in [-\theta_{\rm em}, + \theta_{\rm em}]$. Meanwhile, for gamma-rays to be visible, we impose $90\degree - \alpha \lesssim \zeta \lesssim 90\degree + \alpha $.
A summary of relevant angles in the $(\alpha,\zeta)$ plane is shown in Fig.\ref{fig:separationgammaalphazeta}, related the variation in $\zeta$ to the variation in $\alpha$ for a fixed gamma-ray peak separation~$\Delta$. The orange shaded area delimits the region where radio pulse profiles are detected according to the cone opening angle $\theta_{\rm em}$.
Radio-loud gamma-ray pulsars are located in the upper right part of this shaded area, for angles $\alpha \gtrsim 45\degr$. Actually, for each pulsar with known $\Delta$, we can constraint the obliquity $\alpha$ by setting an interval $[\alpha_{\rm min}, \alpha_{\rm max}]$ as shown in Fig.~\ref{fig:alpha_min_max}. Higher peak separations imply higher obliquities, tending towards 90\degr. The blue points correspond to the results of the fits performed in section~\ref{sec:Results}.

Radio loud single gamma-ray peak pulsars are seen when the observer line of sight is grazing the edges of the current sheet within the striped wind. This occurs whenever $\alpha+\zeta \approx 90\degr$. Moreover his line of sight must cross the radio beam therefore $|\zeta-\alpha|\lesssim \theta_{\rm em}$. This puts severe constraints on $\alpha$, namely $|\alpha-45\degr|\lesssim\theta_{\rm em}/2$. In our case with $\theta_{\rm em}=20\degr$ we get $(\alpha, \zeta) \in [35\degr, 55\degr]^2$ which corresponds to the area around $\Delta=0$ in Fig.~\ref{fig:alpha_min_max}.

\section{Results}
\label{sec:Results}

In this section, we describe our fitting method, the young pulsar population used in our study and eventually discuss the results of the best geometry within the combined gamma-ray striped wind and radio rotating vector model. Implications for the emission sites are also discussed.

To keep the gamma-ray emission model as simple as possible, we use exactly the same size for the current sheet emission for all pulsars, integrating photon emissivity in a spherical shell comprise between the radius $r=\rlight$ and $r=3\,\rlight$. We remind moreover that these gamma-rays are emitted tangentially to the current sheet in its rest frame. But due to Lorentz boosting to the observer frame, this radiation is directed almost radially for that observer.

\subsection{Fitting method}

Our fitting method closely follows the technique used by \cite{benli_constraining_2021}. The important features to be matched are the radio/gamma-ray time lag and the gamma-ray peak separation (if both peaks are visible) and the gamma-ray light-curve profiles. The precise radio pulse profile is irrelevant to our study because we do not investigate in depth the radio emission mechanism. We only require an estimate of its emission altitude and assume a Gaussian shape to accurately localise the radio peak phase taking by definition as phase zero. Most importantly, we fit as properly as possible the time-aligned gamma-ray light curves in accordance with the radio peak synchronisation. According to the pulsar gamma-ray catalogue \citep{abdo_second_2013}, the synchronisation performed by the Fermi/LAT collaboration varies from pulsar to pulsar for several reasons, mainly because the determination of the centre of the radio pulse profile is problematic. In our investigations, we do not suffer from such indeterminacy because we take the plane formed by the magnetic axis and the rotation axis as a fiducial plane which has phase zero by convention. In such a way we get a homogeneous fitting procedure for all pulsars in our sample. The $\rchi^2$ introduced for the gamma-ray light curve fitting is expressed as
\begin{equation}
\label{eq:chi2}
\rchi^{2} = \sum_i \frac{\left(I^\mathrm{obs}_\mathrm{i}-I^\mathrm{model}_\mathrm{i}\right)^{2}}{\sigma^2_\mathrm{i}} 
\end{equation}
where 
$I^\mathrm{obs}_\mathrm{i}$ is the observed gamma-ray intensity, $\sigma^2_\mathrm{i}$ its associated error for the $i^\mathrm{th}$ phase bin, and $I^\mathrm{model}_\mathrm{i}$ the model intensity at the same observational bin. As the observational phase bins do not coincide with the theoretical phase bins, we interpolate the theoretical light-curves at the observational phase bins.

\subsection{Pulsar sample}

Our sample of young and radio-loud gamma-ray pulsars is guided by the existence of good quality gamma-ray light-curves and if possible in conjunction with good radio polarization data in order to fit the polarization position angle (PPA) with the rotating vector model. Our choice implies to pick out pulsars with periods above approximately 30~ms in order to ensure radio photon production at high altitude above the polar cap where the dipole magnetic field approximation holds accurately. The aberration/retardation effect measured in those pulsars indeed constrains the emission height to a fraction of the light-cylinder radius. Bearing in mind all these constraints, we arrive at a reasonable sample of 31~pulsars summarized in Table~\ref{tab:sample}.
The pulsar period ranges from 39~ms to more than 400~ms. Except for a few of them, actually 7, they all show a double gamma-ray pulse profile with $\Delta$ in the range 0.2-0.5. The gamma-ray peak time lag goes from 0.06 to 0.63.

\subsection{Joined RVM and gamma-ray fits}
\label{jrandg}

We start with the sub-sample of pulsars having a reasonable RVM fit to constrain the two angles $\alpha$ and $\beta=\zeta-\alpha$. The gamma-ray light-curves are extracted from the second pulsar catalogue \citep{abdo_second_2013}.

For all these pulsars, we show in a same figure first the radio pulse profile with the best RVM fit, then the radio and gamma-ray $\rchi^2$ fit and eventually the best radio and gamma-ray light-curves predictions compared to observations. Let us shortly go through all these pulsars.

\begin{figure}
	\centering
	\begin{tabular}{l}
	\includegraphics[width=\columnwidth,angle=-90]{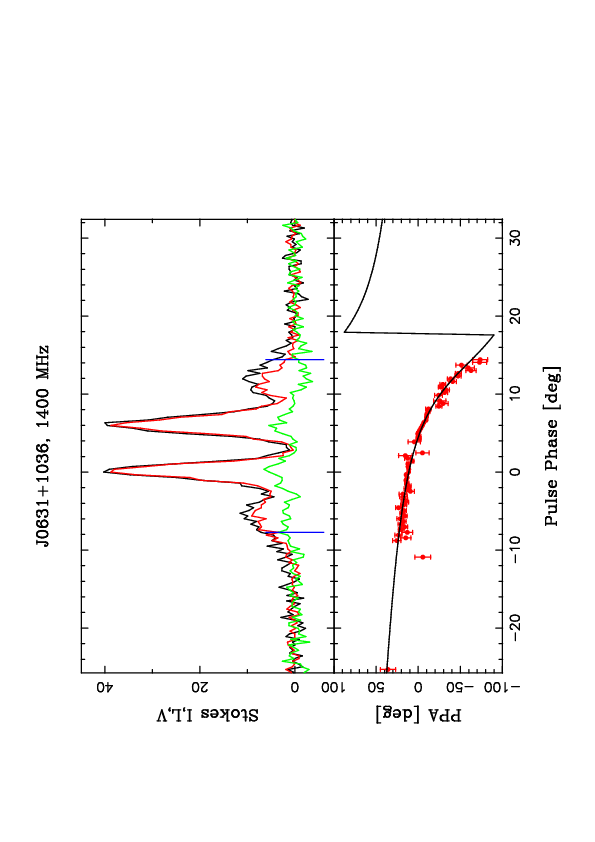} \\
	\includegraphics[width=\columnwidth]{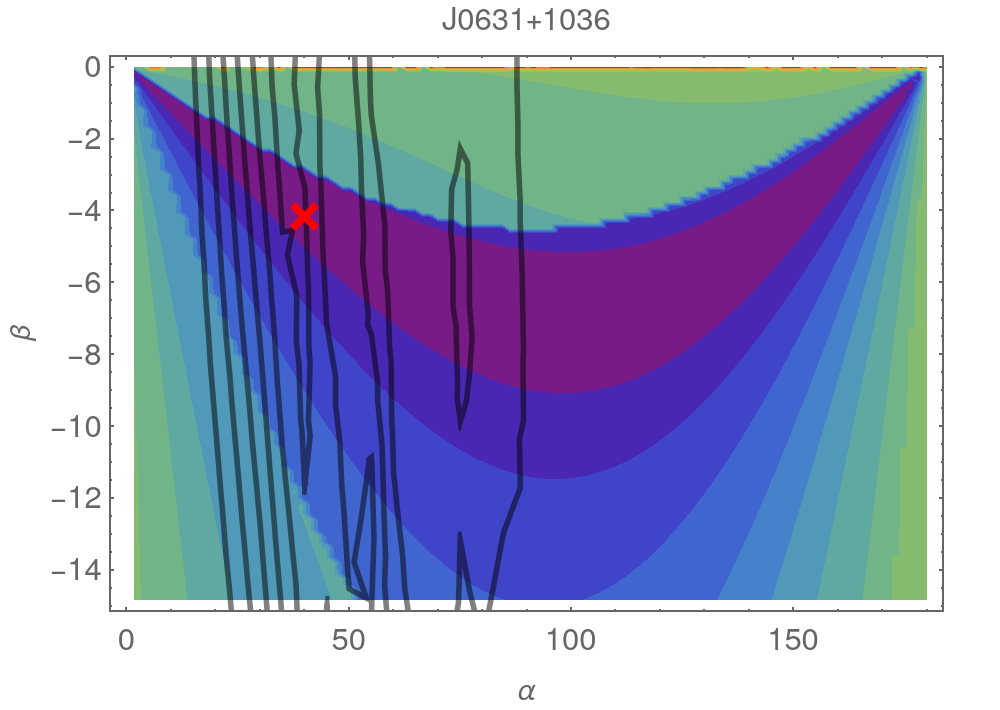} \\	
	\includegraphics[width=\columnwidth]{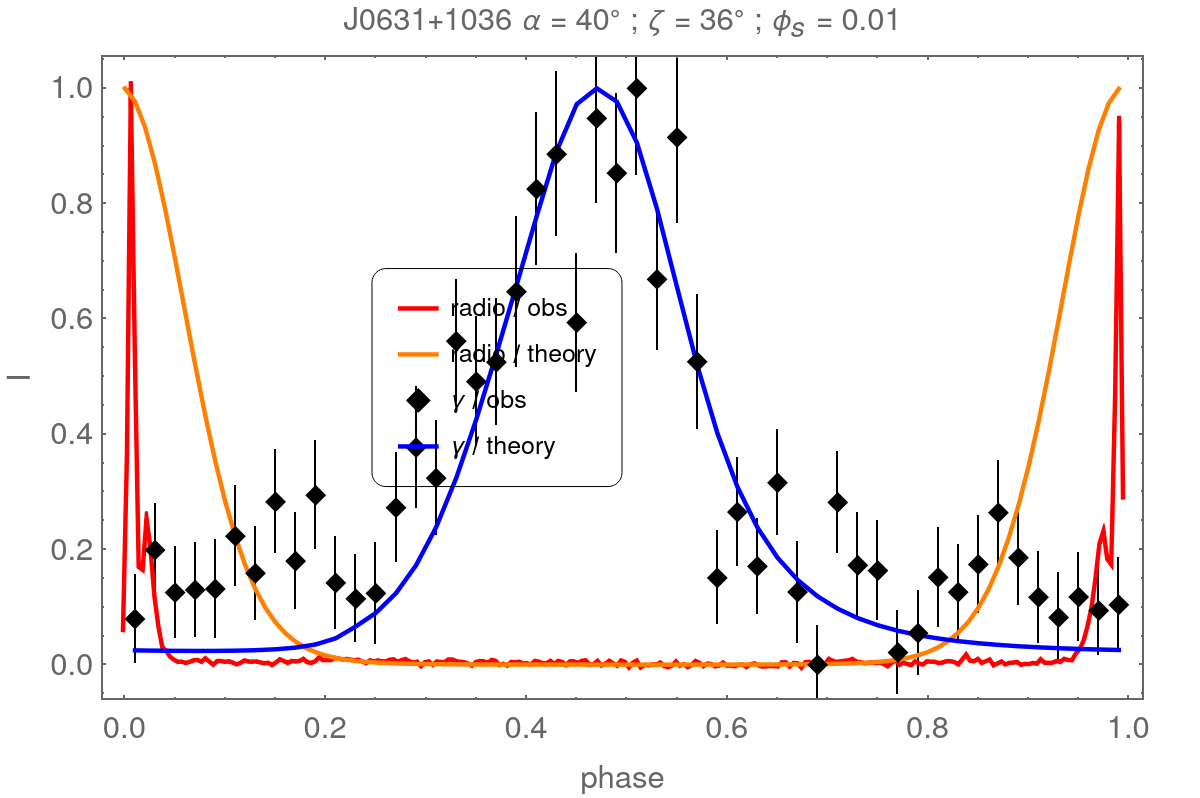}
	\end{tabular}
	\caption{On the top panel the radio polarization data with the best RVM fit of J0631+1036. On the middle panel, the $\log \rchi^2$ contour plots, in colour contours for radio polarization fits, and in solid coloured lines for gamma-ray light-curves. The red cross indicates the best joined radio/gamma-ray fit. On the bottom panel, the associated gamma-ray light-curve for the geometry given by the red cross.}
	\label{fig:J0631+1036chisq}
\end{figure}

\paragraph{PSR~J0631+1036.} With a period of 288~ms, this pulsar shows something like one gamma-ray peak or an unresolved double peak. Its radio pulse profile and the corresponding PPA are shown on the top panel of Fig.~\ref{fig:J0631+1036chisq}. The $\log \rchi^2$ contour plots for radio polarization fits is shown in coloured contours and the gamma-ray light-curve fits in solid lines on the middle panel of Fig.~\ref{fig:J0631+1036chisq}. The red cross indicates the obliquity and the inclination angle for the best joined fit. The corresponding radio and gamma-ray light-curves are overlapped with observations on the bottom panel of Fig.~\ref{fig:J0631+1036chisq}. The gamma-ray best fit light-curve resembles an unresolved double peaked profile. However, a better signal to noise ratio is required to firmly distinguish between an single and a double peak structure. Nevertheless, the joined radio and gamma-ray fit severely constrains the geometry of J0631+1036 because the most likely regions in the $(\alpha,\beta)$ plane are very different for both wavelengths. The gamma-ray fit is good and consistent with radio polarization data. A small offset is required $\phi_s=0.01$ for $\alpha=40\degr$ and $\zeta=36\degr$. We emphasize that other fits are not excluded, because the best radio and best gamma-ray fits are not always strictly compatible. Therefore, depending on the weight of each wavelength for defining a global $\rchi^2$ fit, we arrive at slightly different geometries. For conciseness, we do not plot them.

\begin{figure}
	\centering
	\begin{tabular}{l}
	\includegraphics[width=\columnwidth,angle=-90]{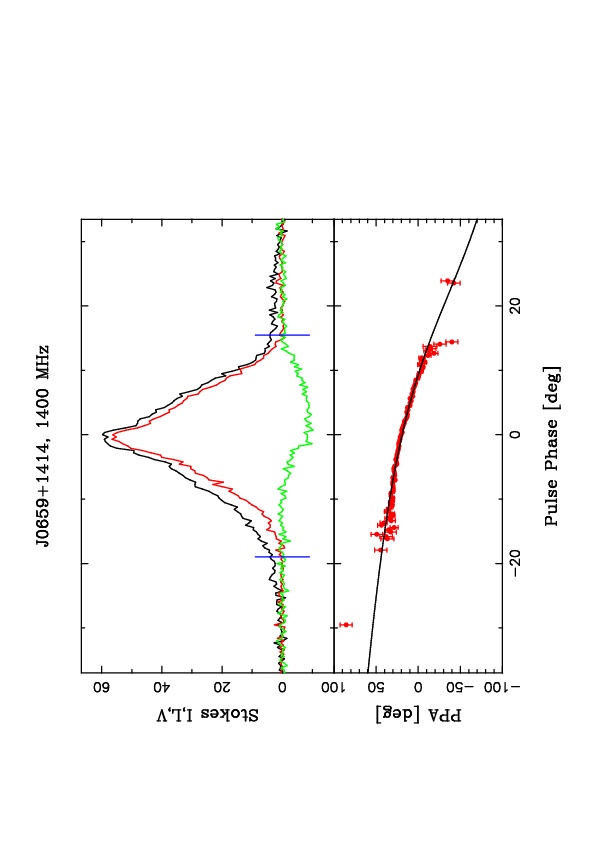} \\
	\includegraphics[width=\columnwidth]{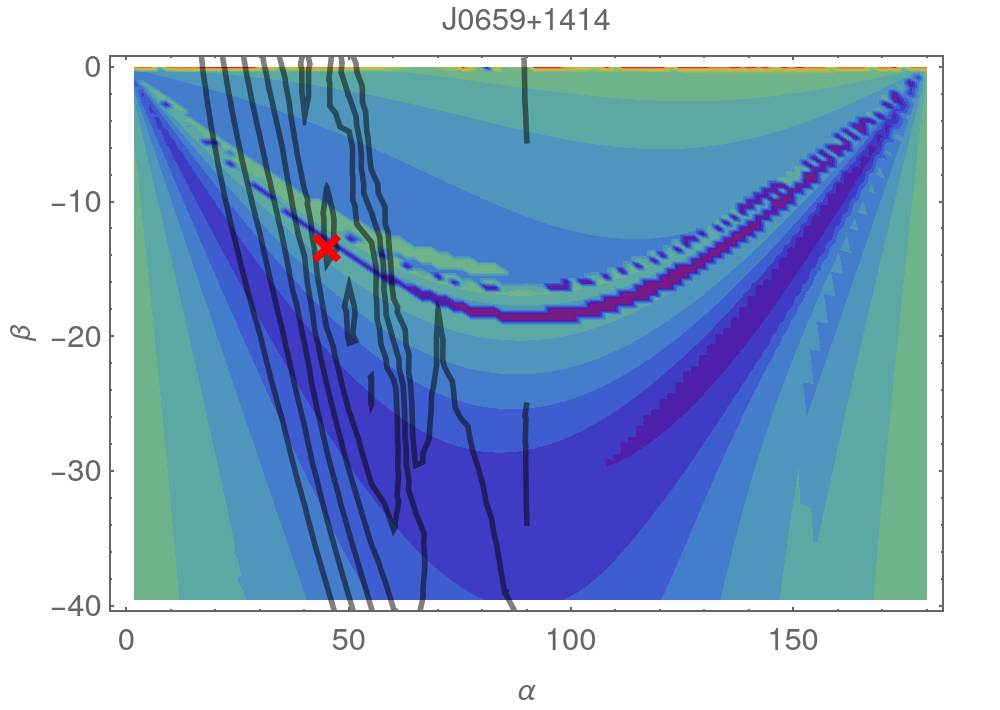} \\
	\includegraphics[width=\columnwidth]{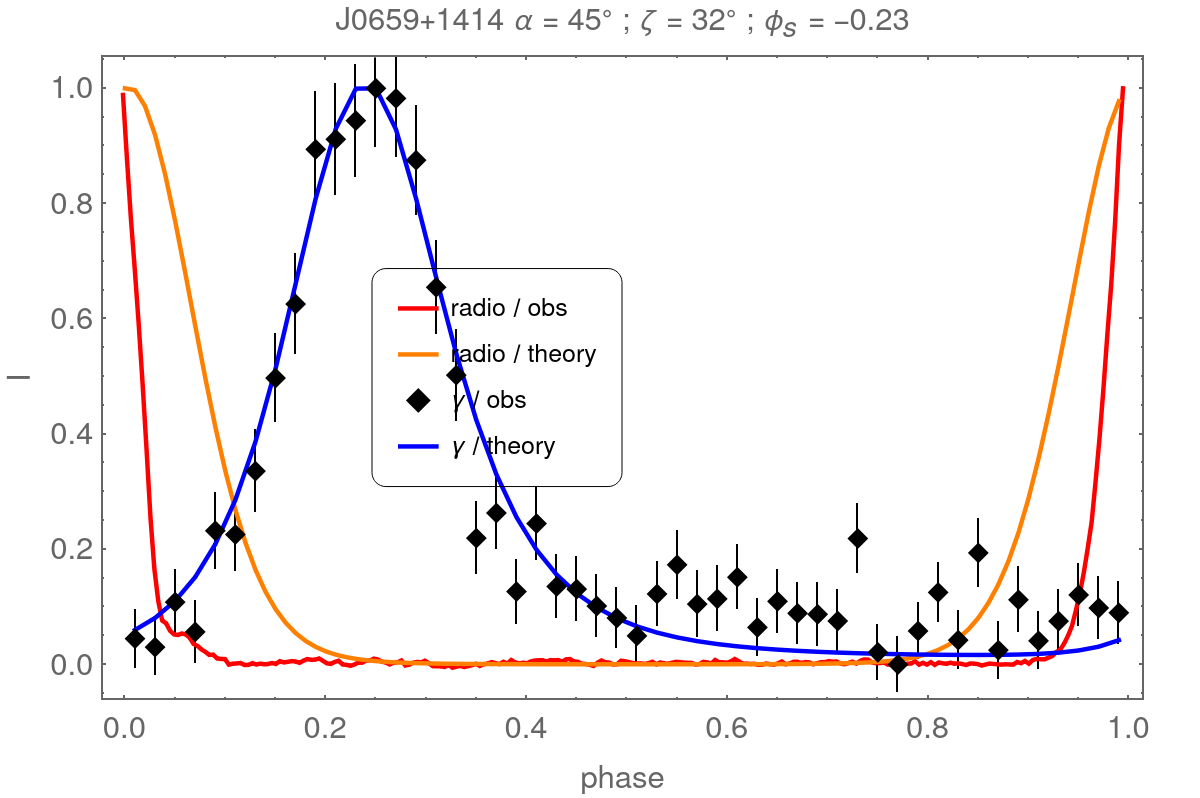} 
	\end{tabular}	
	\caption{Same as Fig.~\ref{fig:J0631+1036chisq} but for J0659+1414. The gamma-ray bets fit coincides with the radio polarization best fit.}
	\label{fig:J0659+1414chisq}
\end{figure}

\paragraph{PSR~J0659+1414.} Fig.~\ref{fig:J0659+1414chisq} shows an example of a clear single gamma-ray peaked pulsar with period 385~ms, namely J0659+1414. Good radio polarization data on the top panel of Fig.~\ref{fig:J0659+1414chisq} enable to constraint the angles through the $\rchi^2$ contour plots of radio and gamma-ray observations as for J0631+1036, middle panel of Fig.~\ref{fig:J0659+1414chisq}. The gamma-ray pulse profile look very symmetric and is well reproduced by our model, showing a symmetrical shape with respect to leading and trailing wings. The best fit shown by the red cross in the middle panel coincides with the radio constrain. However, the additional offset of $\phi_s=-0.23$ is large with a geometry given by $\alpha=45\degr$ and $\zeta=32\degr$.

\begin{figure}
	\centering
	\begin{tabular}{l}
	\includegraphics[width=\columnwidth,angle=-90]{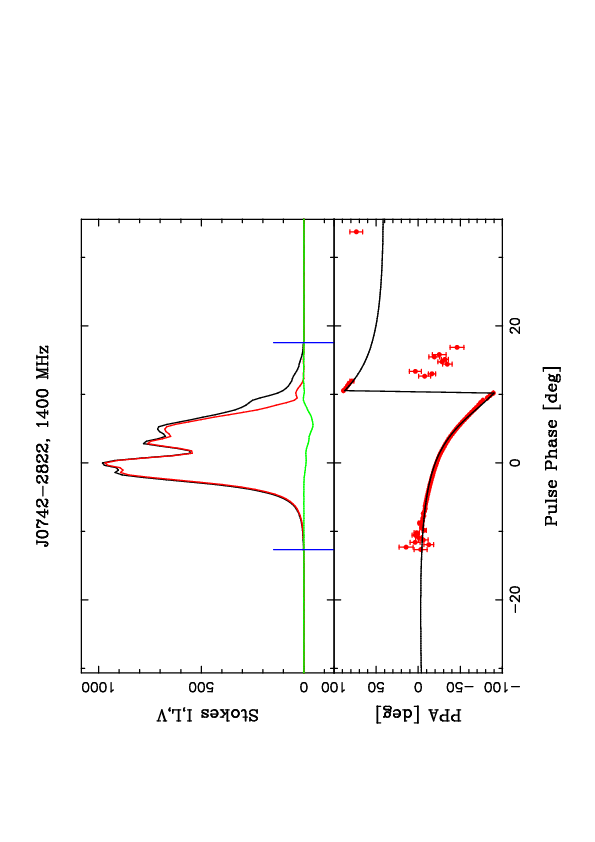} \\
	\includegraphics[width=\columnwidth]{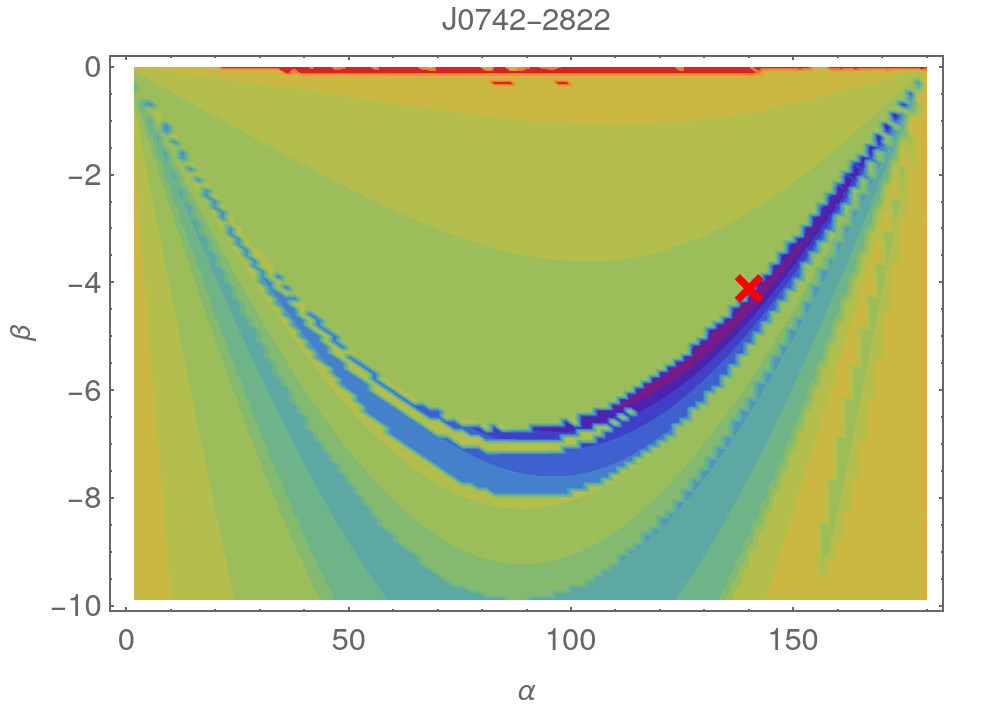} \\
	\includegraphics[width=\columnwidth]{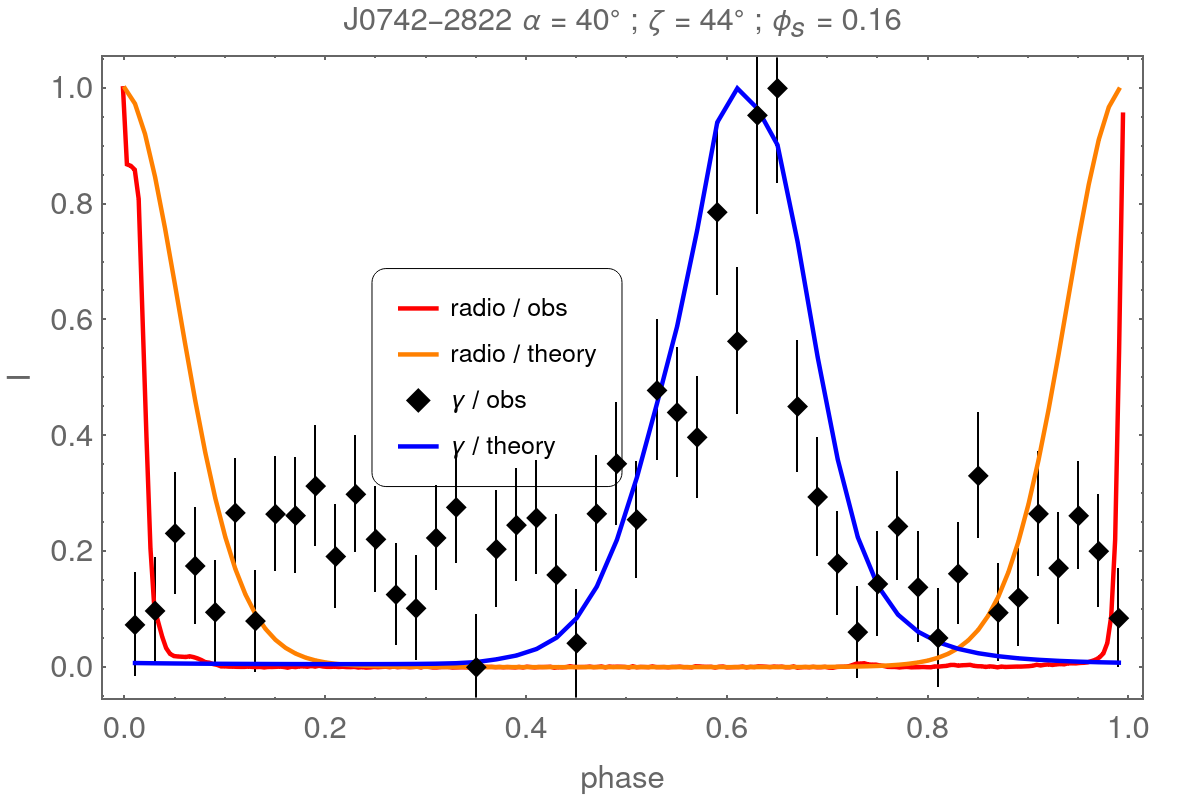} 
	\end{tabular}	
	\caption{Same as Fig.~\ref{fig:J0631+1036chisq} but for J0742-2822.}
	\label{fig:J0742-2822chisq}
\end{figure}

\paragraph{PSR~J0742-2822.} This pulsar has dominantly one gamma-ray pulse with the largest radio time lag of 0.627 and a period of 167~ms. The radio polarization swing is clearly visible on the top panel of Fig.~\ref{fig:J0742-2822chisq}. The contour plots of $\log \rchi^2$ in radio and gamma-rays overlap in a small region as seen in the second panel from the top. It favours an obliquity~$\alpha$ larger than 90\degr. Strictly speaking, we have not performed simulations for $\alpha>90\degr$ but we can use the symmetry of the gamma-ray light-curves to find the high energy profiles for $\alpha>90\degr$. Indeed, our striped wind model is symmetric about the equatorial plane, meaning that the configuration $(\alpha,\zeta)$ gives exactly the same light curves as the configuration~$(\pi - \alpha, \pi - \zeta)$. In other words, the radio fit $(\alpha, \beta)$ gives the same results as the fits for $(\pi - \alpha, -\beta)$. Therefore, for the gamma-ray light curve, we use a kind of reciprocal to the $\rchi^2$ obtained from the original radio data by changing $\alpha$ to $\pi-\alpha$ and $\beta$ to $-\beta$. Doing this we get the middle panel of Fig.~\ref{fig:J0742-2822chisq} showing the best gamma-ray fit coincident with radio polarization. It corresponds to $\alpha=40\degr$ and $\beta=4\degr$.
Reversing the symmetry argument, the real best fit is given by an offset equal to $\phi_s=0.16$ for $\alpha=140\degr$ (180\degr-40\degr) and $\zeta=136\degr$ ($\beta=-4\degr$). 

\begin{figure}
	\centering
	\begin{tabular}{l}
	\includegraphics[width=\columnwidth,angle=-90]{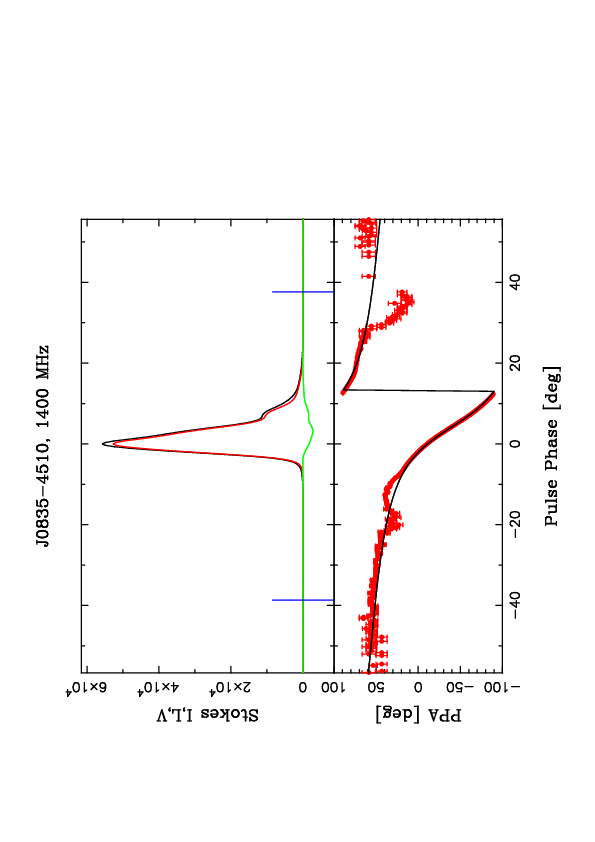} \\		
	\includegraphics[width=\columnwidth]{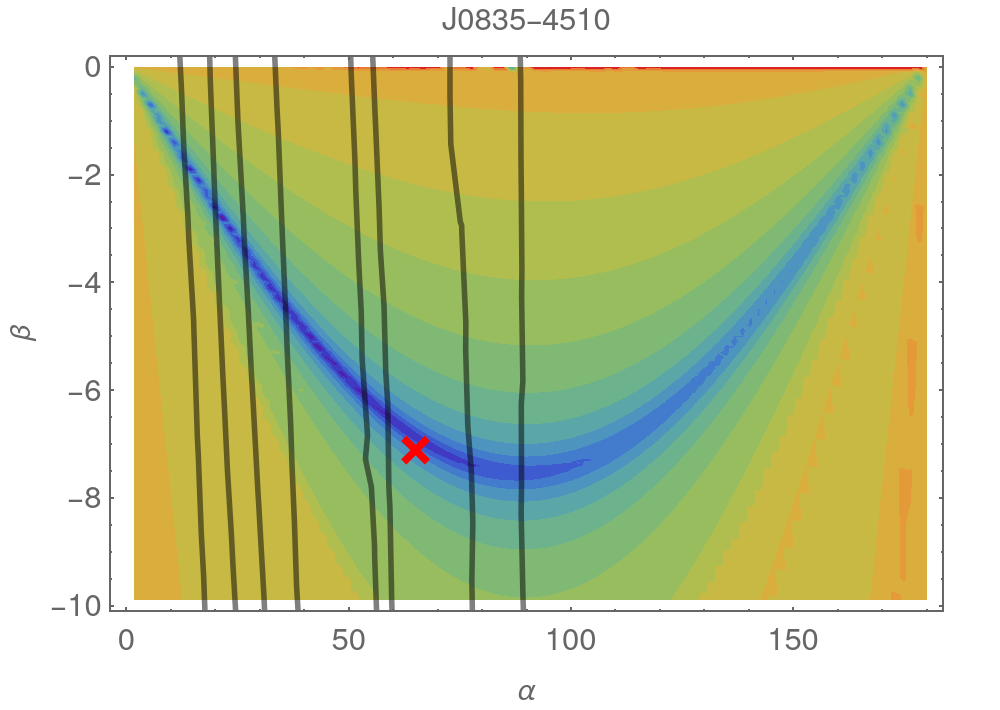} \\	
	\includegraphics[width=\columnwidth]{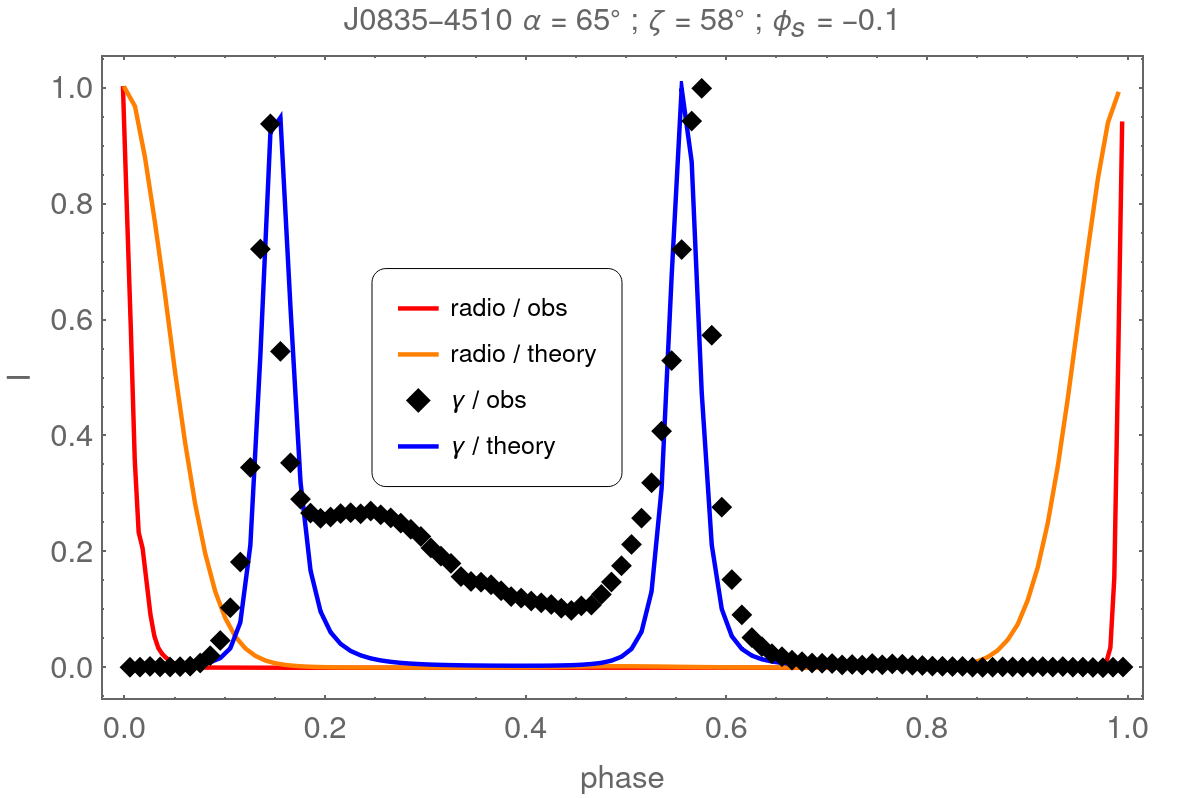}
	\end{tabular}	
	\caption{Same as Fig.~\ref{fig:J0631+1036chisq} but for J0835-451. The third peak is not taken into account.}
	\label{fig:J0835-451chisq}
\end{figure}

\paragraph{PSR~J0835-451.} The Vela pulsar with period 89~ms shows two prominent and well defined gamma-ray peaks surrounding a weaker third peak wandering in phase with energy, bottom panel of Fig.~\ref{fig:J0835-451chisq}. Our model can only produce two peaks so we discard the third peak. The radio polarization can be reasonably fitted with the RVM model but only around the steepest gradient, top panel of Fig.~\ref{fig:J0835-451chisq}. The middle panel shows the $\log \rchi^2$ contour plots for radio polarization and gamma-ray light-curves with the red cross lying slightly apart from the RVM constrain. The two prominent gamma-ray peaks are well fitted with the geometry shown on the bottom panel. The offset is $\phi_s=-0.1$ for $\alpha=65\degr$ and $\zeta=58\degr$. 

\begin{figure}
	\centering
	\begin{tabular}{l}
	\includegraphics[width=\columnwidth,angle=-90]{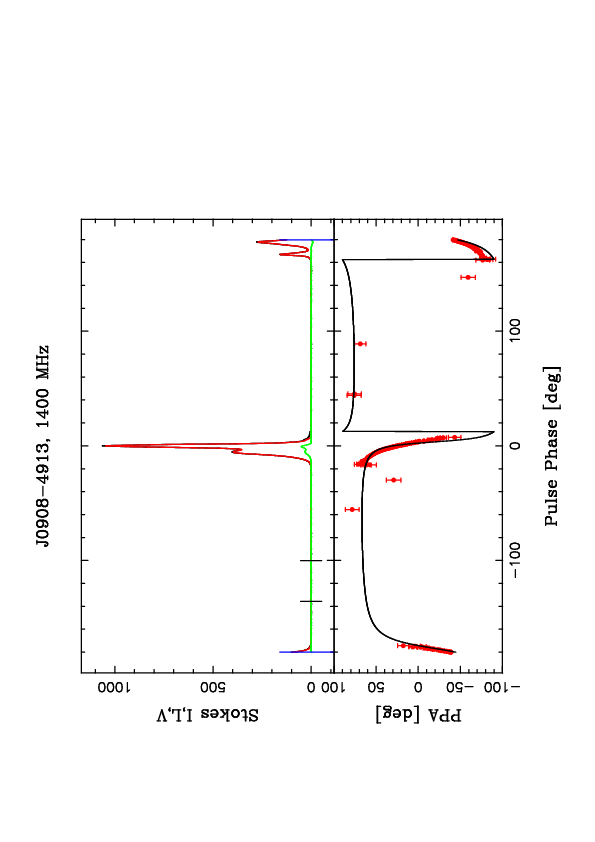} \\		
	\includegraphics[width=\columnwidth]{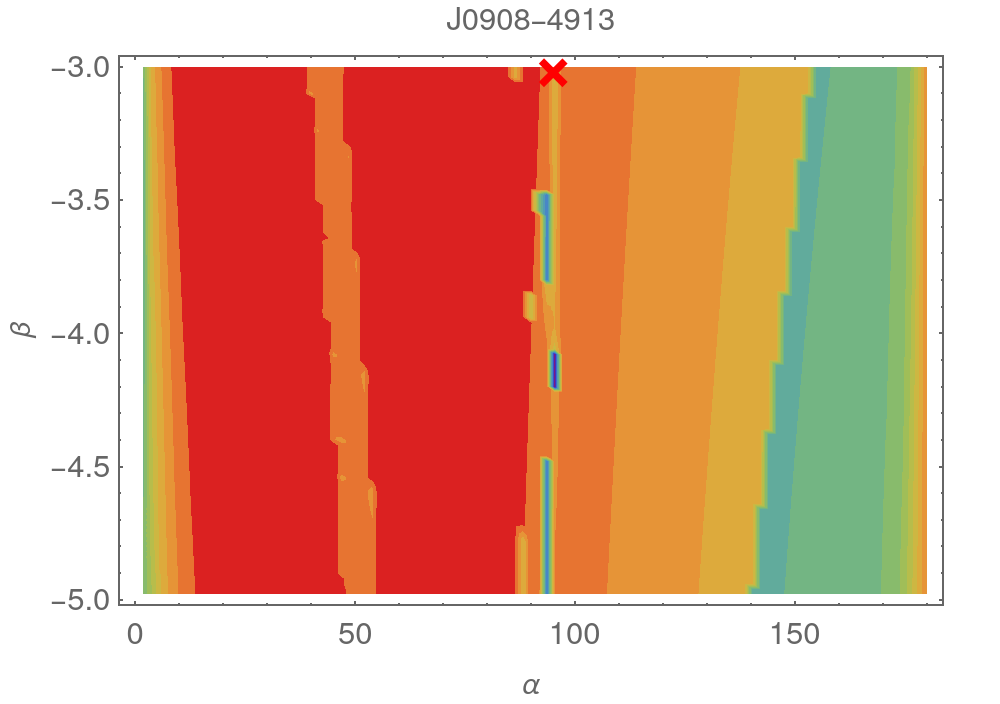} \\
	\includegraphics[width=\columnwidth]{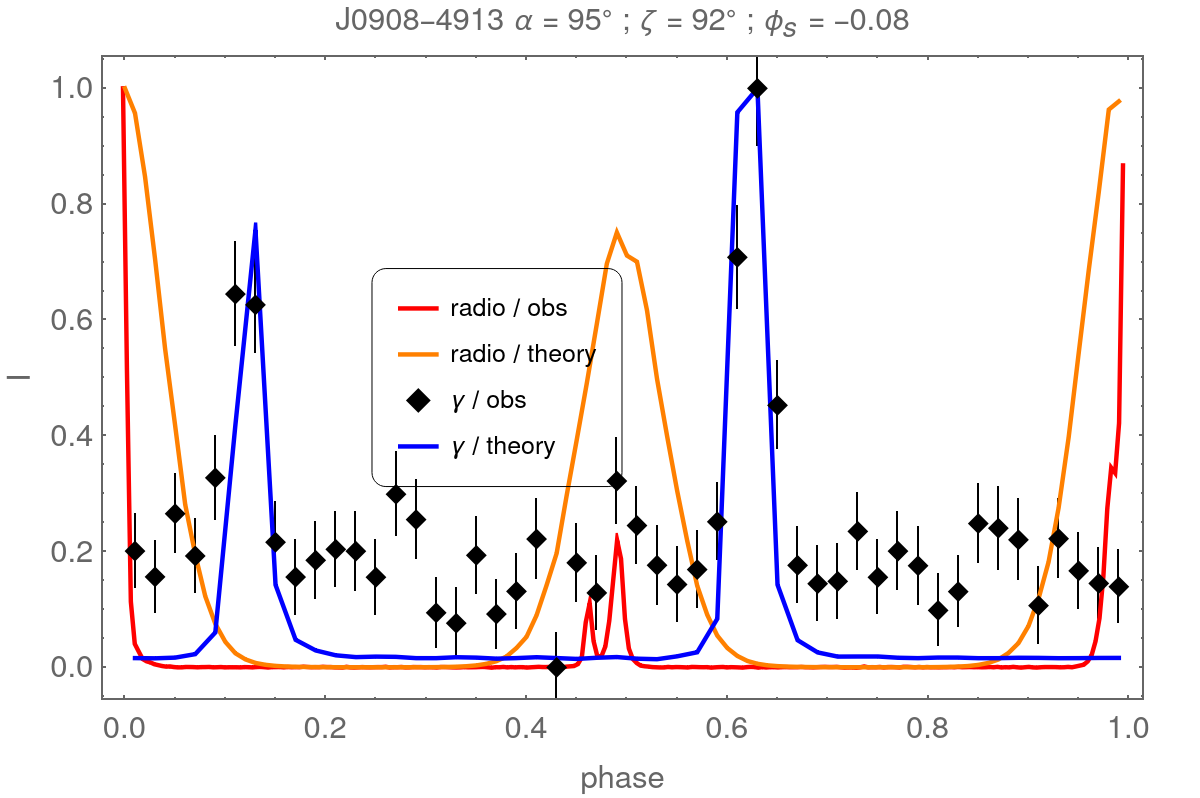}
	\end{tabular}	
	\caption{Same as Fig.~\ref{fig:J0631+1036chisq} but for J0908-4913.}
	\label{fig:J0908-4913chisq}
\end{figure}

\paragraph{PSR~J0908-4913.} This pulsar of 107~ms is another example of double peaked gamma-ray pulsar, although noisy, bottom panel of Fig.~\ref{fig:J0908-4913chisq}. It also shows a less prominent interpulse in radio at phase 0.5, suggesting it to be close to an orthogonal rotator, top panel. Indeed, the RVM constrain are shown in the middle panel of Fig.~\ref{fig:J0908-4913chisq}, clearly highlighting the orthogonal nature of the pulsar with a line of sight passing close to the magnetic axis because $-3\degr<\beta<-5\degr$. See also \cite{kramer_high-precision_2008} for similar conclusions. As for J0742-2822, the obliquity $\alpha$ is larger than 90\degr. We use again the symmetry argument to find the best gamma-ray fit with $\alpha'=85\degr$ and $\zeta'=88\degr$. The two peak maximum intensity are different and not fully reproduced by our model. Nevertheless, the two radio peaks are visible for an offset of $\phi_s=-0.08$ and the real angles are $\alpha=95\degr$ and $\zeta=92\degr$, depicted by the red cross lies on the RVM fit contour, thus indeed being an orthogonal rotator. Our predicted radio interpulse intensity is much higher than the observer flux. A proper understanding of this effect, if not geometric, requires knowledge of the radio emission mechanism which is out of our scope.

\begin{figure}
	\centering
	\begin{tabular}{l}
	\includegraphics[width=\columnwidth,angle=-90]{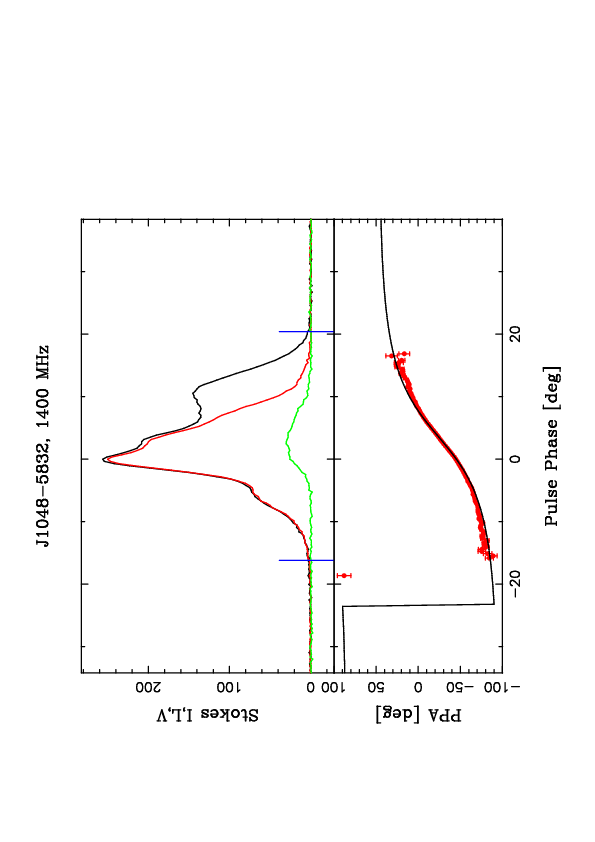} \\		
	\includegraphics[width=\columnwidth]{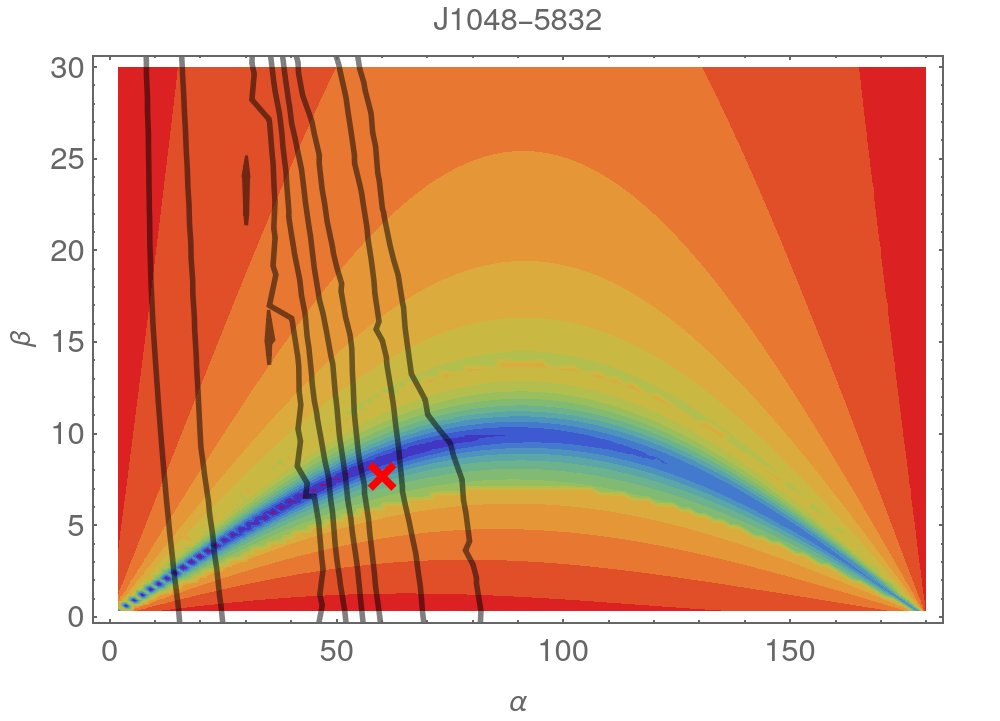} \\
	\includegraphics[width=\columnwidth]{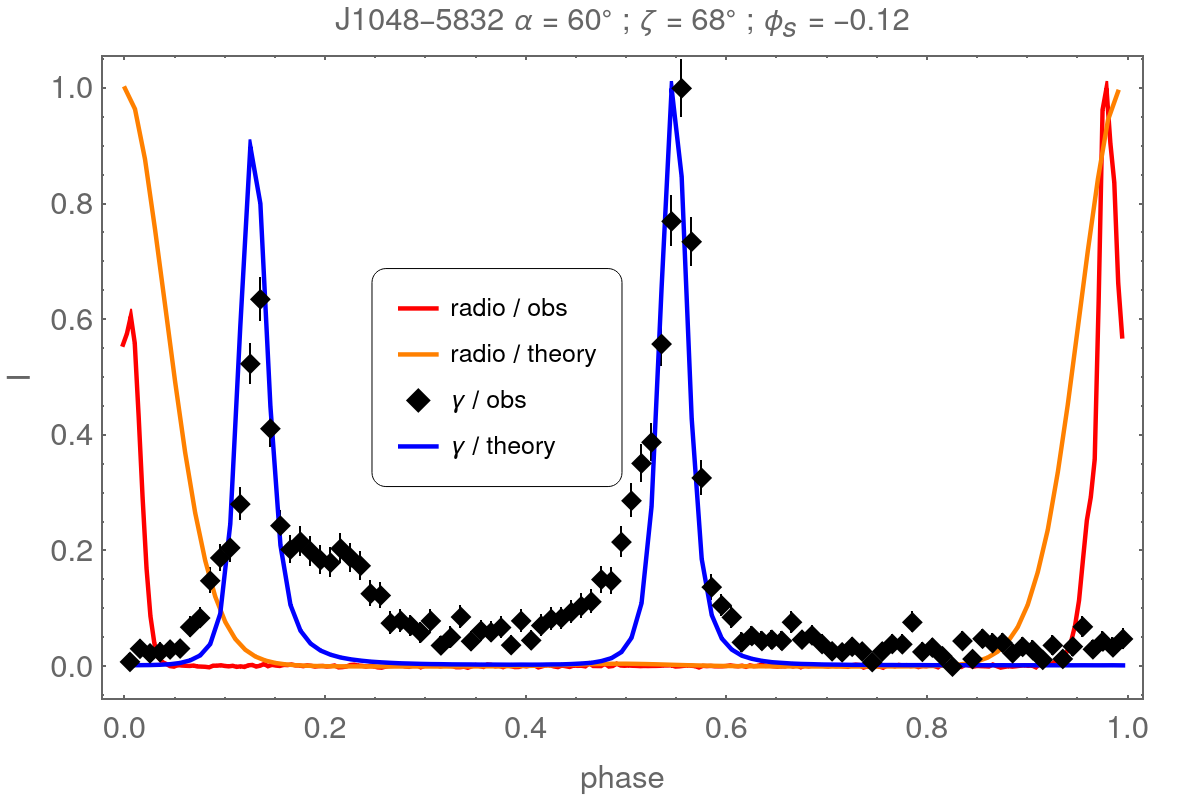} 
	\end{tabular}	
	\caption{Same as Fig.~\ref{fig:J0631+1036chisq} but for J1048-5832.}
	\label{fig:J1048-5832chisq}
\end{figure}

\paragraph{PSR~J1048-5832.} This is another bright gamma-ray pulsar of period 124~ms, possessing very good radio polarization data, top panel of Fig.~\ref{fig:J1048-5832chisq}, leading to an accurate $\rchi^2$ plot as seen in the middle panel of Fig.~\ref{fig:J1048-5832chisq}. It shows two narrow and prominent gamma-ray pulses well fitted by the red cross area coincident with radio polarization constraints. The overlapping region therefore severely pins down the geometry of J1048-5832 to be around $\alpha = 60\degr$ and $\zeta=68\degr$ with an offset of $\phi_s=-0.12$.

\begin{figure}
	\centering
	\begin{tabular}{l}
		\includegraphics[width=\columnwidth,angle=-90]{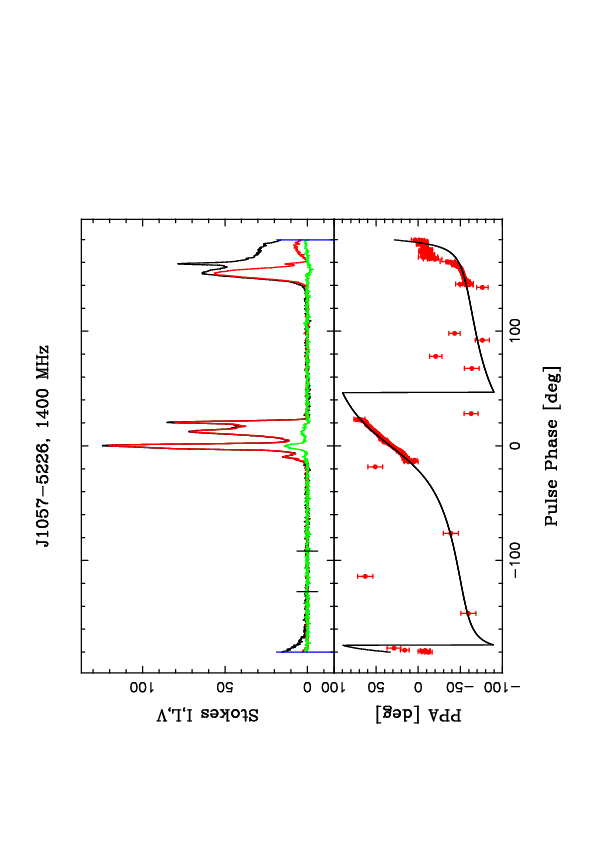} \\		
		\includegraphics[width=\columnwidth]{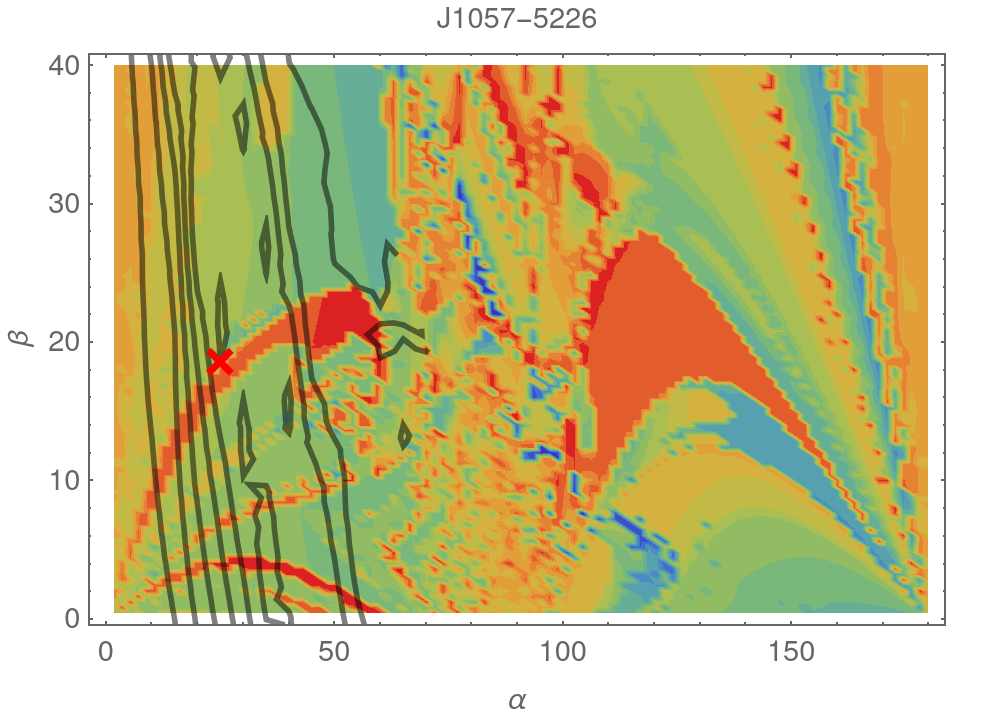} \\
		\includegraphics[width=\columnwidth]{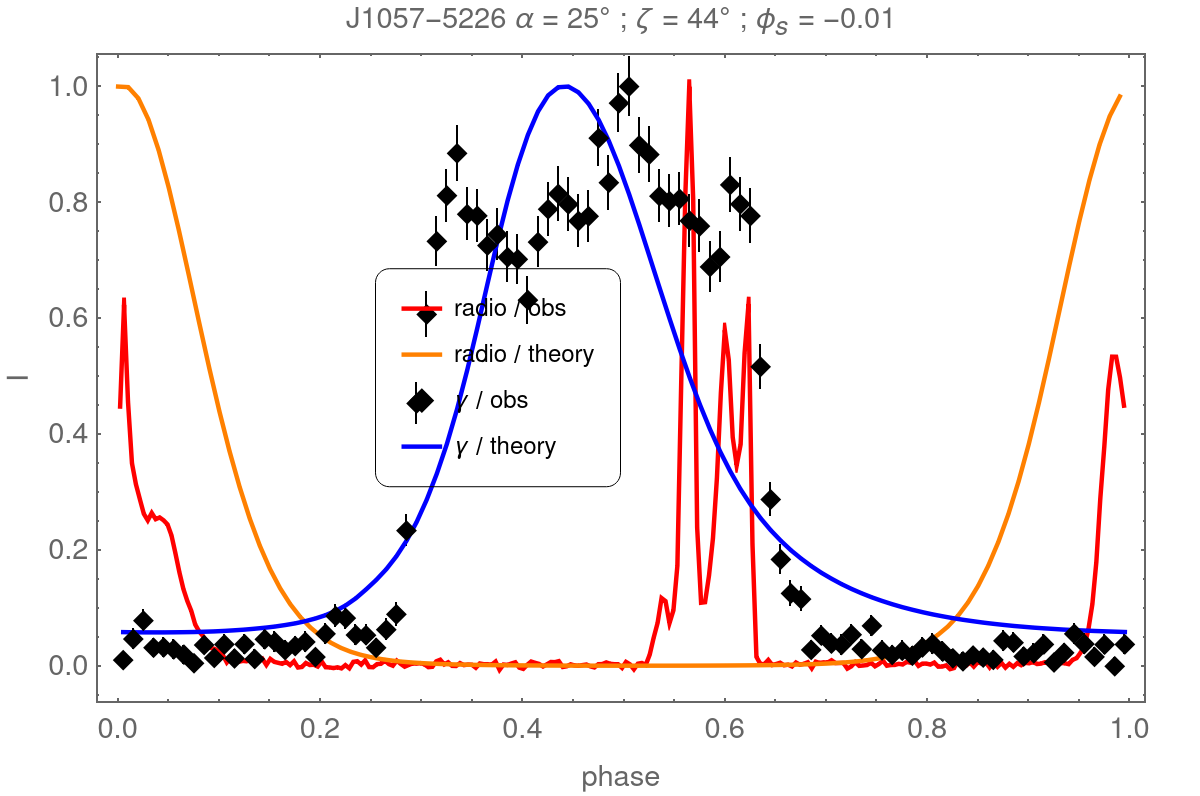} 
	\end{tabular}	
	\caption{Same as Fig.~\ref{fig:J0631+1036chisq} but for J1057-5226.}
	\label{fig:J1057-5226chisq}
\end{figure}

\paragraph{PSR~J1057-5226.} A single gamma-ray peak with a kind of large plateau or an unresolved double gamma-ray peak is visible for this pulsar, Fig.~\ref{fig:J1057-5226chisq}. A radio pulse as well as an interpulse is seen making it possibly an almost orthogonal rotator. However, we found a relatively low obliquity of only $\alpha \approx 25\degr$ with $\zeta=44\degr$ and $\phi_s=-0.01$. No radio interpulse is predicted by this geometry. This pulsar does not easily accommodate with our picture of a combined polar cap striped wind emission model. 
The RVM fit to the radio polarization however is consistent with $\alpha \approx 75\degr$ and $\zeta \approx 110\degr$, which is also the result obtained by \cite{weltevrede_mapping_2009} and is shown in top panel of \ref{fig:J1057-5226chisq}. This is the only example of our sample that does not fit into the joint radio and gamma-ray fitting procedure.

\begin{figure}
	\centering
	\begin{tabular}{l}
	\includegraphics[width=\columnwidth,angle=-90]{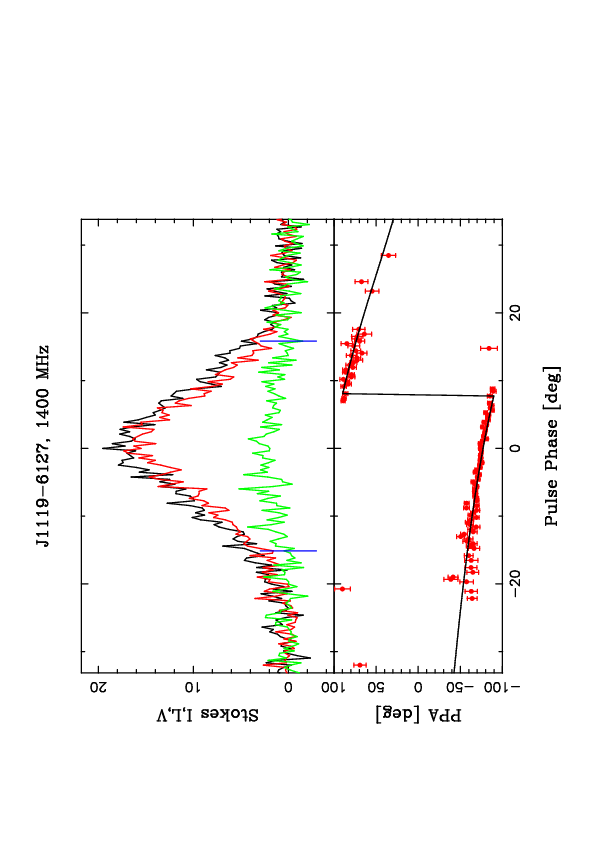} \\		
	\includegraphics[width=\columnwidth]{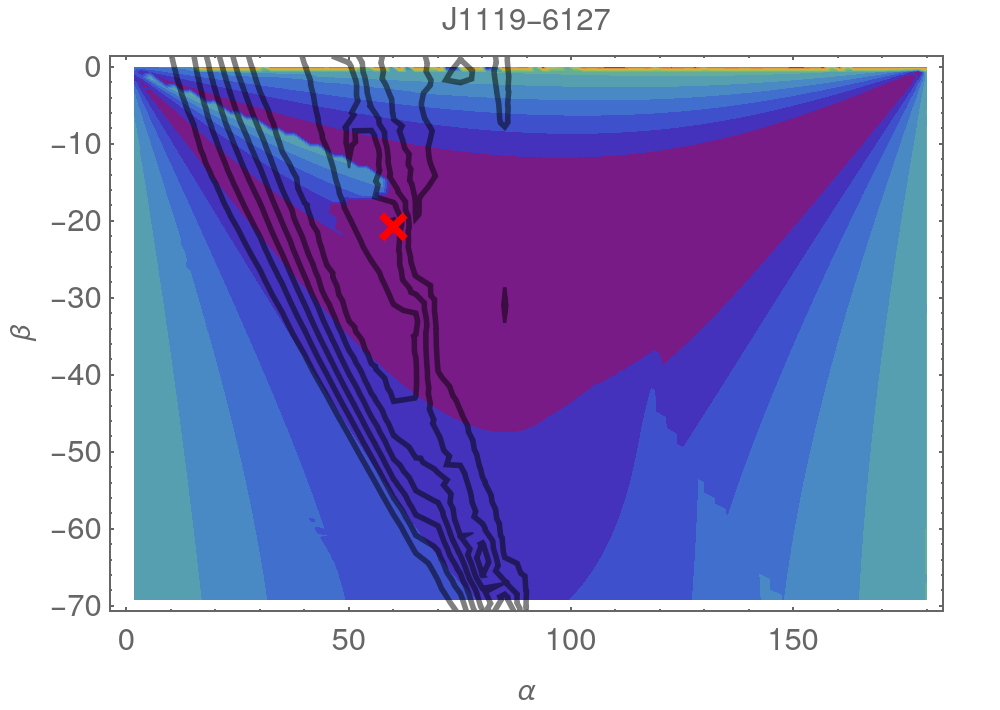} \\
	\includegraphics[width=\columnwidth]{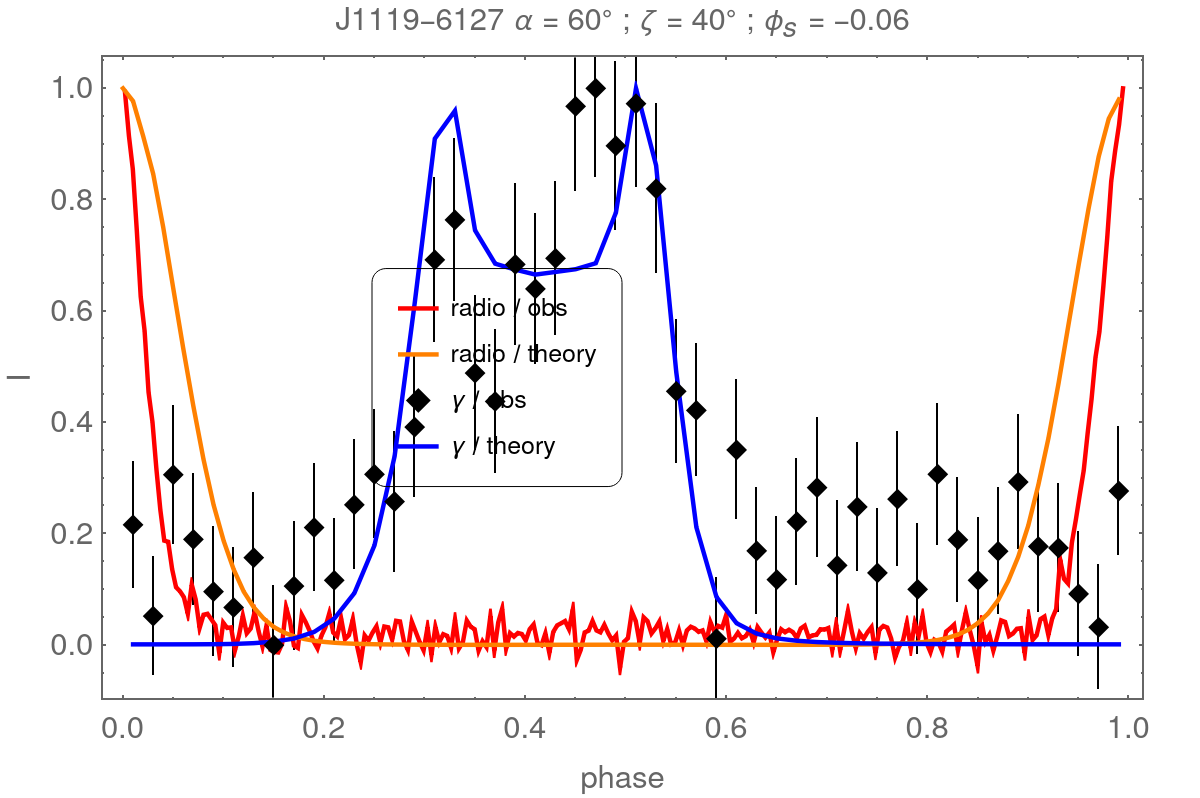} 
	\end{tabular}	
	\caption{Same as Fig.~\ref{fig:J0631+1036chisq} but for J1119-6127.}
	\label{fig:j1119-6127chisq}
\end{figure}

\paragraph{PSR~J1119-6127.} This pulsar with a period of 408~ms shows a weakly double peaked gamma-ray profile. Its radio polarization data are noisy, top panel of  Fig.~\ref{fig:j1119-6127chisq}, implying a large area for the PPA constrain, middle panel. Also, two distinct joined radio gamma-ray best fits are possible. One fit leads to a single gamma-ray profile, not shown, and one to an unresolved double gamma-ray light-curve, bottom panel and red cross in the middle panel. We had to add an additional phase shift of about $\phi_s=-0.06$ for $\alpha = 60\degr$ and $\zeta=40\degr$. Better quality gamma-ray data will certainly favour this second option.

\begin{figure}
	\centering
	\begin{tabular}{l}
	\includegraphics[width=\columnwidth,angle=-90]{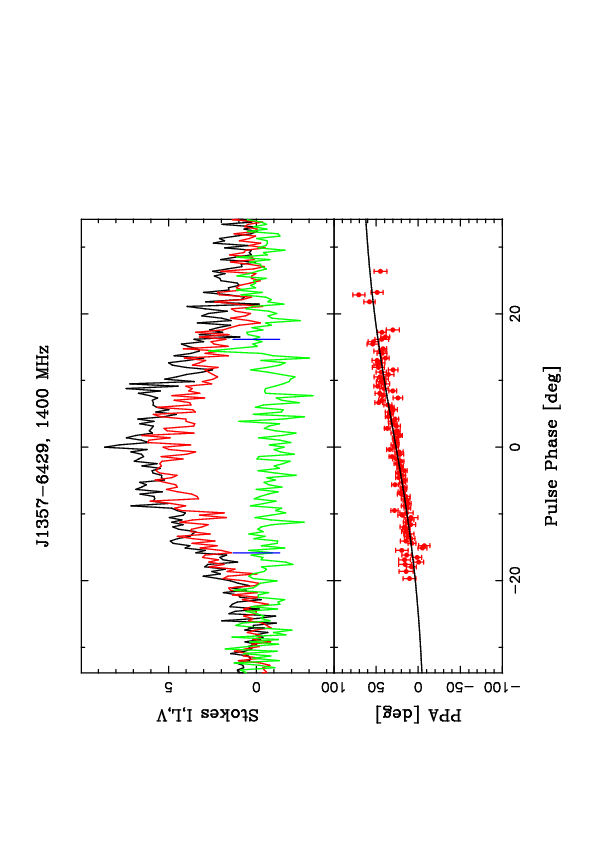} \\		
	\includegraphics[width=\columnwidth]{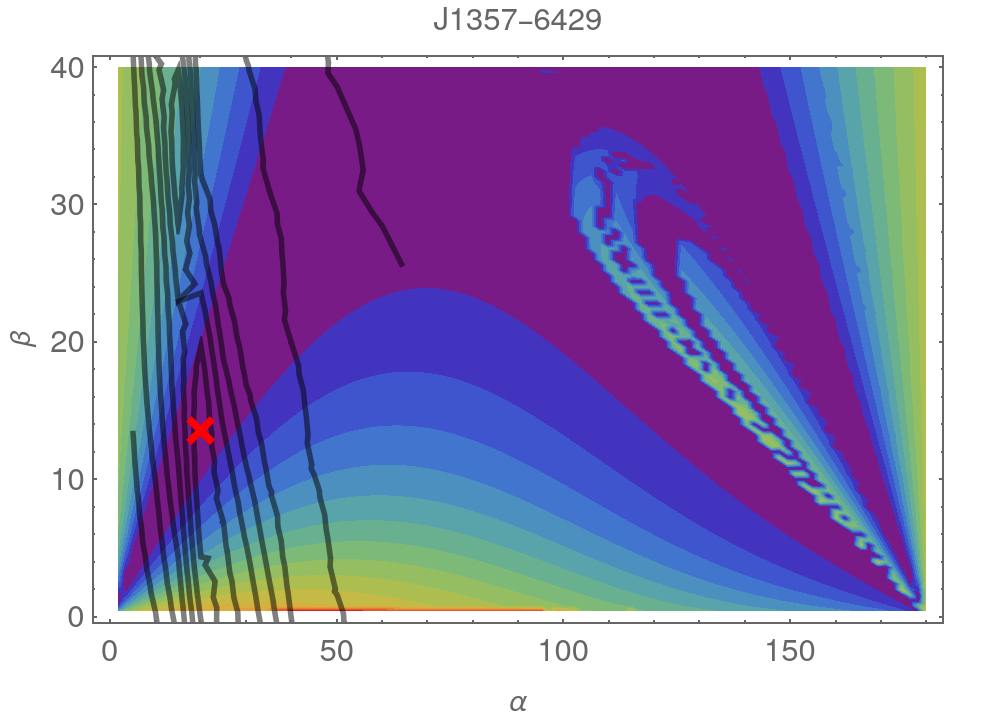} \\
	\includegraphics[width=\columnwidth]{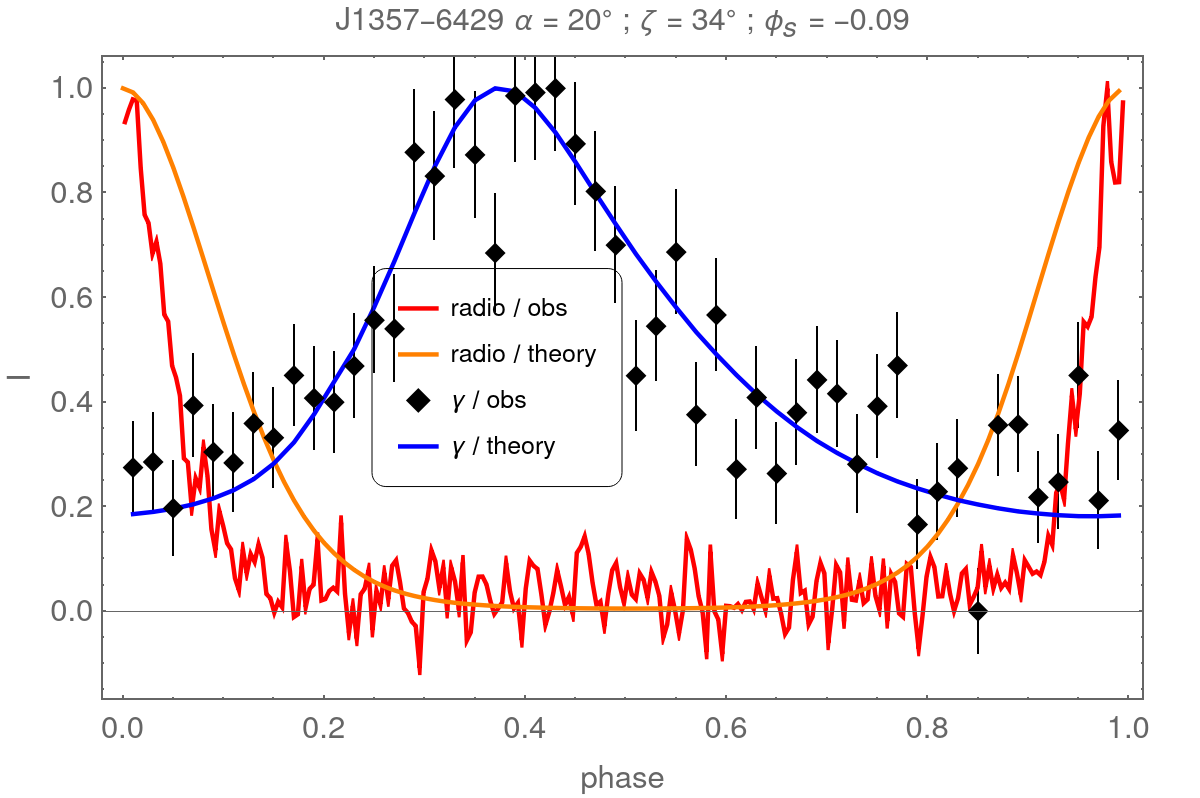} 
	\end{tabular}	
	\caption{Same as Fig.~\ref{fig:J0631+1036chisq} but for J1357-6429.}
	\label{fig:j1357-6429chisq}
\end{figure}

\paragraph{PSR~J1357-6429.} Similar to the previous pulsar, PSR~J1357-6429 is noisy in radio, top panel of Fig.~\ref{fig:j1357-6429chisq} with large uncertainties in the RVM constrain, middle panel. Two options are given by either an unresolved double gamma-ray peak, not shown, or a single peak gamma-ray, bottom panel. It is another example of a single gamma-ray peak pulsar fitted with a small obliquity. We added an additional phase shift of $\phi_s=-0.09$ for $\alpha = 20\degr$ and $\zeta=34\degr$ which also seems the most likely.

\begin{figure}
	\centering
	\begin{tabular}{l}
	\includegraphics[width=\columnwidth,angle=-90]{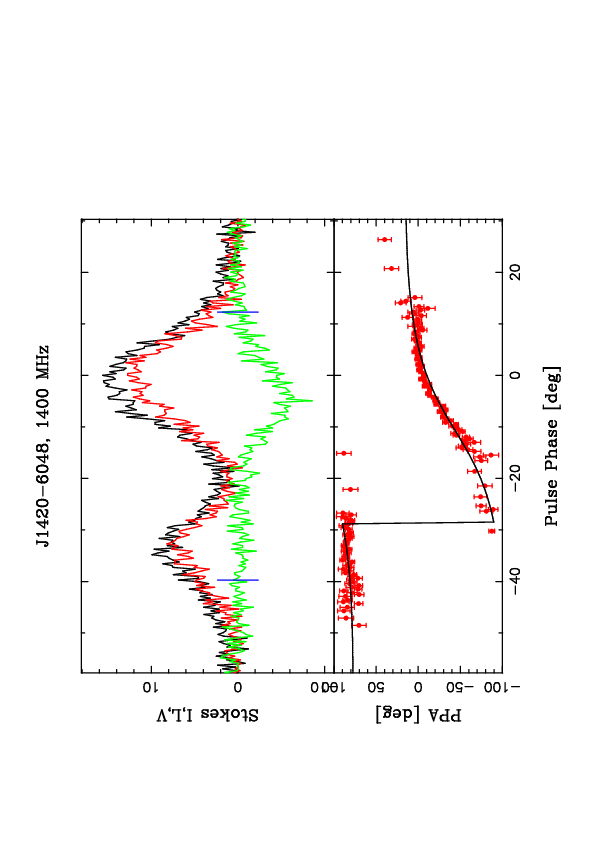} \\		
	\includegraphics[width=\columnwidth]{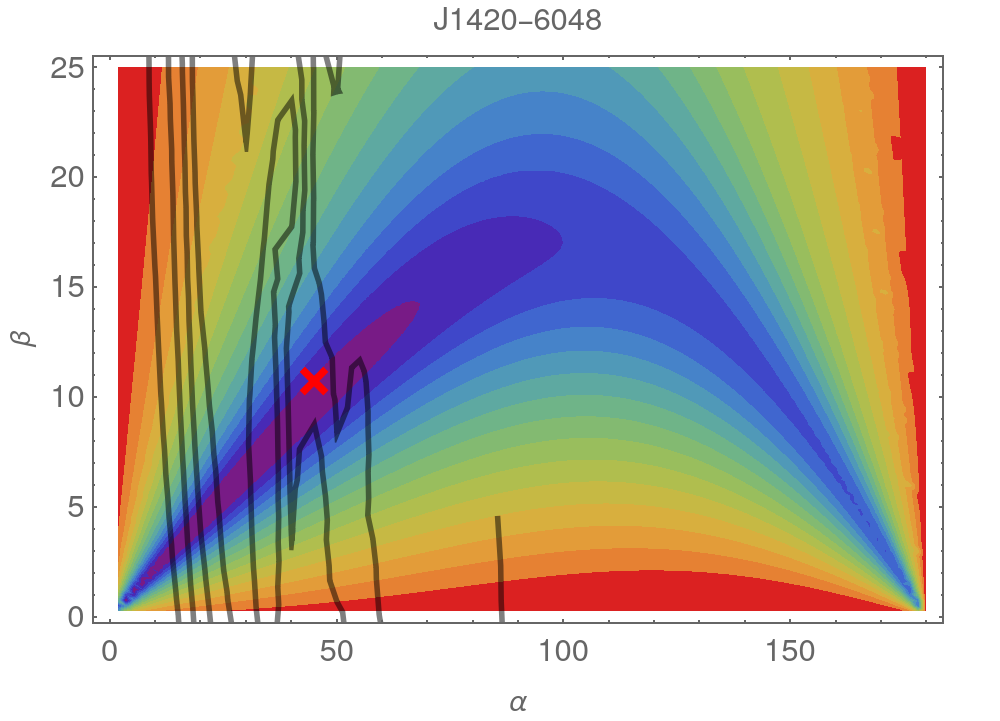} \\
	\includegraphics[width=\columnwidth]{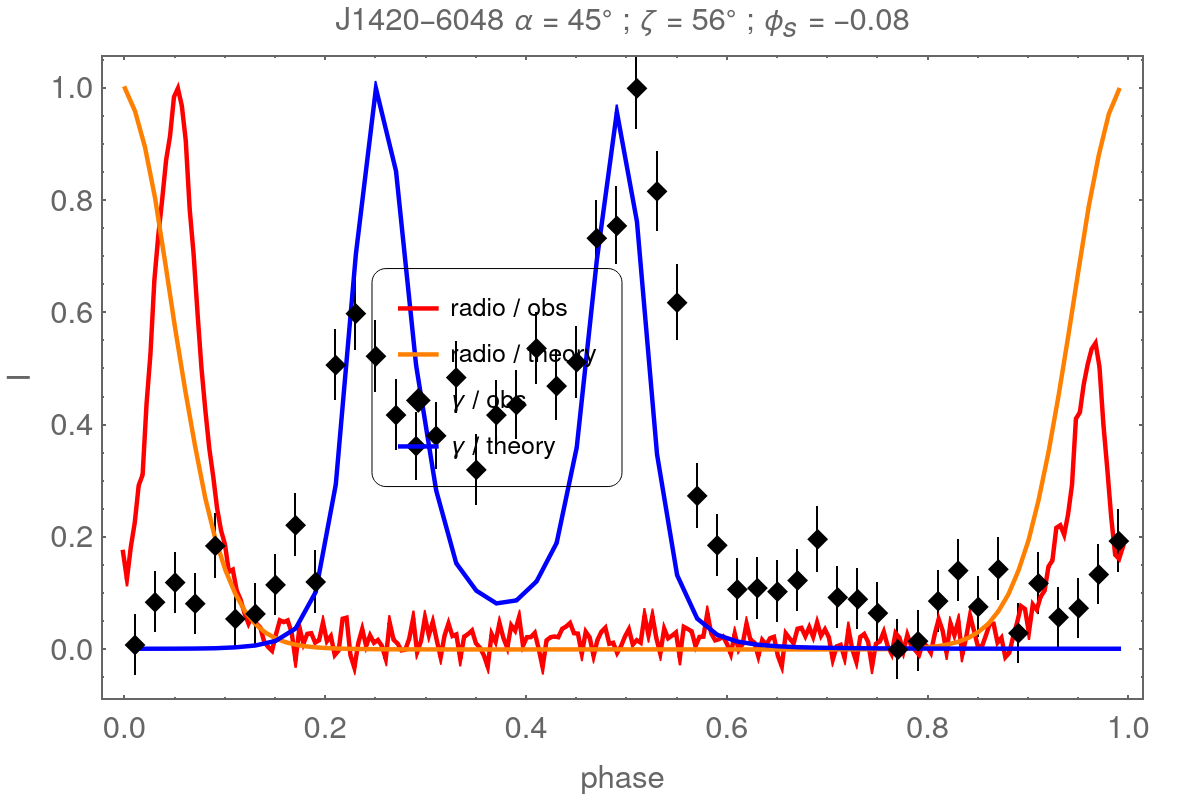} 
	\end{tabular}	
	\caption{Same as Fig.~\ref{fig:J0631+1036chisq} but for J1420-6048.}
	\label{fig:j1420-6048chisq}
\end{figure}

\paragraph{PSR~J1420-6048.} This double peaked gamma-ray pulsar possesses an unresolved double gamma-ray peak profile with an asymmetry in their peak intensity, bottom panel of Fig.~\ref{fig:j1420-6048chisq}. 
The good radio polarization data, top panel, furnishes good RVM constrain, middle panel. 
Most likely is the geometry given by the red cross leading to the double peak profile visible in the bottom panel. The best fit configuration has a phase shift of $\phi_s=-0.08$ for $\alpha = 45\degr$ and $\zeta=56\degr$.

\begin{figure}
	\centering
	\begin{tabular}{l}
	\includegraphics[width=\columnwidth,angle=-90]{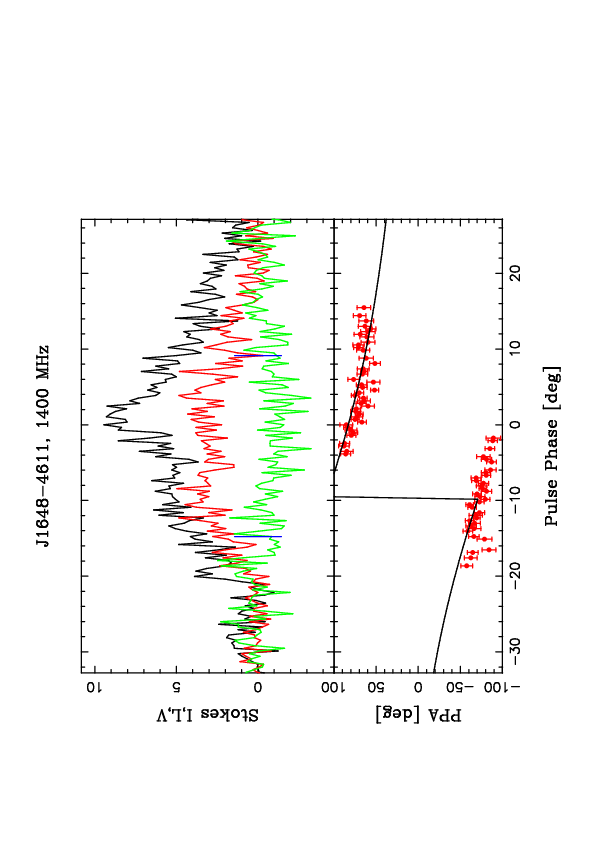} \\		
	\includegraphics[width=\columnwidth]{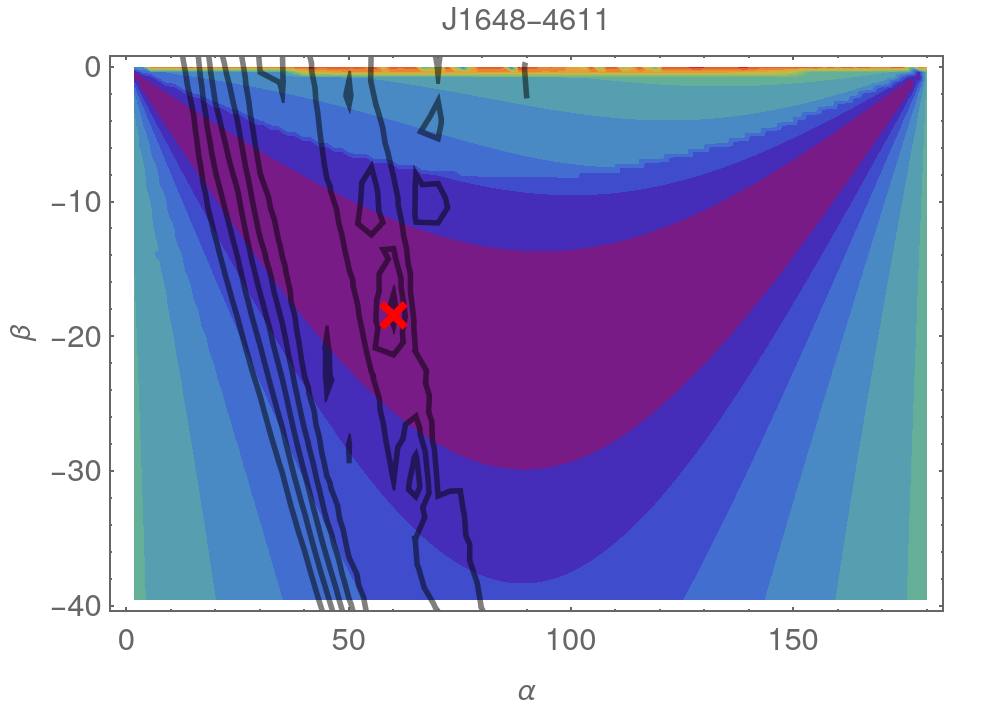} \\
	\includegraphics[width=\columnwidth]{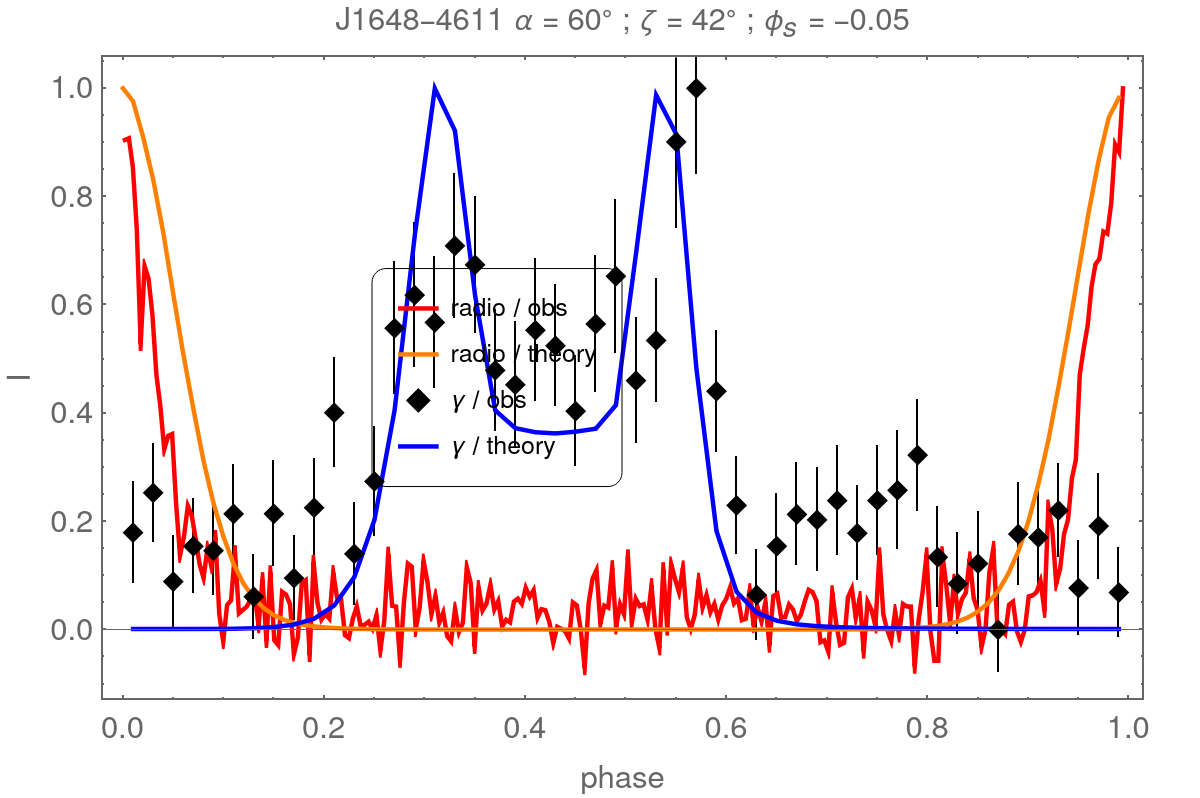}
	\end{tabular}	
	\caption{Same as Fig.~\ref{fig:J0631+1036chisq} but for J1648-4611.}
	\label{fig:j1648-4611chisq}
\end{figure}

\paragraph{PSR~J1648-4611.} The situation for this pulsar is more clear cut. Although the radio polarization data are noisy, top panel of Fig.~\ref{fig:j1648-4611chisq}, the joined radio gamma-ray fit leads to a well defined geometry shown in the middle panel. The gamma-ray pulse profile resembles to a double peaked curve with a plateau, bottom panel. The corresponding phase shift is $\phi_s=-0.05$ for $\alpha = 60 \degr$ and $\zeta = 42\degr$.

\begin{figure}
	\centering
	\begin{tabular}{l}
	\includegraphics[width=\columnwidth,angle=-90]{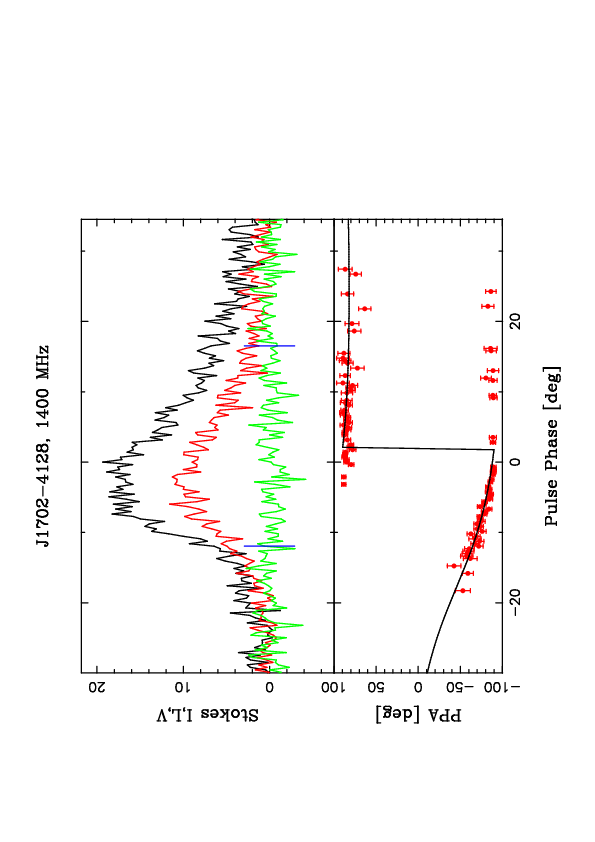} \\		
	\includegraphics[width=\columnwidth]{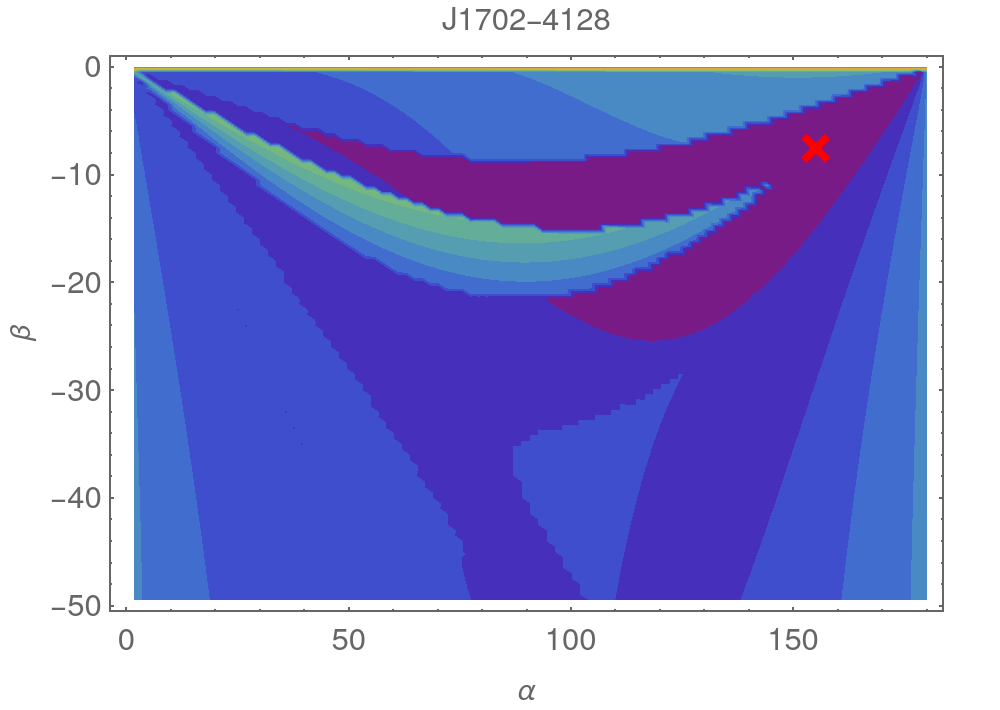} \\
	\includegraphics[width=\columnwidth]{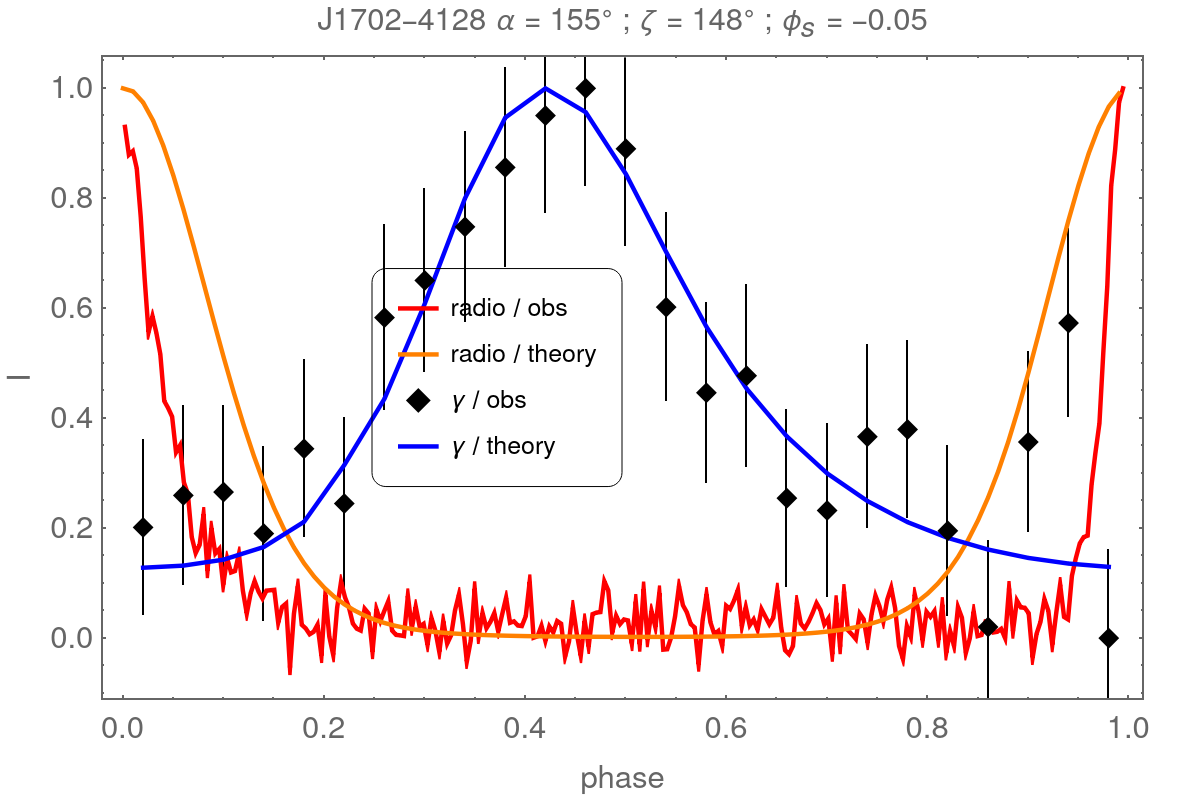}
	\end{tabular}	
	\caption{J1702-4310 $\log \rchi^2$ contour plots, for radio polarization fits in colour, and in solid black lines for gamma-ray light-curves.}
	\label{fig:J1702-4310_chisq}
\end{figure}

\paragraph{PSR~J1702-4310.} The top panel of Fig.~\ref{fig:J1702-4310_chisq} show the radio polarization data of PSR~J1702-4310. The associated RVM constrain are given in the middle panel and not very constraining. This is another example of $\alpha>90\degr$. One best fitting geometry, depicted by the red cross produces a single gamma-ray peak, bottom panel. The phase shift is $\phi_s=-0.05$ for $\alpha = 25\degr$ and $\zeta = 32\degr$. We do not expect such fitting to be very reliable because the gamma-ray statistics is weak. Switching back to the real geometry, we get $\alpha = 155\degr$ and $\zeta = 148\degr$.

\begin{figure}
	\centering
	\begin{tabular}{l}
	\includegraphics[width=\columnwidth,angle=-90]{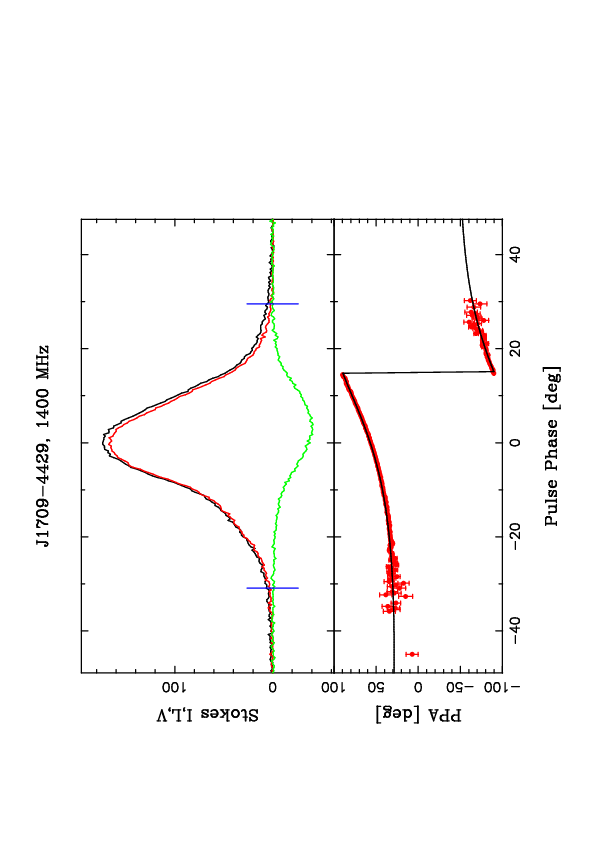} \\		
	\includegraphics[width=\columnwidth]{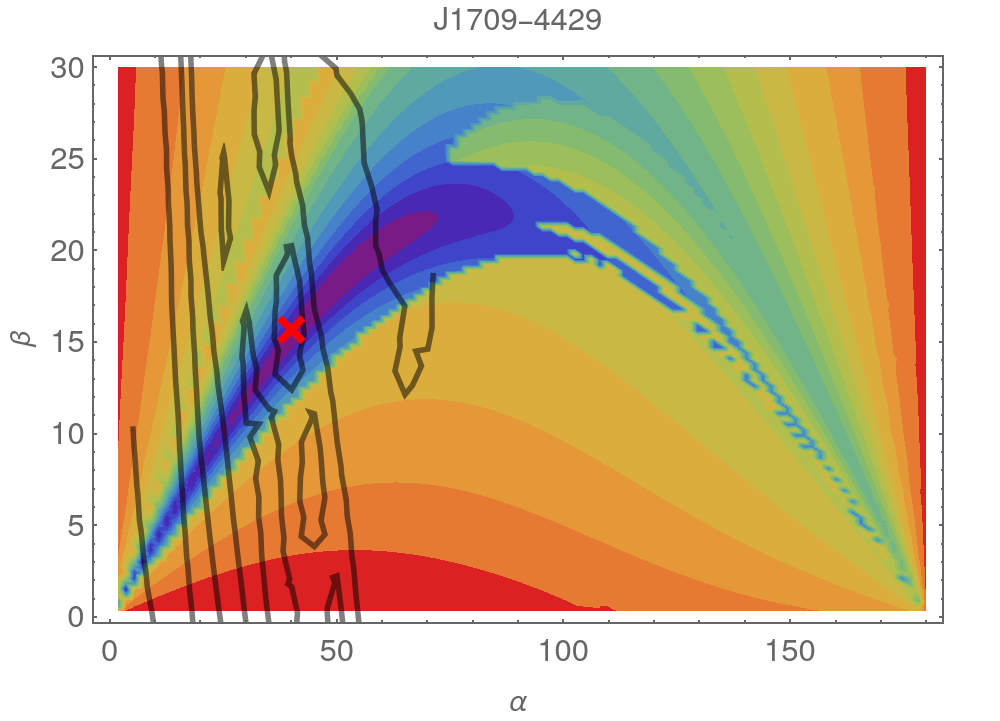} \\
	\includegraphics[width=\columnwidth]{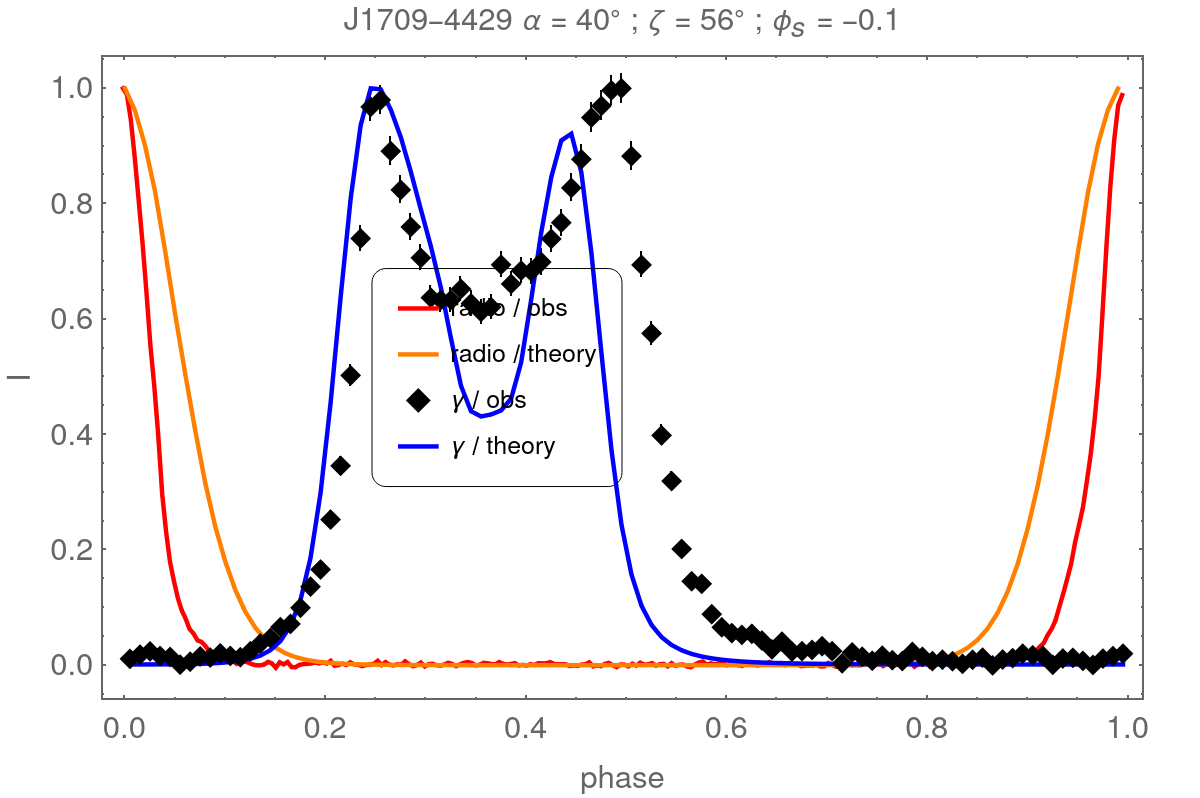}
	\end{tabular}	
	\caption{Same as Fig.~\ref{fig:J0631+1036chisq} but for J1709-4429.}
	\label{fig:J1709-4429_chisq}
\end{figure}

\paragraph{PSR~J1709-4429.} Very good PPA data are available for this pulsar, top panel of Fig.~\ref{fig:J1709-4429_chisq}. Here also, one geometrical configuration is highlighted, coincident with both radio and gamma-ray, middle panel. It produces a double peaked gamma-ray light-curve, bottom panel. The phase shift is $\phi_s=-0.1$ for $\alpha = 40\degr$ and $\zeta = 56\degr$.

\begin{figure}
	\centering
	\begin{tabular}{l}
	\includegraphics[width=\columnwidth,angle=-90]{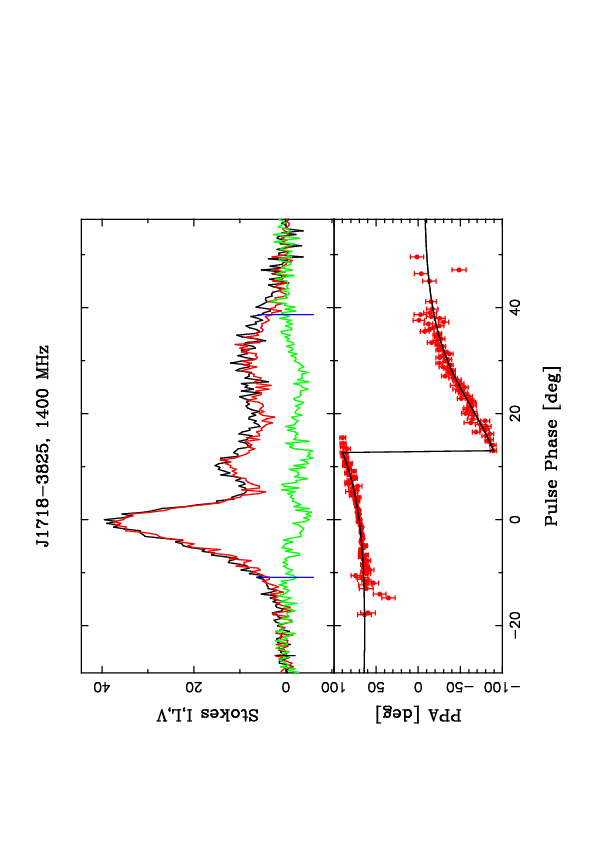} \\		
	\includegraphics[width=\columnwidth]{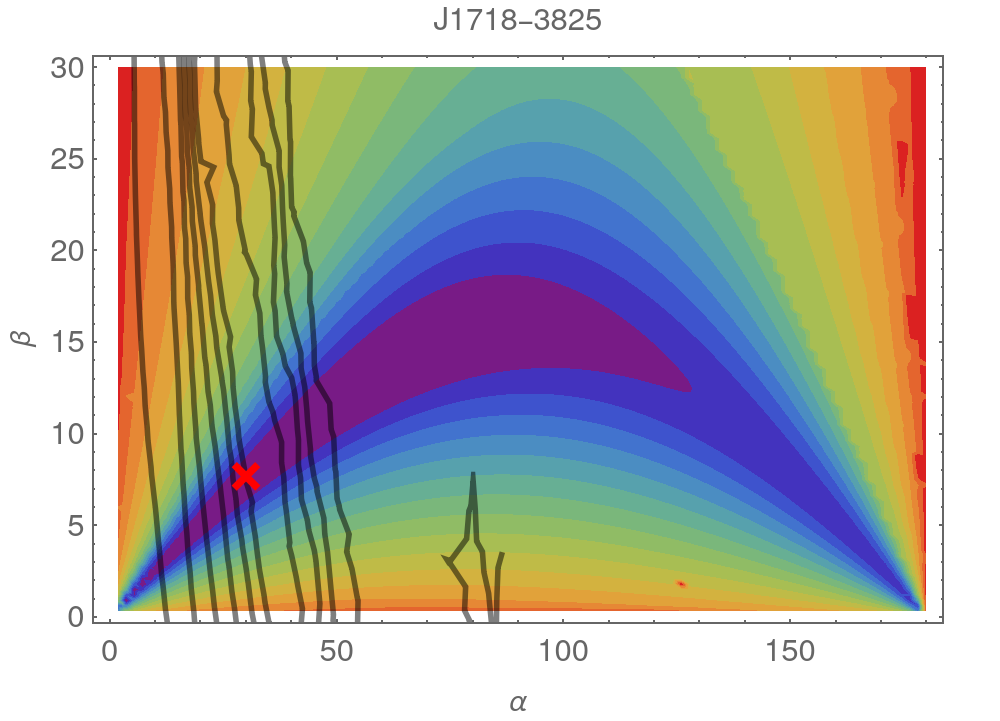} \\
	\includegraphics[width=\columnwidth]{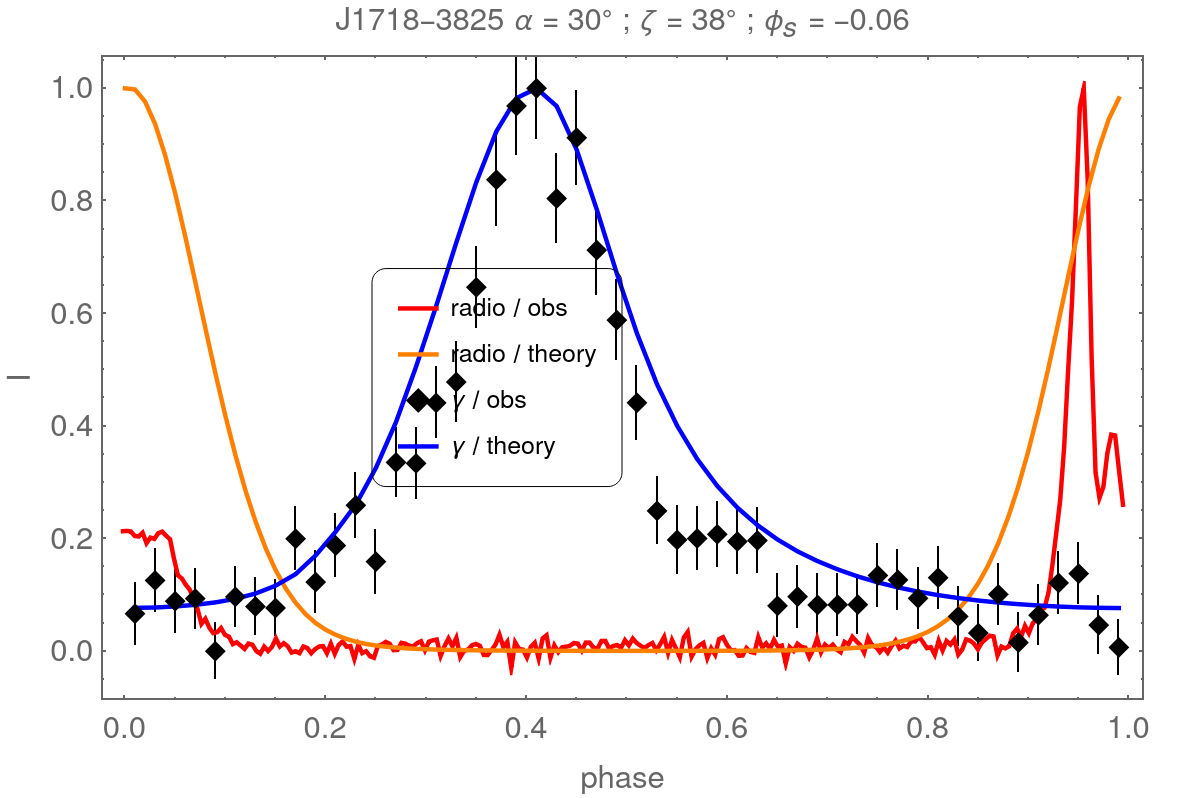}
	\end{tabular}	
	\caption{Same as Fig.~\ref{fig:J0631+1036chisq} but for J1718-3825.}
	\label{fig:J1718-3825_chisq}
\end{figure}

\paragraph{PSR~J1718-3825.} In the same vain as for the previous pulsar, the top panel of Fig.~\ref{fig:J1718-3825_chisq} show the PPA evolution leading to a well defined radio constrained geometry, in the middle panel, picking out one configuration with the red cross coincident with both wavelengths, producing a single peaked gamma-ray profile. The phase shift is $\phi_s=-0.06$ for $\alpha = 30\degr$ and $\zeta = 38\degr$.

\begin{figure}
	\centering
	\begin{tabular}{l}
	\includegraphics[width=\columnwidth,angle=-90]{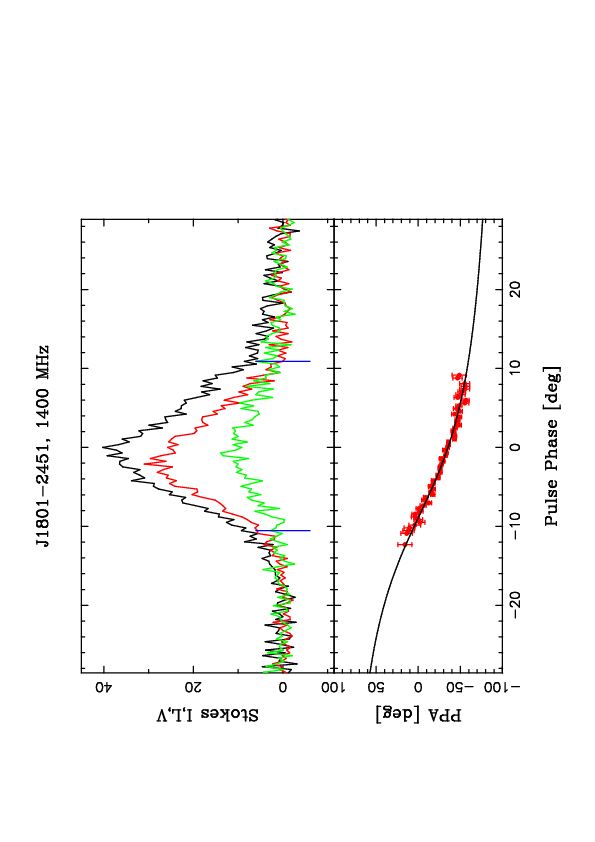} \\		
	\includegraphics[width=\columnwidth]{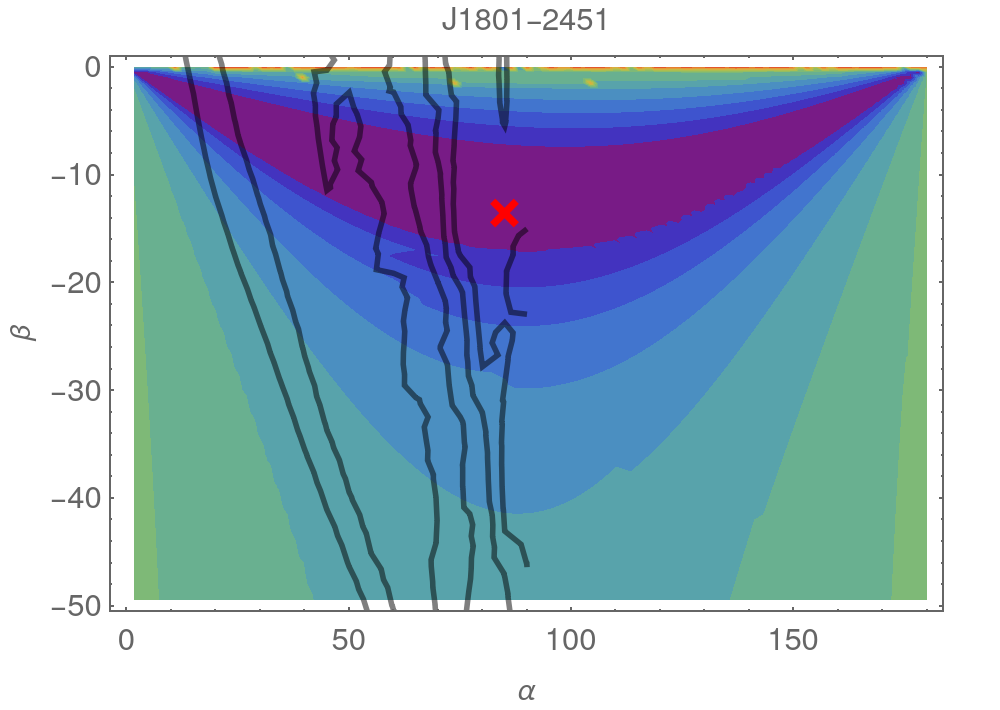} \\
	\includegraphics[width=\columnwidth]{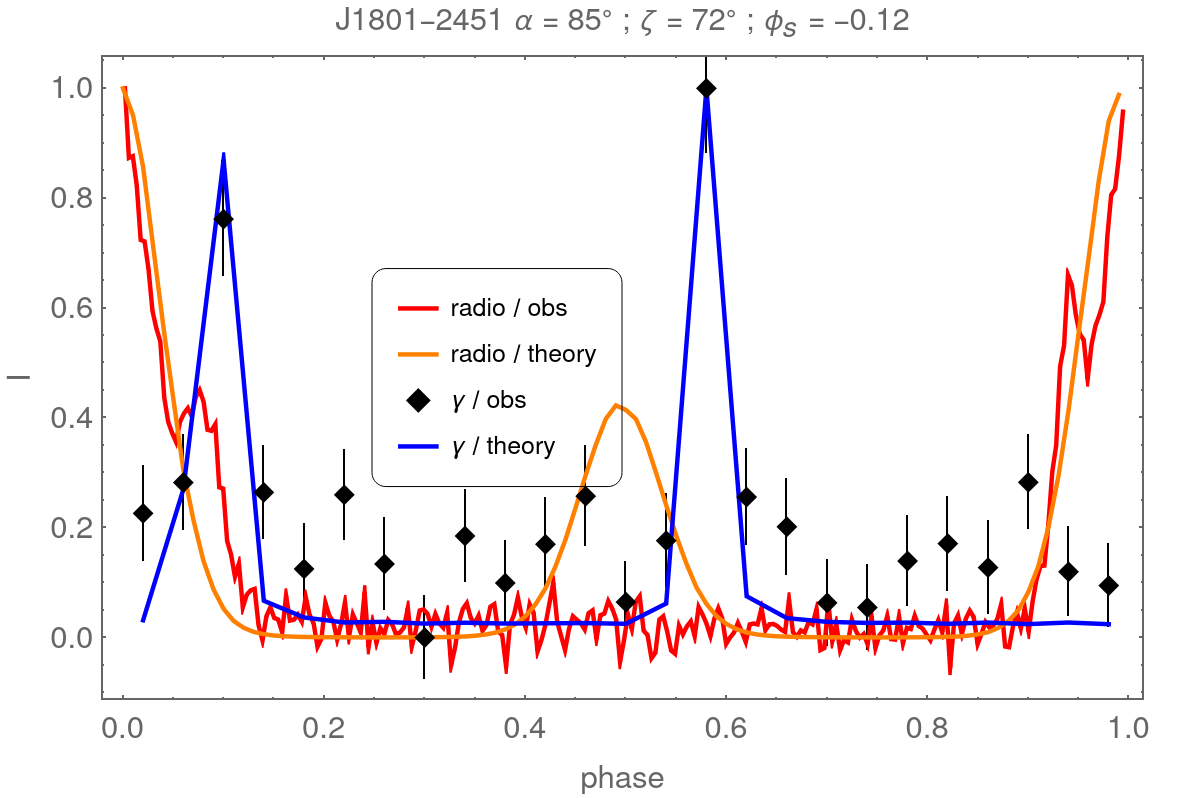}
	\end{tabular}	
	\caption{Same as Fig.~\ref{fig:J0631+1036chisq} but for J1801-2451.}
	\label{fig:j1801-2451chisq}
\end{figure}

\paragraph{PSR~J1801-2451.} This pulsar shows a strongly double peaked gamma-ray profile with one radio pulse, bottom panel of Fig.~\ref{fig:j1801-2451chisq}. The good radio PPA data, top panel, furnish a reasonable constrain on the geometry, middle panel. The best fit is nearly an orthogonal rotator with a line of sight almost located in the equatorial plane. The phase shift is $\phi_s=-0.12$ for $\alpha = 85\degr$ and $\zeta = 72\degr$. This configuration shows a second but weak radio peak, the interpulse, not seen in the data. We therefore conclude that the true geometry must slightly deviate from our choice, in addition to the fact that we do not model the radio emission cone. This second pulse would disappear if a smaller cone of emission is used.

\begin{figure}
	\centering
	\begin{tabular}{l}
	\includegraphics[width=\columnwidth,angle=-90]{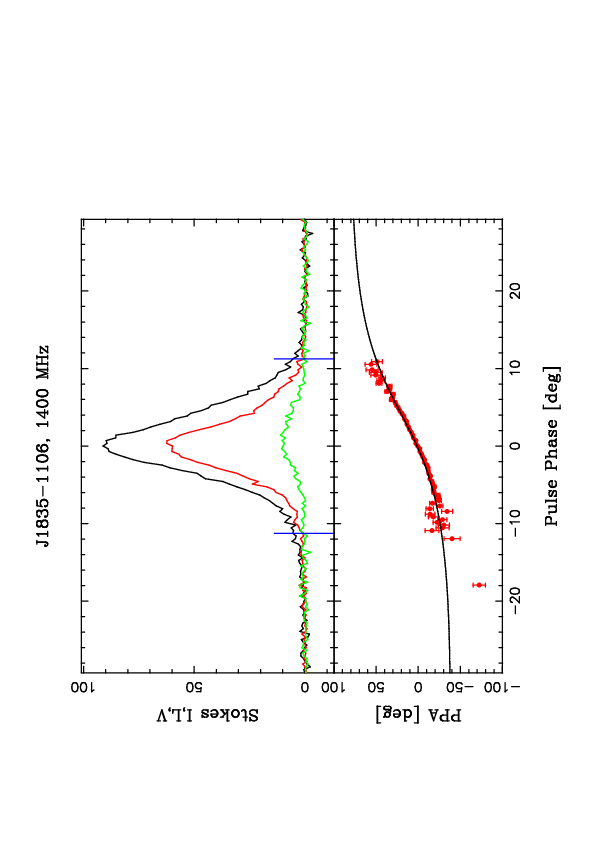} \\		
	\includegraphics[width=\columnwidth]{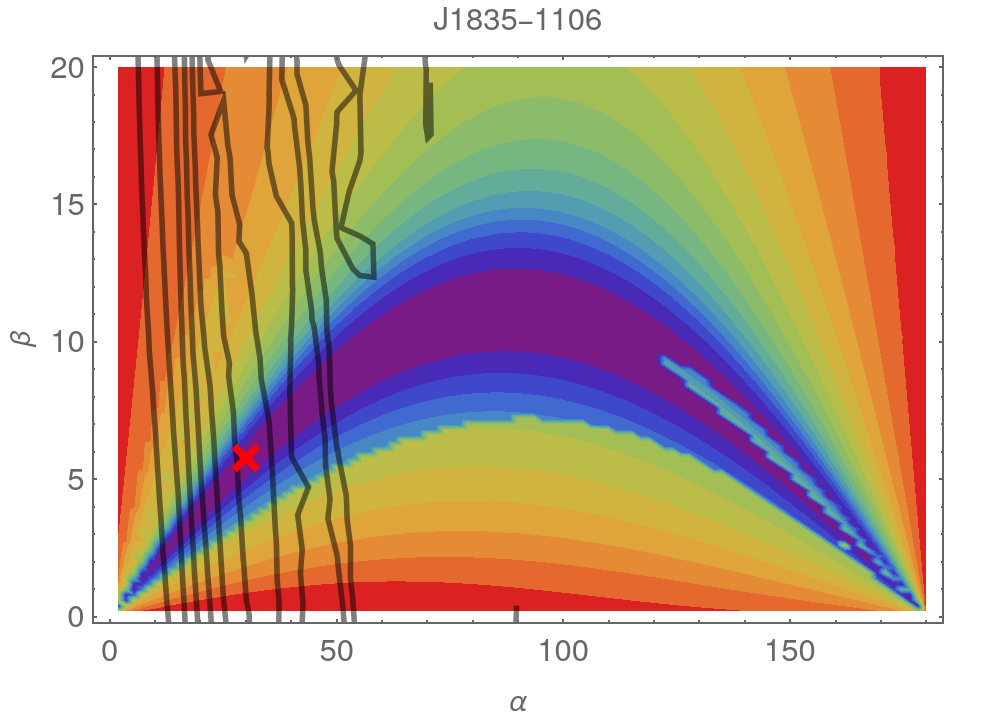} \\
	\includegraphics[width=\columnwidth]{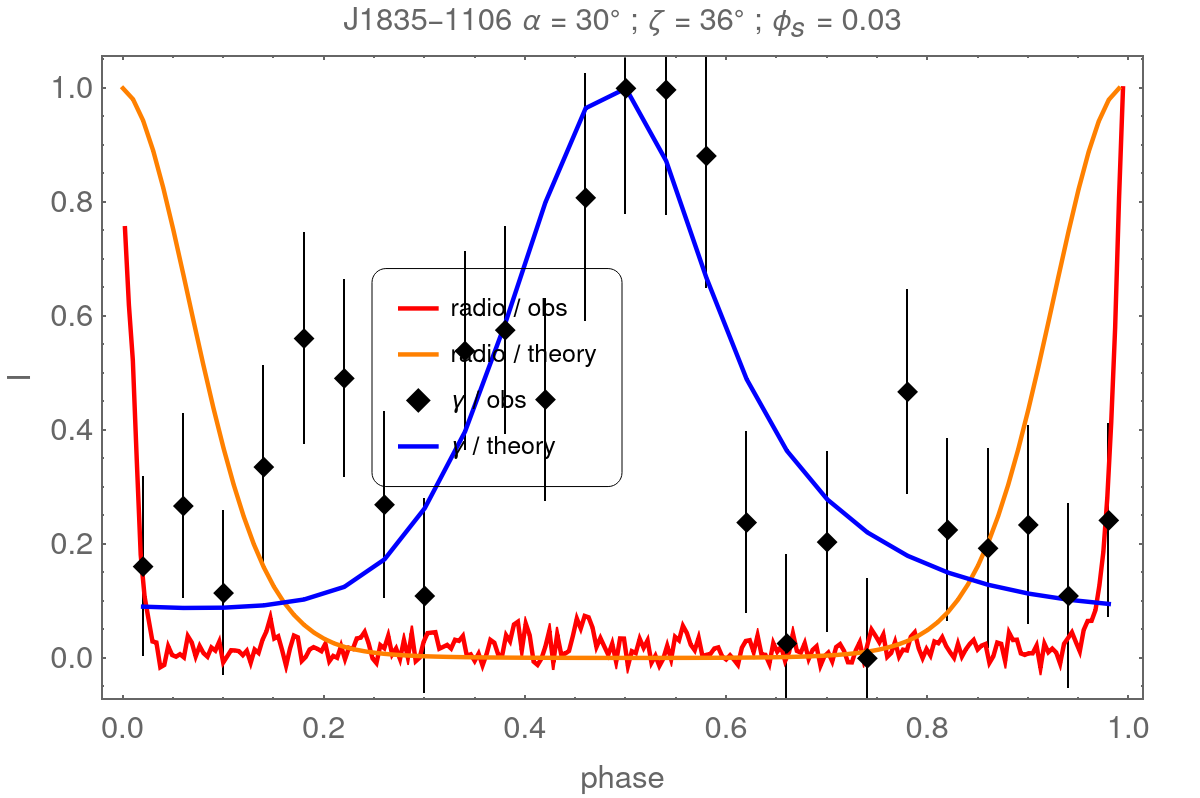}
	\end{tabular}	
	\caption{Same as Fig.~\ref{fig:J0631+1036chisq} but for J1835-1106.}
	\label{fig:j1835-1106chisq}
\end{figure}

\paragraph{PSR~J1835-1106.} The radio data of this pulsar are shown in the top panel of Fig.~\ref{fig:j1835-1106chisq}. The RVM constrain are well defined in the middle panel. We found a best fit with the red cross producing only one gamma-ray peak with a phase shift of $\phi_s=0.03$ for $\alpha = 30\degr$ and $\zeta = 36\degr$. The second gamma-ray pulsar catalogue reports the presence of two peaks. Therefore here again, the joined radio gamma-ray constrain seems to lead to some inconsistency with data. These discrepancies must be carefully analysed, but we are waiting for better observations from Fermi/LAT supposed to be published in a third gamma-ray pulsar catalogue before exploring the implication for the emission mechanism.

\begin{figure}
	\centering
	\begin{tabular}{l}
	\includegraphics[width=\columnwidth,angle=-90]{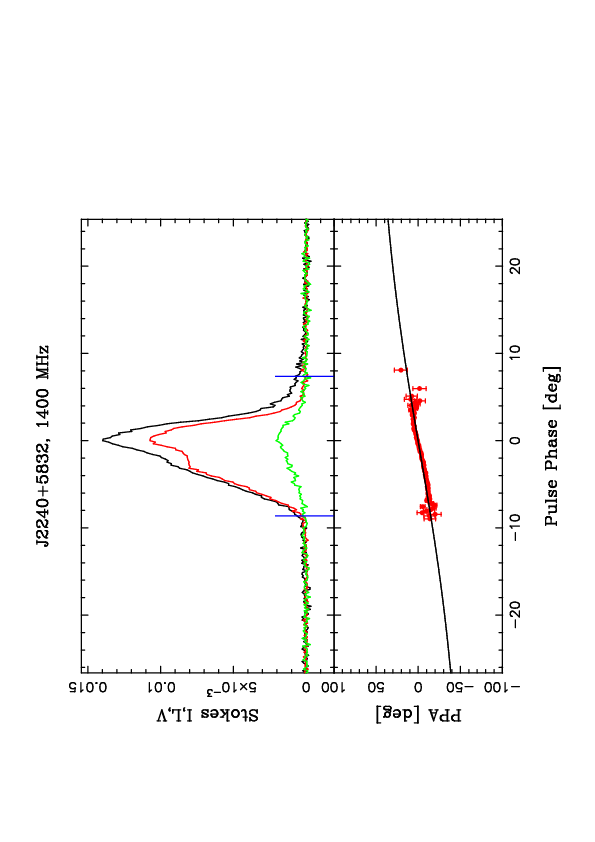} \\		
	\includegraphics[width=\columnwidth]{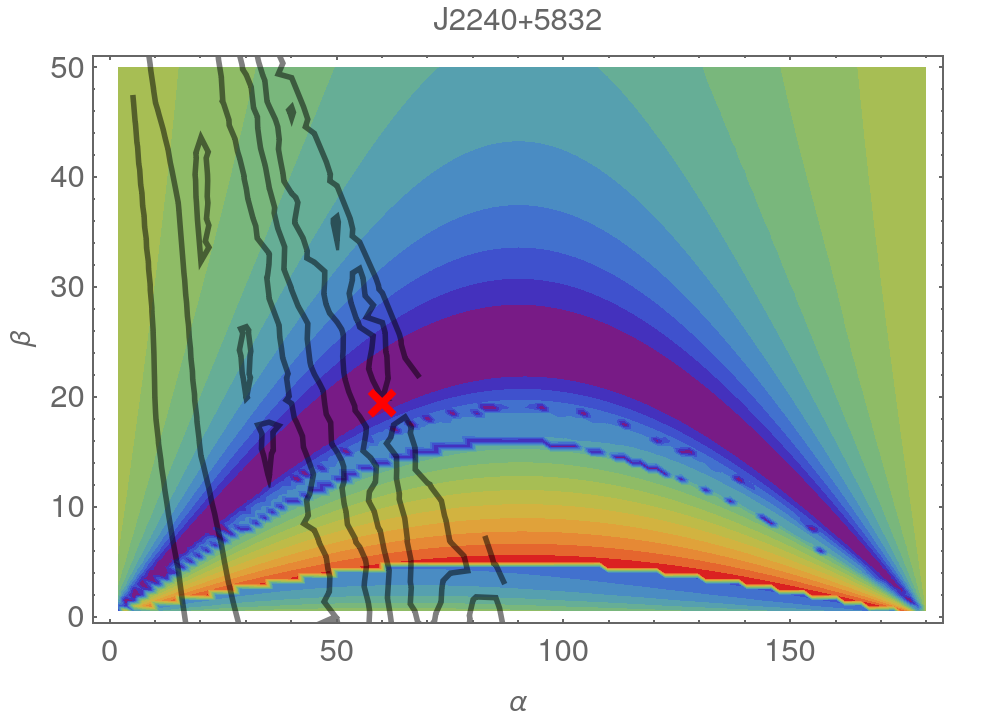} \\
	\includegraphics[width=\columnwidth]{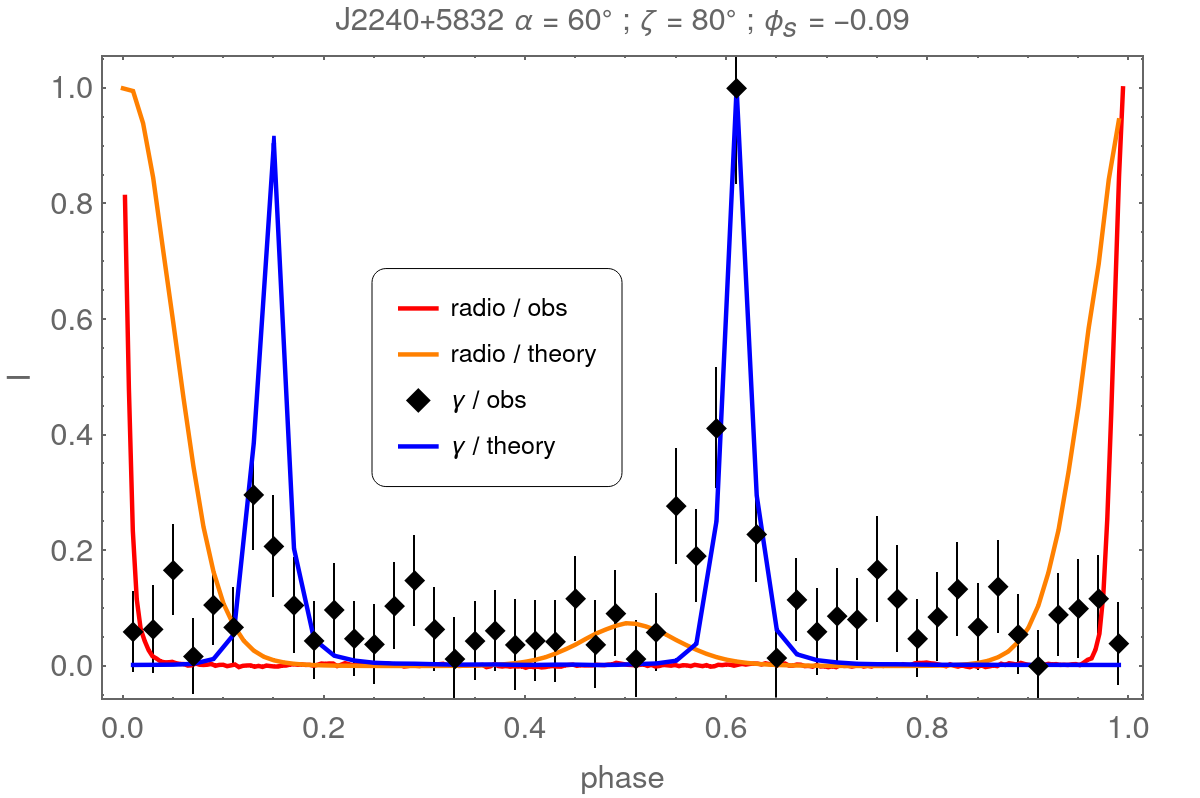}
	\end{tabular}	
	\caption{Same as Fig.~\ref{fig:J0631+1036chisq} but for J2240+5832.}
	\label{fig:j2240+5832chisq}
\end{figure}

\paragraph{PSR~J2240+5832.} This is the last example of our sample showing good radio PPA data, top panel of Fig.~\ref{fig:j2240+5832chisq} \citep{theureau_psrs_2011}. The associated RVM constrain are good, middle panel, the radio gamma-ray overlapping region leading to the best geometry depicted by the red cross. The corresponding gamma-ray light-curve is shown in the bottom panel for a phase shift of $\phi_s=-0.09$ with $\alpha=60\degr$ and $\zeta=80\degr$. The weak radio interpulse is predicted, but due to the large opening of the emission cone. According to the narrow width of the radio pulse, this emission cone is largely overestimated and should disappear when shrinking to the real size of radio observations, in red solid line.

\subsection{Only gamma-ray fits}

\begin{figure*}
	\centering
	\begin{tabular}{ccc}
	\includegraphics[width=0.33\textwidth]{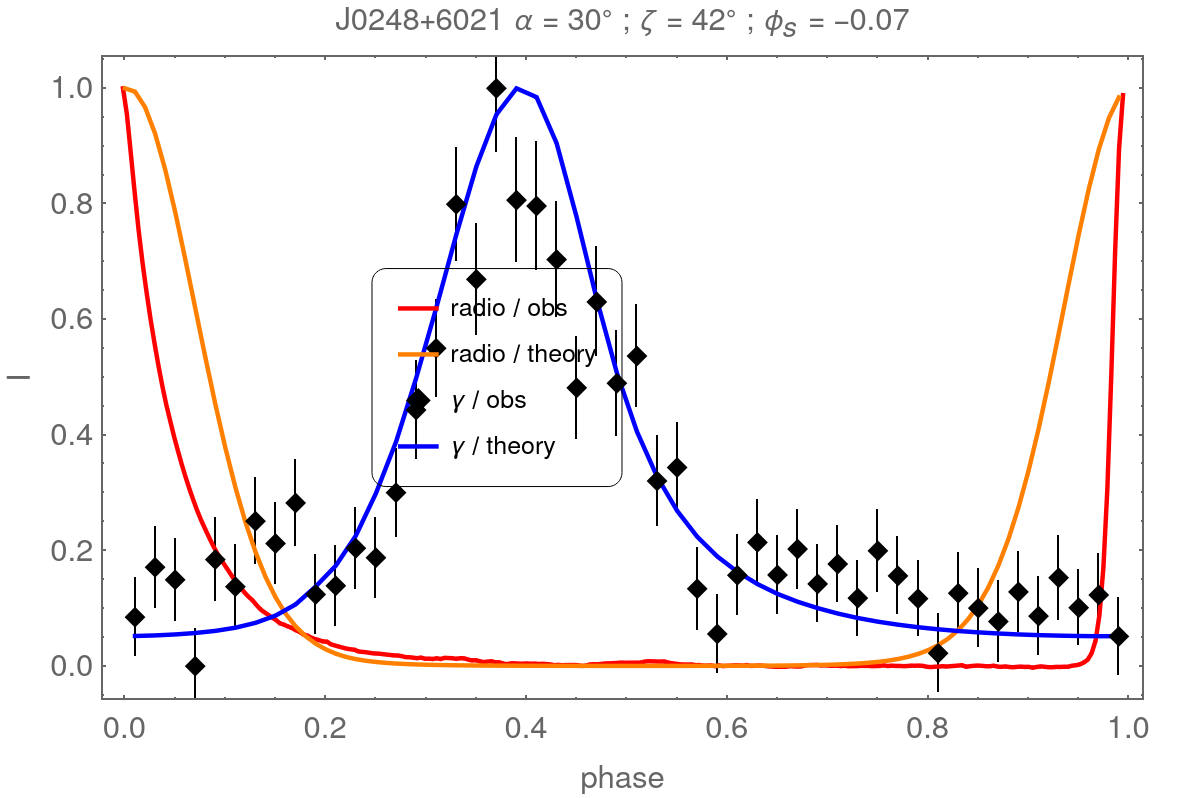} &
	\includegraphics[width=0.33\textwidth]{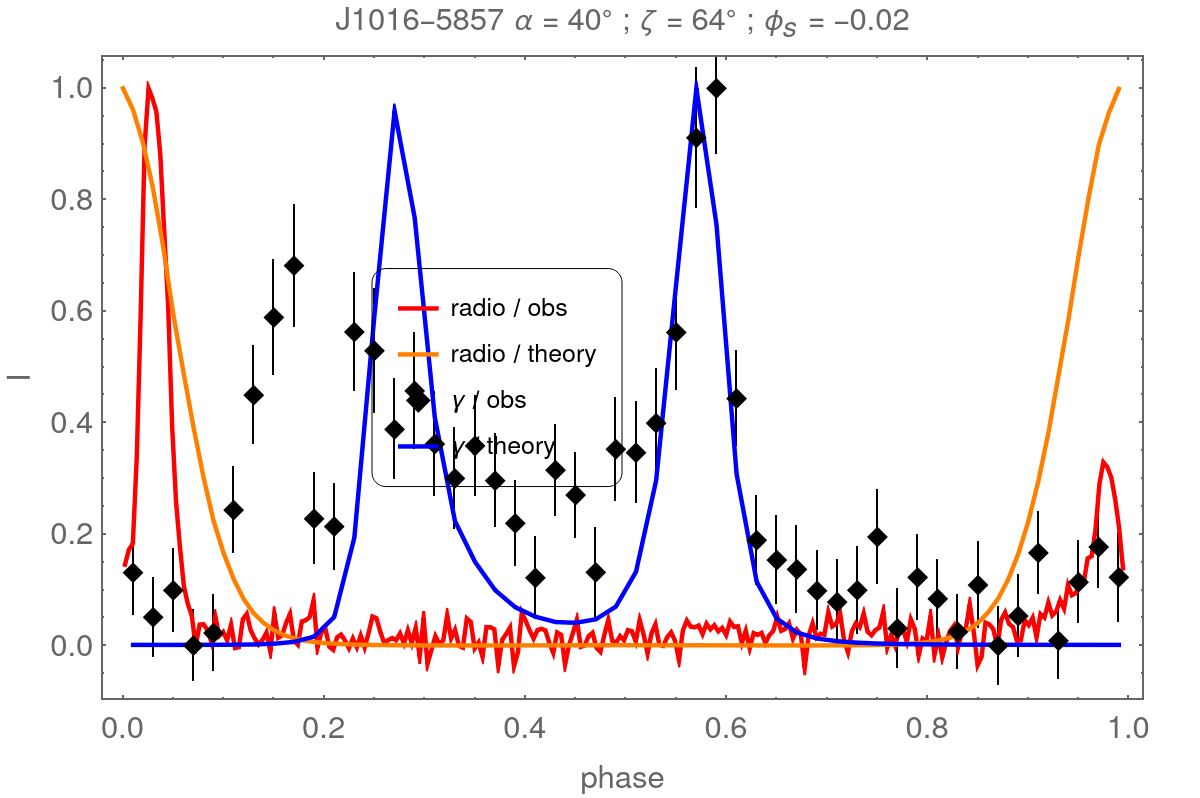} &
	\includegraphics[width=0.33\textwidth]{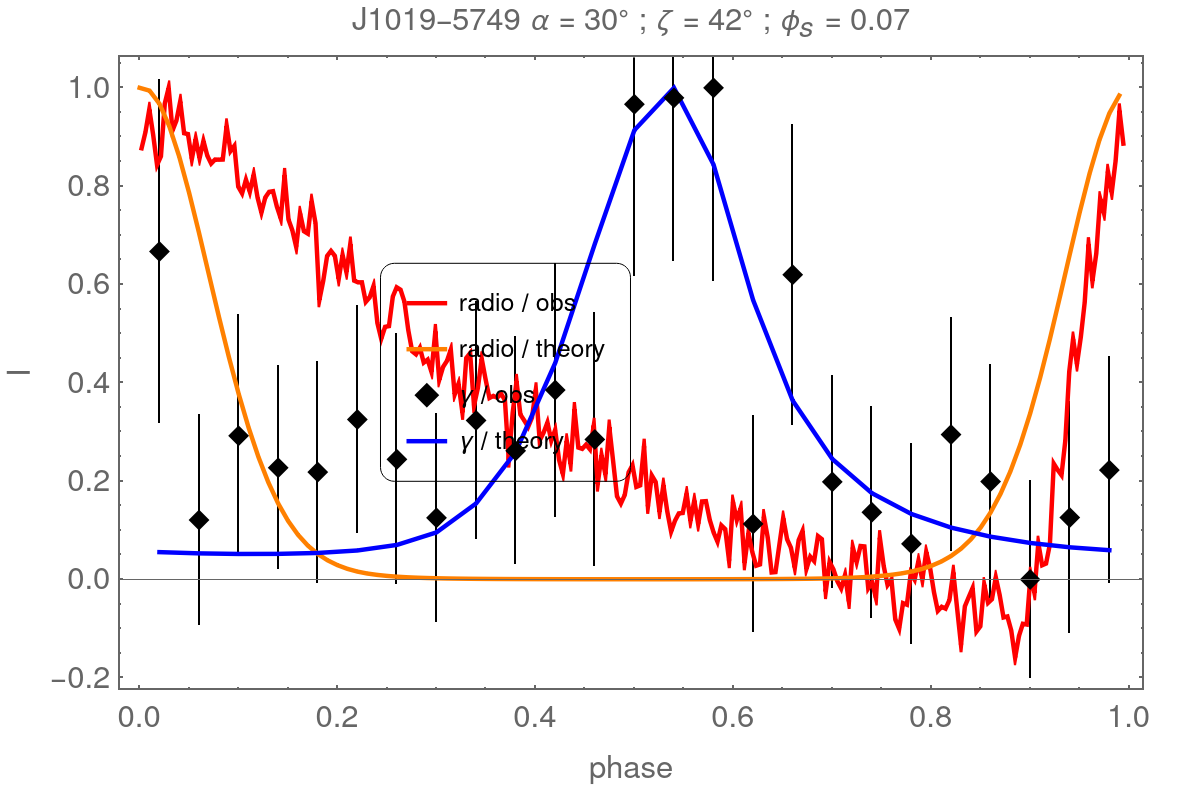} \\
	\includegraphics[width=0.33\textwidth]{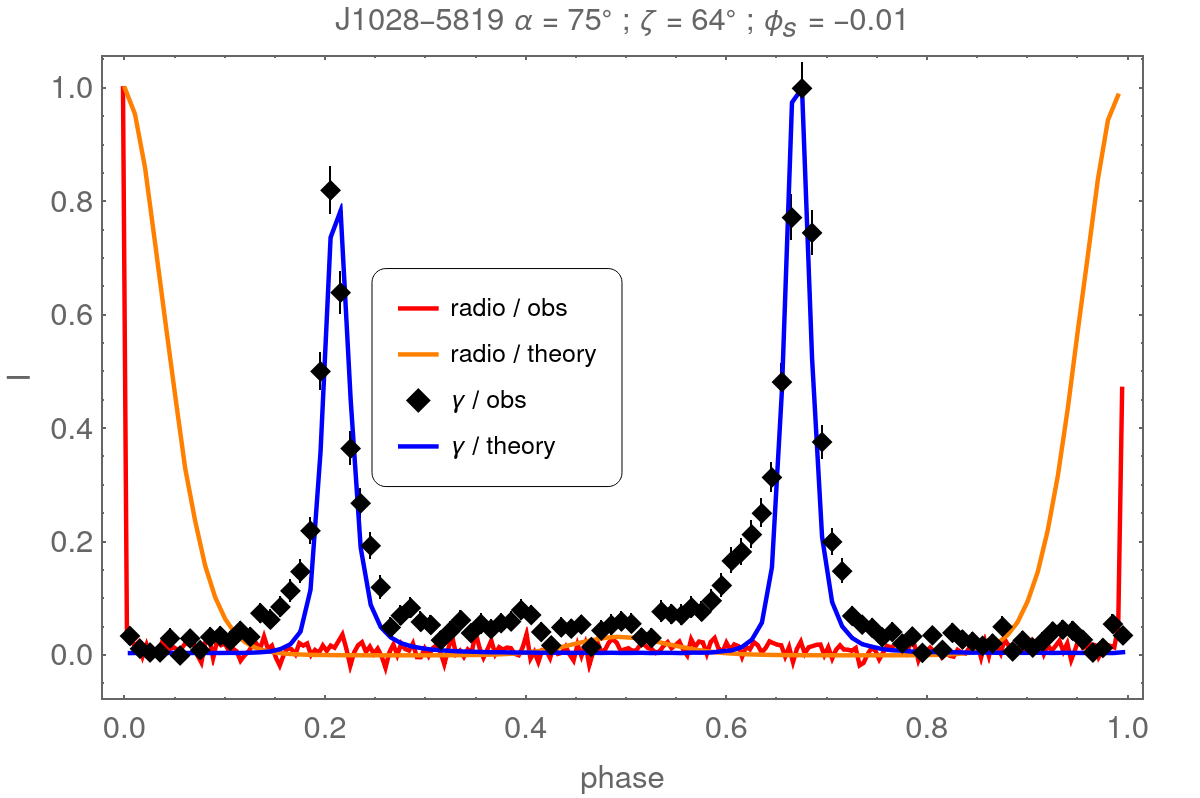} & 
	\includegraphics[width=0.33\textwidth]{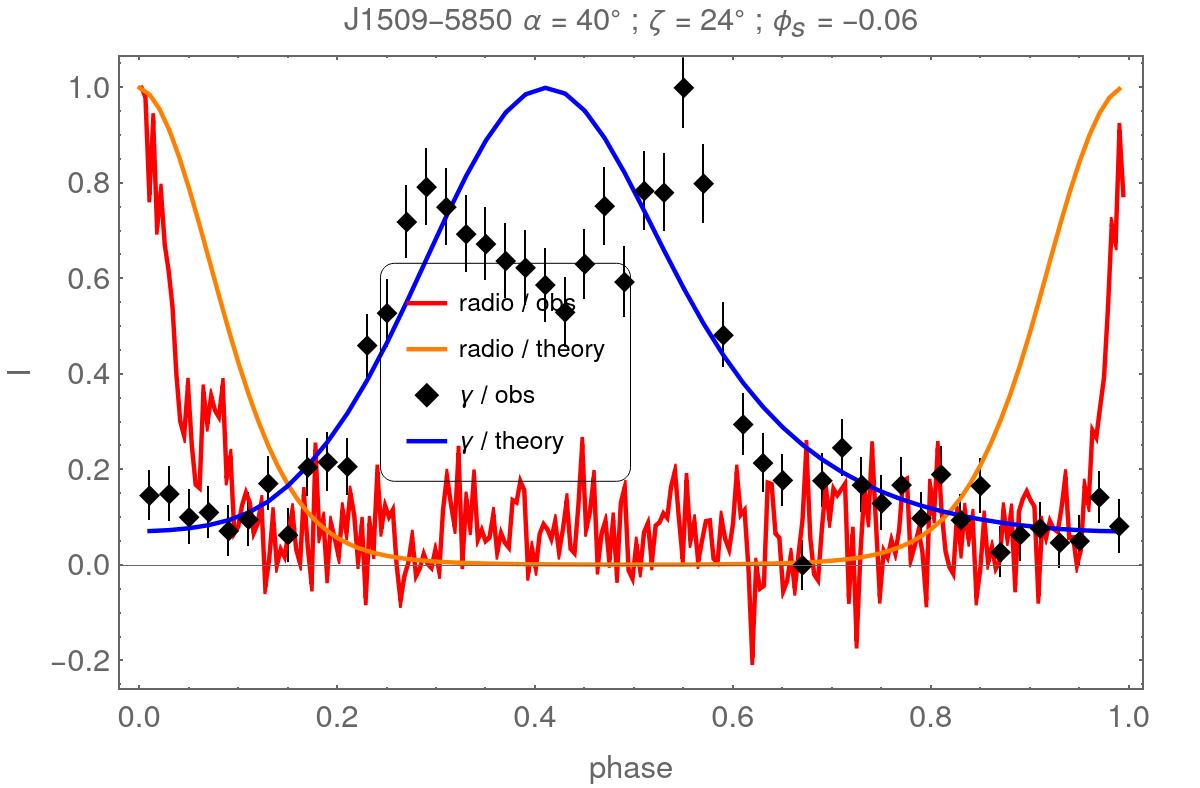} &
	\includegraphics[width=0.33\textwidth]{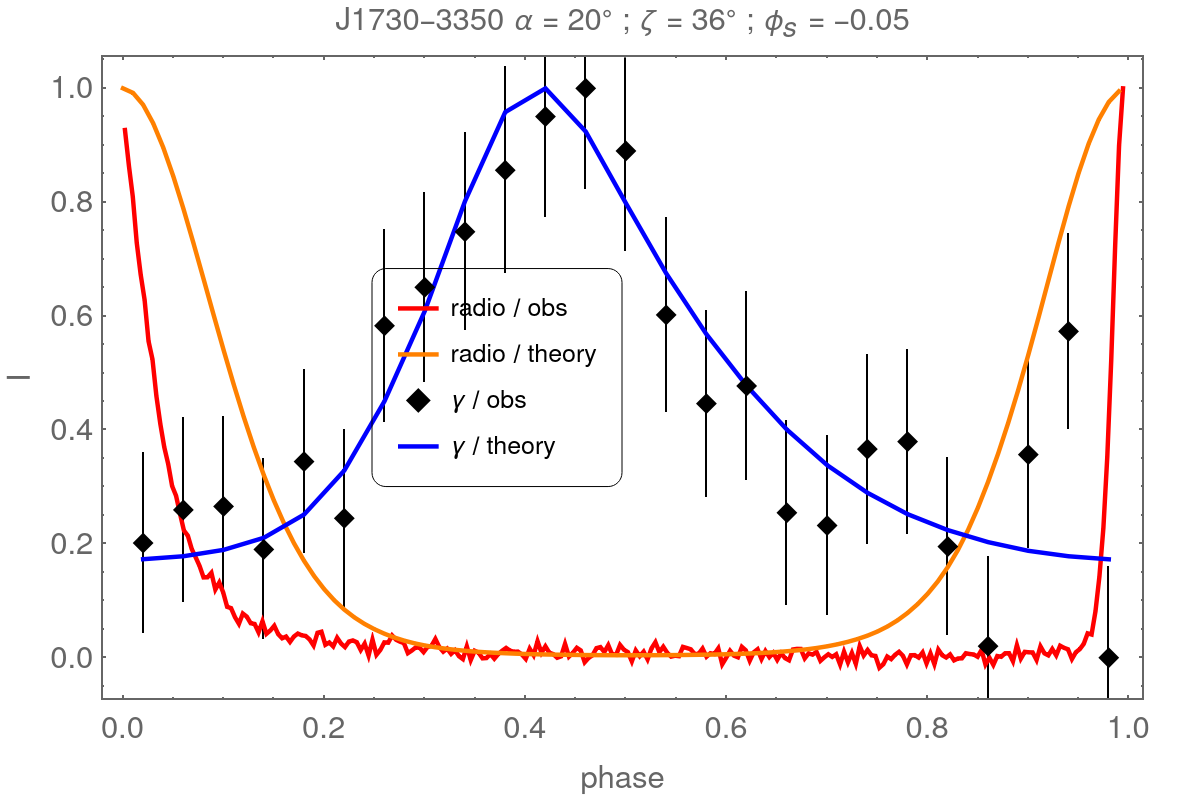} \\
	\includegraphics[width=0.33\textwidth]{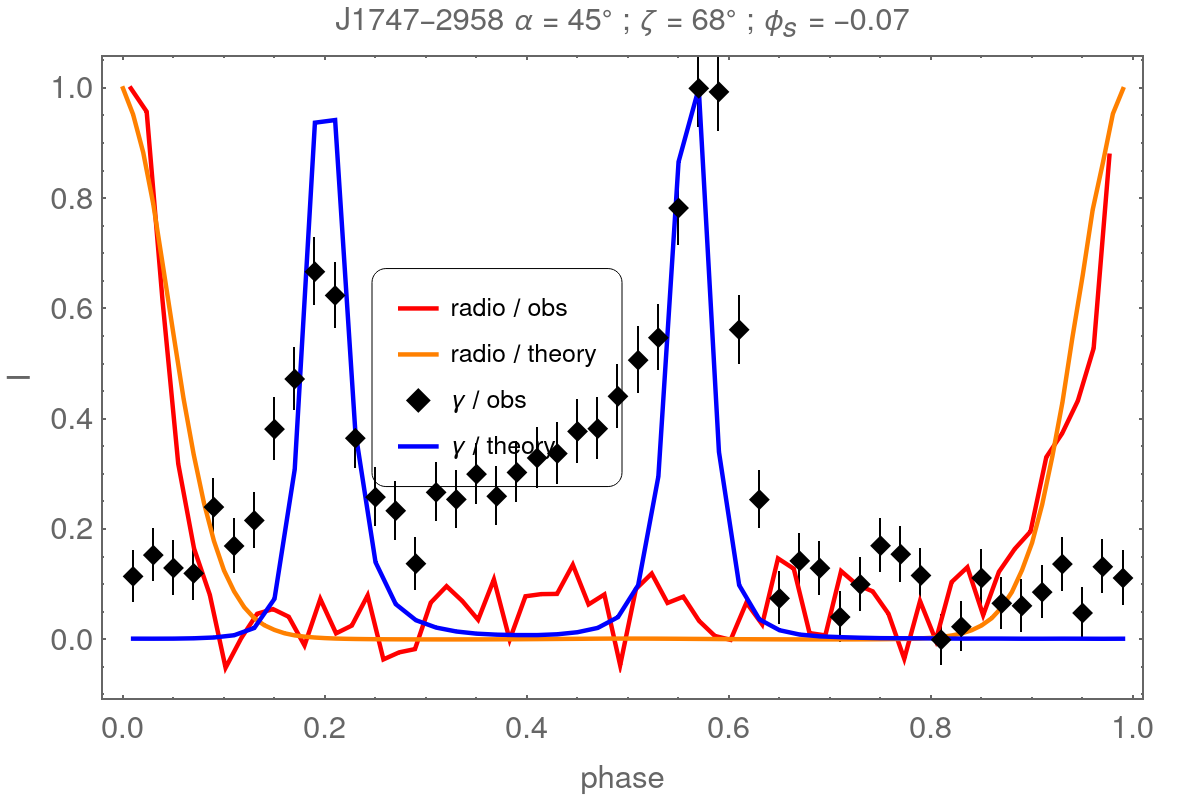} &
	\includegraphics[width=0.33\textwidth]{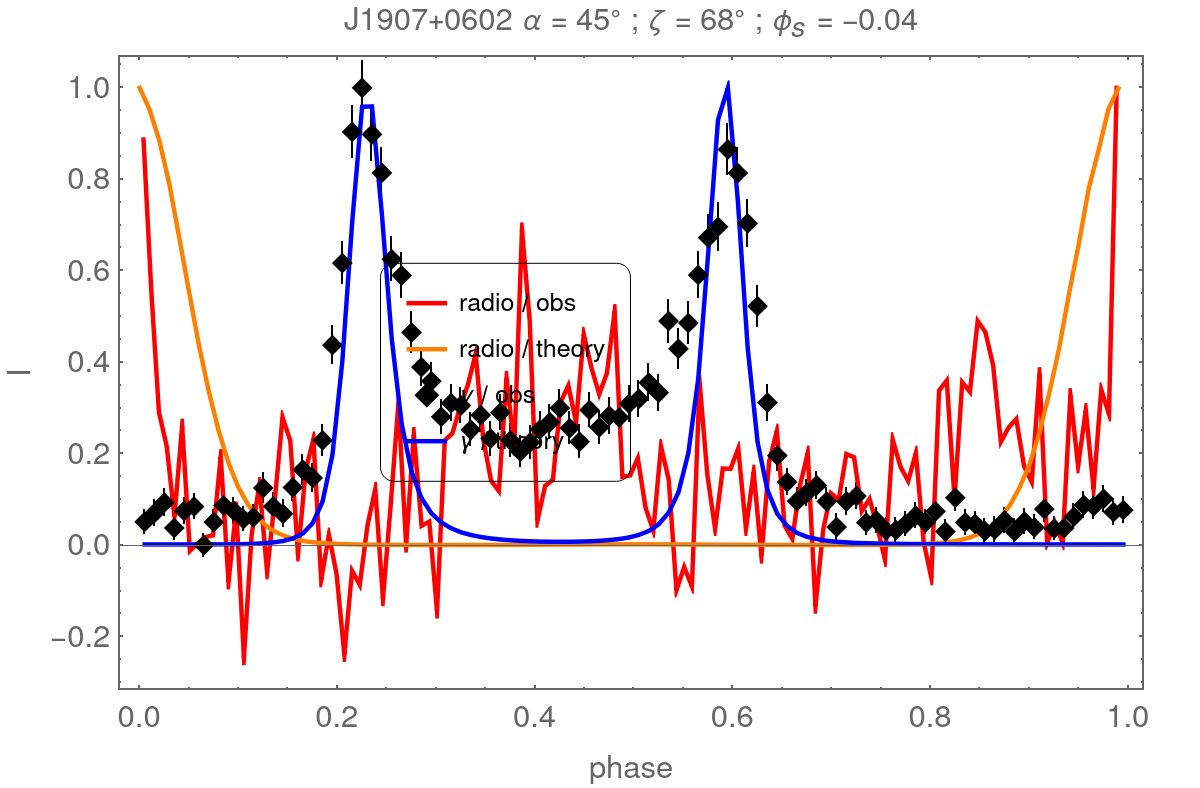} &
	\includegraphics[width=0.33\textwidth]{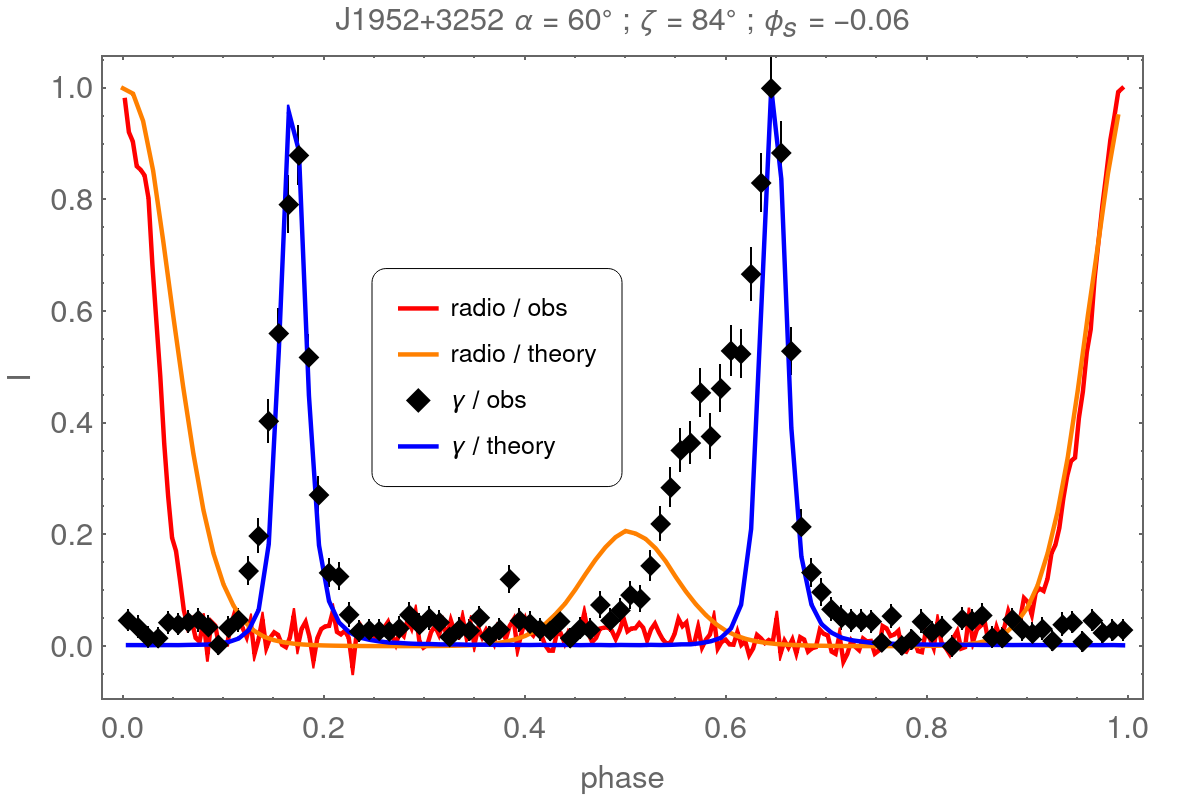} \\
	\includegraphics[width=0.33\textwidth]{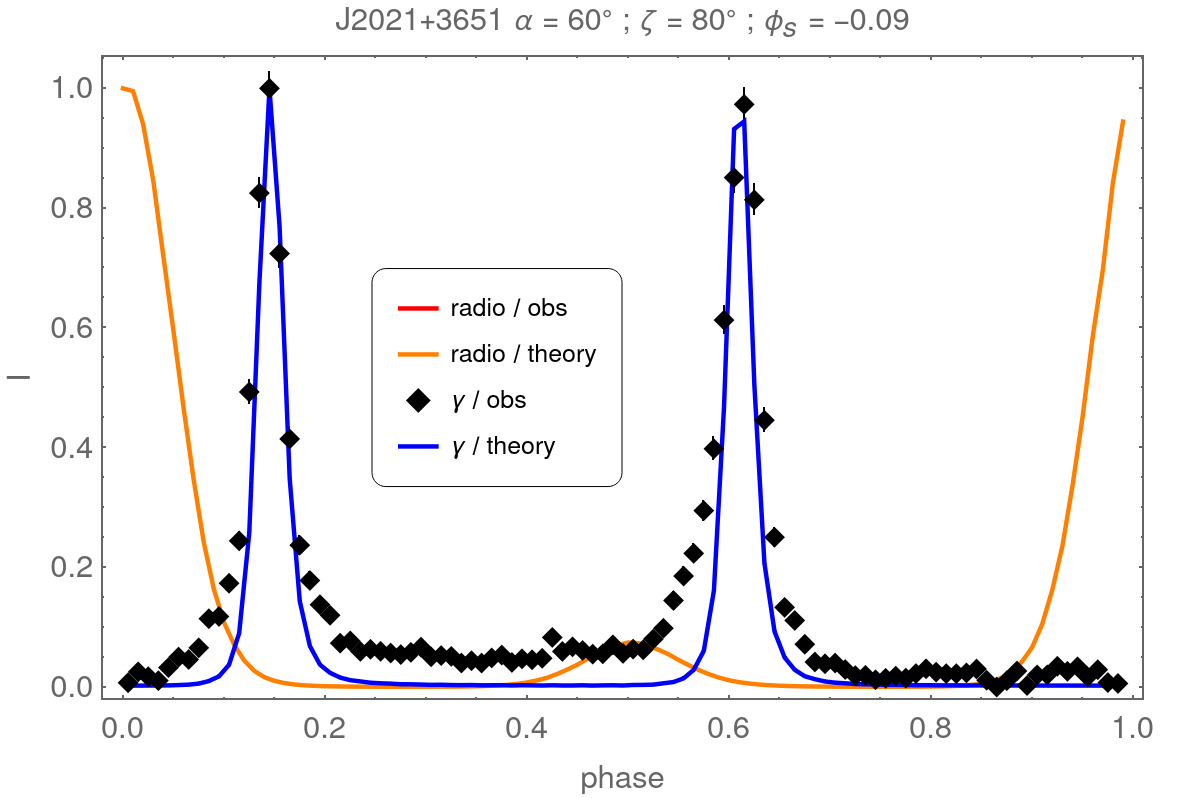} &
	\includegraphics[width=0.33\textwidth]{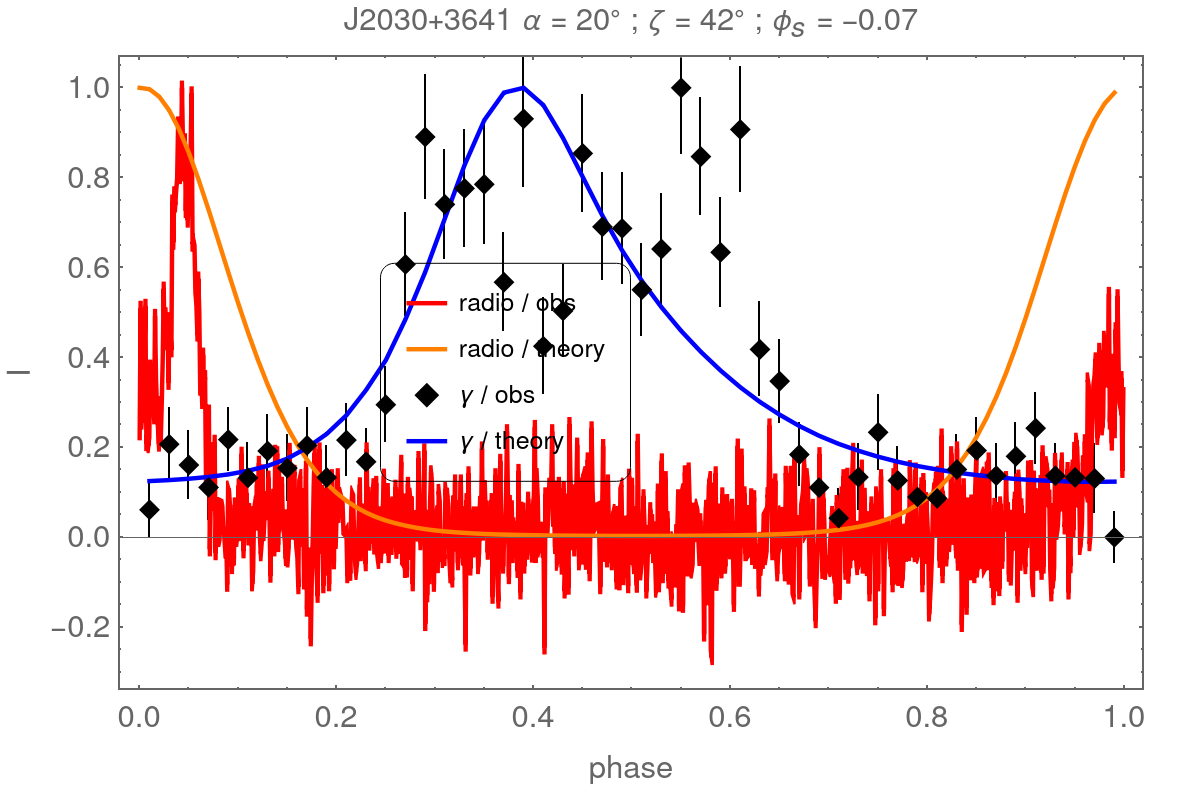} &
	\includegraphics[width=0.33\textwidth]{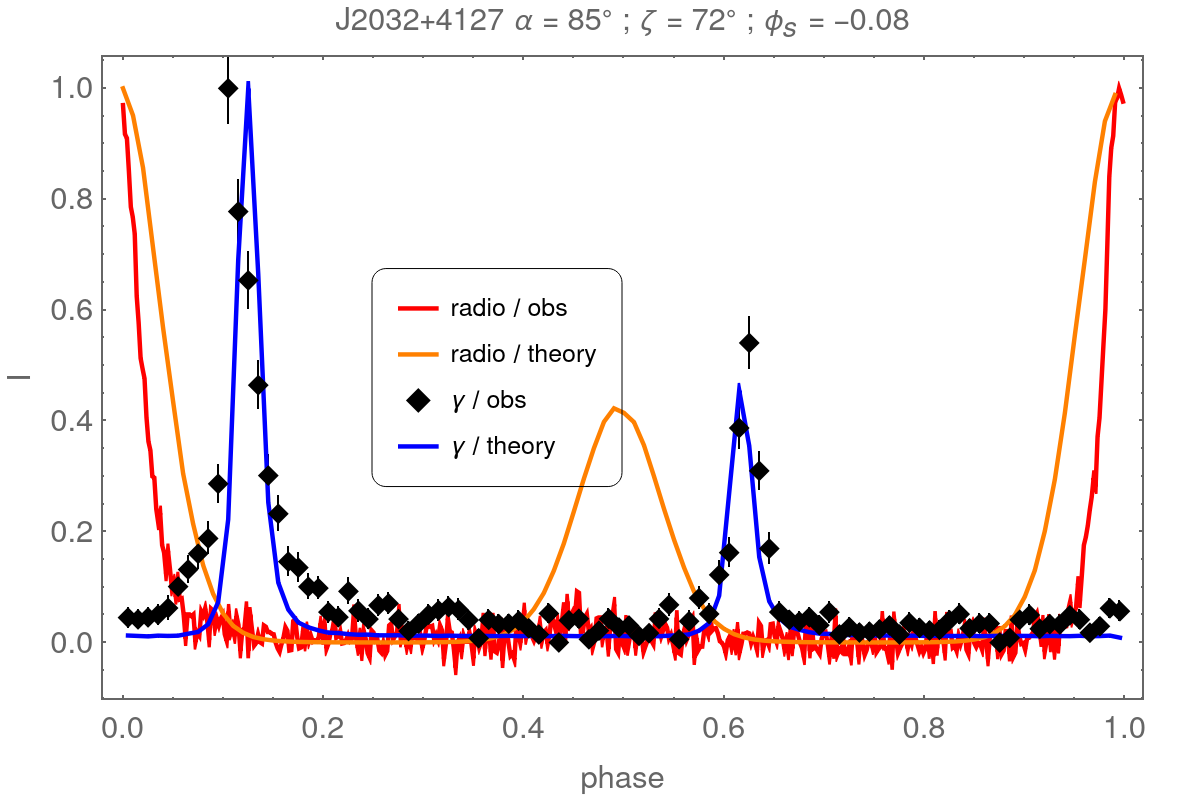} \\
	\includegraphics[width=0.33\textwidth]{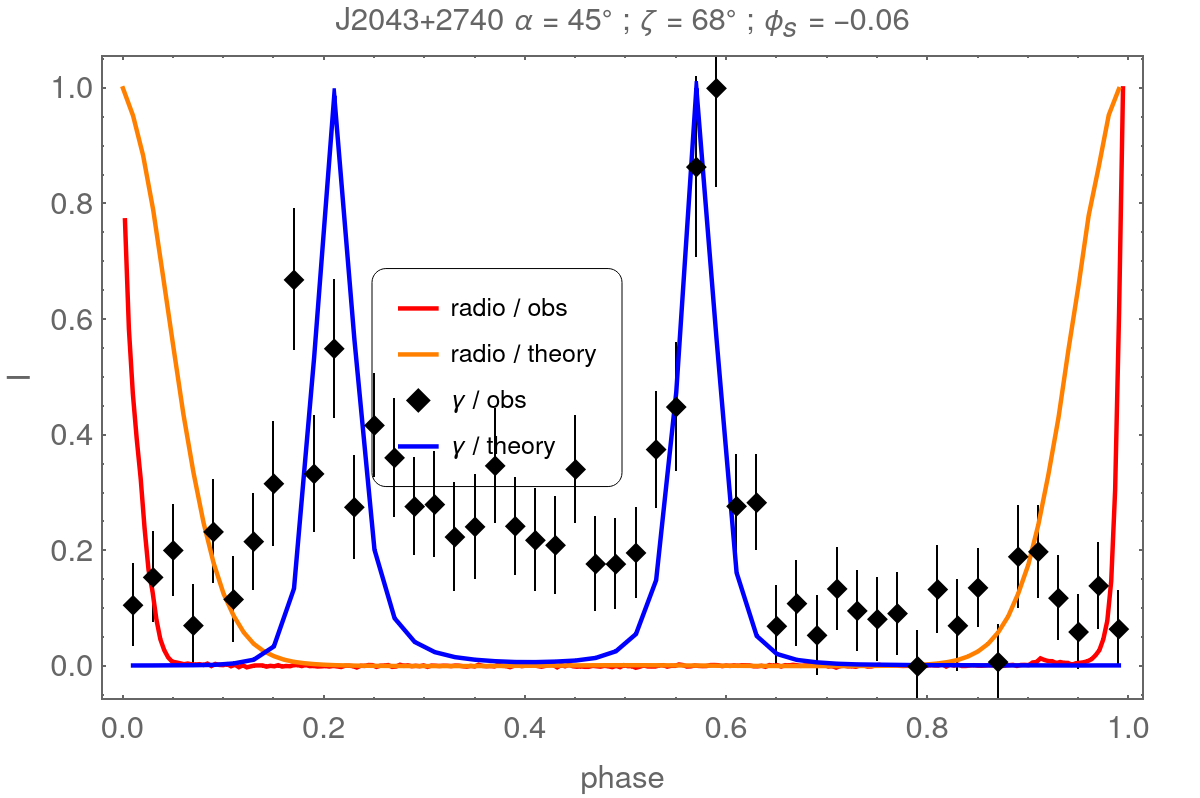} &
	\includegraphics[width=0.33\textwidth]{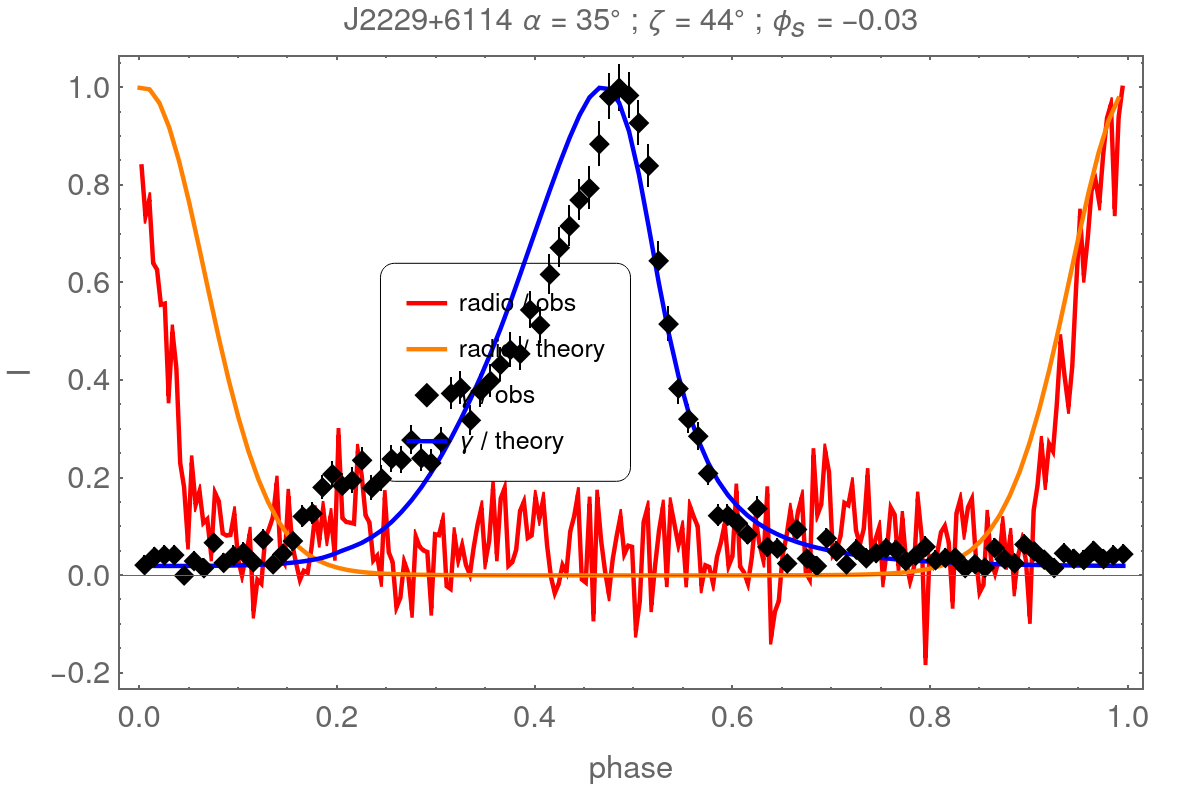} 
	\end{tabular}
	\caption{Best fit parameters and gamma-ray light-curves for the second part of the young radio loud gamma-ray pulsar sample not having usable RVM fits.}
	\label{fig:gamma_only}
\end{figure*}

The second part of the sample includes only gamma-ray pulsars not showing a radio signal loud enough for performing a reasonable RVM fit as done in the previous section. Nevertheless, gamma-ray light-curve fitting alone can already help to constrain the geometry of many individual pulsars. Below, we summarize the best fit for some of these young radio gamma-ray pulsars. The figure~\ref{fig:gamma_only} gives an overview of our fitting results.

\paragraph{PSR~JJ0248+6021.} It is a single peaked radio and gamma-ray pulsar. By only fitting the gamma-ray light-curve and its delay compared to the radio profile, we arrive at the best geometry given by a phase shift of $\phi_s=-0.07$ for $\alpha = 30\degr$ and $\zeta = 42\degr$. Some other very similar angles also give reasonable fits but they are not shown.

\paragraph{PSR~J1016--5857.} This pulsar shows two sharp gamma-ray peaks separated by a kind of bridge emission. We are able to fit these two peaks but not the bridge emission. The phase shift of $\phi_s=-0.02$ for $\alpha = 40 \degr$ and $\zeta = 64\degr$.

\paragraph{PSR~J1019-5749.} The radio pulse of this pulsar spans almost the entire period but this is certainly an artefact due to its large dispersion measure. Having also only one gamma-ray peak, we indeed found a small obliquity of $\alpha \approx30\degr$ with $\zeta=42\degr$ and an additional offset of $\phi_s=0.07$. Larger obliquities are also permissible with slightly less good fits. They are not shown.

\paragraph{PSR~J1028-5819.} This pulsar shows one radio pulse and two narrow strongly peaked gamma-ray pulses. The peak separation and shape are well fitted by the striped wind model with $\alpha \approx 75\degr$ and $\zeta=64\degr$ with a small offset of $\phi_s=-0.01$.

\paragraph{PSR~J1509-5850.} Similar two the previous case but without interpulse emission, PSR~J1509-5850 shows two not well separated gamma-ray peaks. The best fit is associated with a phase shift of $\phi_s=-0.06$ for $\alpha = 40\degr$ and $\zeta=24\degr$. We did not find any better geometry reproducing two unresolved gamma-ray peaks.

\paragraph{PSR~J1730-3350.} This is a single radio and gamma-ray pulse profile pulsar. Best fitting parameters are a phase shift of $\phi_s=-0.05$ for $\alpha = 20\degr$ and $\zeta = 36\degr$.

\paragraph{PSR~J1747-2958.} This pulsar is similar to PSR~J1016--5857, showing the same radio and gamma-ray profiles with a weak bridge emission. Its fitting parameters are therefore close to the one used for PSR~J1016--5857 with a phase shift of $\phi_s=-0.07$ for $\alpha = 45\degr$ and $\zeta = 68\degr$.

\paragraph{PSR~J1907+0602.} This pulsar has two separated gamma-ray peaks with a bridge emission and very noisy radio pulse with possibly an interpulse emission. Our best fit sets $\alpha=45\degr$ and $\zeta=68\degr$ and the phase shift to $\phi_s=-0.04$. Nevertheless from the radio interpulse, we would expect an orthogonal rotator.

\paragraph{PSR~J1952+3252.} This pulsar has two well separated but asymmetrical gamma-ray peaks and one radio pulse. The asymmetry cannot be explained by our symmetrical striped wind model. The parameters used in the plot are a phase shift of $\phi_s=-0.06$ for $\alpha = 60\degr$ and $\zeta=84\degr$.

\paragraph{PSR~J2021+3651.} This is again an interesting example of prominent and symmetric gamma-ray pulse profiles and a clear single radio pulse. Best fit parameters are a phase shift of $\phi_s=-0.09$ for $\alpha = 60\degr$ and $\zeta=80\degr$.

\paragraph{PSR~J2030+3641.} This is a very noisy radio and gamma-ray pulsar. It has been fitted by a single gamma-ray profile such that the phase shift is $\phi_s=-0.07$ for $\alpha = 20\degr$ and $\zeta = 42\degr$.

\paragraph{PSR~J2032+4127.} Another example of two narrow gamma-ray peaks with a single radio pulse. It has been fitted by an almost orthogonal rotator with a phase shift of $\phi_s=-0.08$ for $\alpha = 85\degr$ and $\zeta= 72\degr$.

\paragraph{PSR~J2043+2740.} A noisy gamma-ray pulsar with two pronounced gamma-ray pulses and bridge emission. The phase shift is $\phi_s=-0.06$ for $\alpha = 45\degr$ and $\zeta = 68\degr$.

\paragraph{PSR~J2229+6114.} Finally, the single gamma-ray peak pulsar with asymmetrical leading and trailing edge has been fitted with phase shift of $\phi_s=-0.03$ for $\alpha = 35\degr$ and $\zeta = 44\degr$.

\subsection{Summary}

Gathering all the results from the two previous sections, our best fit values for the angles $\alpha$ and $\zeta$ and for the phase offset $\phi_s$ are summarized in Table~\ref{tab:meilleur_fit}.
\begin{table}
\begin{tabular}{cccc}
	\hline
	PSR	& $\alpha$ (in \degr) & $\zeta$ (in \degr) & $\phi_s$ \\
	\hline\hline
J0248+6021 &  30 & 42 & -0.07 \\
J0631+1036$^{\ast}$ &  40 & 36 &  0.01 \\
J0659+1414$^{\ast}$ &  45 & 32 & -0.23 \\
J0742-2822$^{\ast}$ & 140 & 136 & 0.16 \\
J0835-4510$^{\ast}$ &  65 & 58 & -0.1 \\
J0908-4913$^{\ast}$ &  95 & 92 & -0.08 \\
J1016-5857 &  40 & 64 & -0.02 \\
J1019-5749 &  30 & 42 & 0.07 \\
J1028-5819 &  75 & 64 & -0.01 \\
J1048-5832$^{\ast}$ &  60 & 68 & -0.12 \\
J1057-5226$^{\ast}$ &  25 & 44 & -0.01 \\
J1119-6127$^{\ast}$ &  60 & 40 & -0.06 \\
J1357-6429$^{\ast}$ &  20 & 34 & -0.09 \\
J1420-6048$^{\ast}$ &  45 & 56 & -0.08 \\
J1509-5850 &  40 & 24 & -0.06 \\
J1648-4611$^{\ast}$ &  60 & 42 & -0.05 \\
J1702-4128$^{\ast}$ &  155 & 148 & -0.05 \\
J1709-4429$^{\ast}$ &  40 & 56 & -0.1 \\
J1718-3825$^{\ast}$ &  30 & 38 & -0.06 \\
J1730-3350 &  20 & 36 & -0.05 \\
J1747-2958 &  45 & 68 & -0.07 \\
J1801-2451$^{\ast}$ &  85 & 72 & -0.12 \\
J1835-1106$^{\ast}$ &  30 & 36 & 0.03 \\
J1907+0602 &  45 & 68 & -0.04 \\
J1952+3252 &  60 & 84 & -0.06 \\
J2021+3651 &  60 & 80 & -0.09 \\
J2030+3641 &  20 & 42 & -0.07 \\
J2032+4127 &  85 & 72 & -0.08 \\
J2043+2740 &  45 & 68 & -0.06 \\
J2229+6114 &  35 & 44 & -0.03 \\
J2240+5832$^{\ast}$ &  60 & 80 & -0.09 \\
	\hline
	\end{tabular}
	\caption{Best fit values for the geometry of each pulsar according to the analysis of their gamma-ray light curves. Pulsars with superscript $\ast$ are ones which have radio polarization RVM fits (see Table~\ref{tab:sample}) using the $\alpha$ and $\zeta$ values given above (except for PSR J1057-5226 as discussed in section~\ref{jrandg}).}
\label{tab:meilleur_fit}
\end{table}

\begin{figure}
	\centering
	\includegraphics[width=\columnwidth]{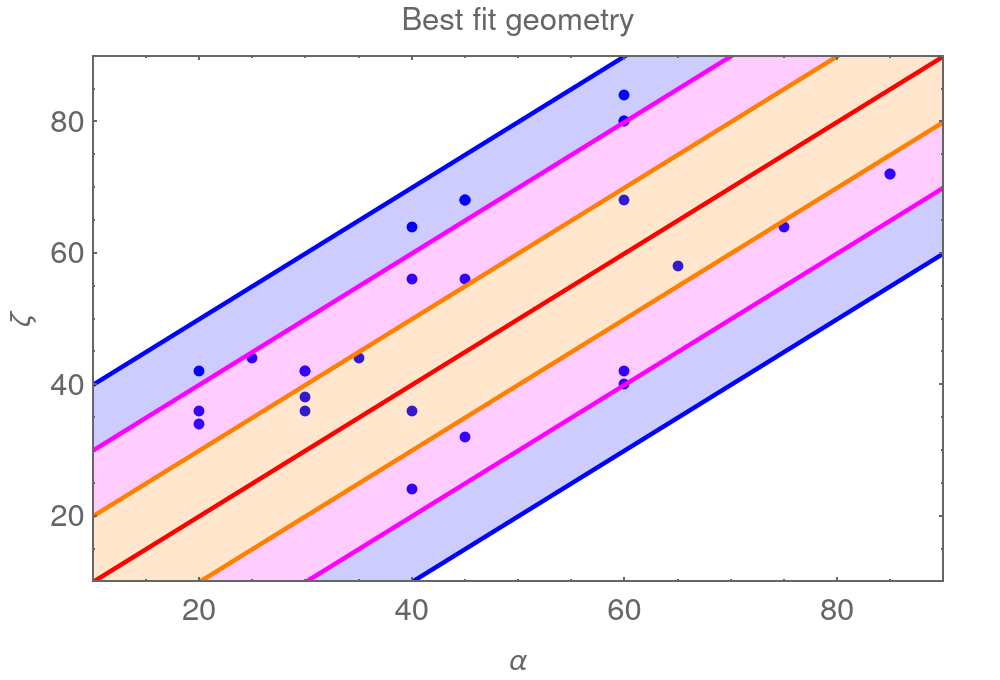}
	\caption{Correlation between the angles~$\alpha$ and $\zeta$ for our best fits are shown in blue dots according to table~\ref{tab:meilleur_fit}. The red line shows $\alpha=\zeta$ the orange lines are offset by $\pm10\degr$, the magenta lines by $\pm20\degr$ and the blue lines by $\pm30\degr$.}
	\label{fig:anglegeometrie}
\end{figure}

\begin{figure}
	\centering
	\includegraphics[width=\columnwidth]{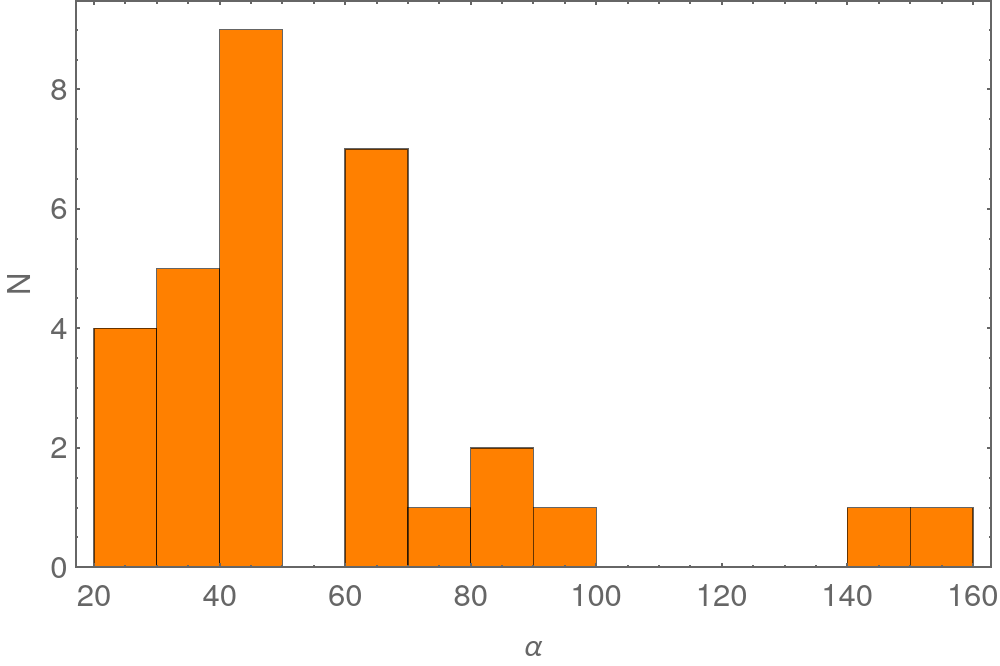}	\\
	\includegraphics[width=\columnwidth]{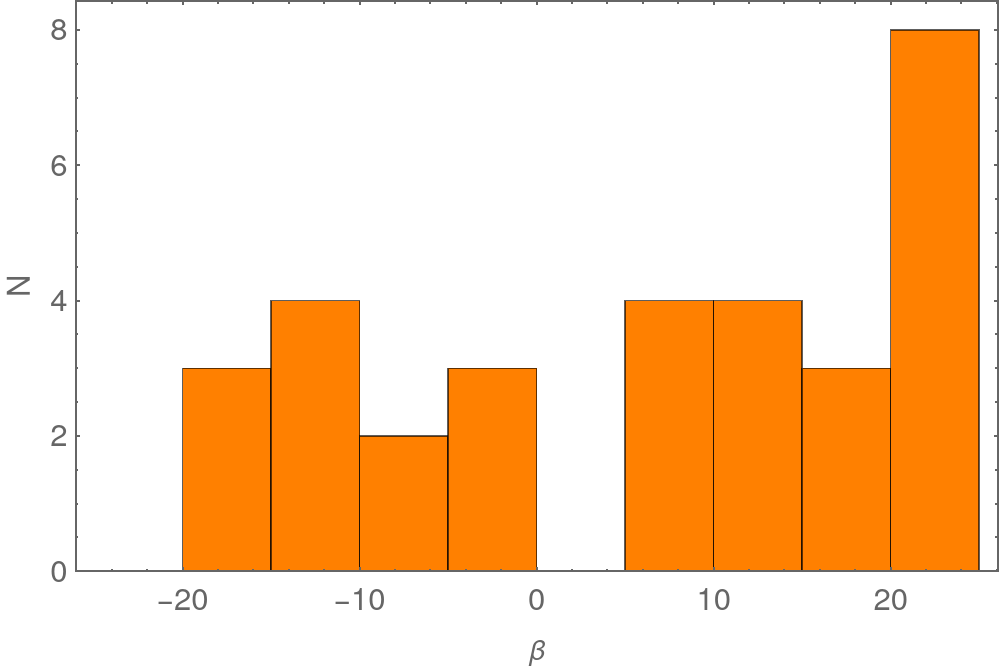} \\
	\includegraphics[width=\columnwidth]{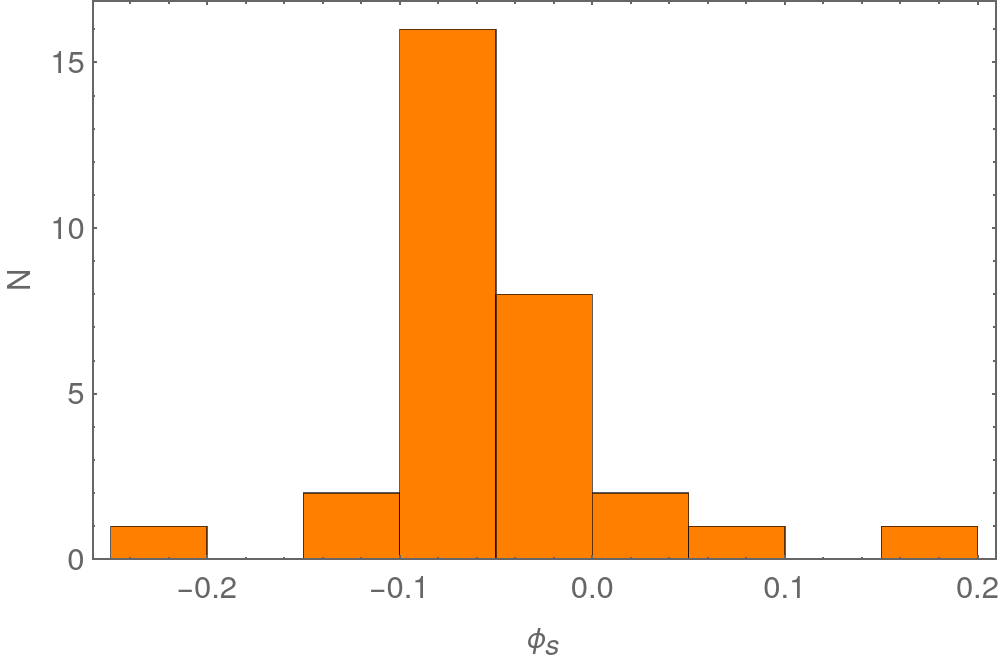}
	\caption{Histogram of best fit parameters given by the obliquity~$\alpha$ the line of sight $\beta$ and the phase offset~$\phi_s$ for the best fits given in table~\ref{tab:meilleur_fit}.}
	\label{fig:offset}
\end{figure}

Fig.~\ref{fig:anglegeometrie} summarizes the best fit angles $\alpha$ and $\zeta$, showing that they follow the relation $|\zeta - \alpha | \lesssim 30\degr$ which is slightly larger than what we have expected from the constrain in Sec.~\ref{sec:FFE_Magnetosphere}. This means that according to our model, some pulsars could have an emission height above the fiducial altitude of $0.05\,\rlight$.
Jointly, the distribution of angles~$\alpha$, $\beta$ and offsets~$\phi_s$ according to the same best fit values are shown in the histograms of Fig.~\ref{fig:offset}. Half of the sample has an obliquity less than 45\degr. The line of sight angle $\beta$ is distributed approximately symmetrically with respect to the magnetic axis. Interestingly, we found an important cluster of offsets around a phase $\phi_s \approx -0.05$ equitably distributed on both sides of this value with some marginal outliers lying more than 0.15 away from this median value of $-0.05$. This clustering indicates that some systematics has not been included in our study. The first gamma-ray peak is expected to come early than predicted by our model. One possibility would be that the striped wind emission is delayed, not peaking right at the light cylinder but at larger distances, a fraction of a light-cylinder radii~$\Delta r$ away from the light-cylinder. This repelling to larger distances automatically shifts the gamma-ray profile closer to the radio pulse by a phase $\phi \approx \tfrac{\Delta r}{2\,\pi\,\rlight}$ where $\Delta r$ measures this additional distance. Setting $\Delta r \approx \rlight/2$ leads to an additional phase shift of $\phi \approx 0.08$, sufficient to explain the histogram. Another possibility would be the forward beaming of the current sheet emission at the light-cylinder, forward with respect to the rotation direction, due to an azimuthal velocity close to the speed of light in this region. Such aberration effects also shorten the time lag between radio and gamma-rays. Some additional work is needed to accurately pin down the geometry. Only careful individual pulsar analysis will be able to tune these parameters firmly to irrelevant uncertainties.

\section{Conclusions}
\label{sec:Conclusion}

Multi-wavelength observations of neutron star pulsed emission offers a precious tool to explore the emission location within the pulsar magnetosphere and wind. We indeed showed that fitting simultaneously the radio and gamma-ray pulse profile of radio-loud gamma-ray pulsars severely constrains the geometry of the dipole magnetic field and observer line of sight with respect to the rotation axis. Moreover, when radio polarization data are available, additional constrains arise from fitting the rotating vector model. We showed that the RVM alone cannot be used to constrain the geometry of radio pulsar simply by minimizing the RVM $\rchi^2$ fit. Some additional knowledge from other wavelengths is requires. In most cases, the gamma-ray and radio fitting regions possess and overlapping area consistent with the gamma-ray light-curves, reducing the uncertainties in both angles, obliquity and line of sight. We applied our method to a large sample of radio loud gamma-ray pulsars with small error boxes, except for rare cases. As a good proxy, radio emission emanates from altitudes around $0.05-0.1\,\rlight$, where the magnetic field is dominantly dipolar whereas gamma-ray photons are produced at the light cylinder or slightly further away, extending to several $\rlight$ around the striped wind current sheet.

From a theoretical point of view, this study was only based on the impact of geometrical considerations on the radio and light-curve association, in the limit of a force-free magnetosphere. There is no mention about neither energetic nor particle dynamics. Particle acceleration and radiation needs to go beyond the force-free approximation by adding some dissipation like resistivity or radiation reaction damping. When acceleration and radiation sites will be accurately localised by these self-consistent models, we will be able to produce multi-wavelength phase resolved spectra and light-curves to pin down even better and more faithfully the magnetosphere geometry and its internal electrodynamics. Observational signatures of such dissipative magnetosphere needs to be performed to further support our emission model.

From an observational point of view, some pulsars would greatly benefit from better signal to noise ratio of the radio and gamma-ray pulse profiles. The upcoming third pulsar catalogue in gamma-rays and the construction of the square kilometer array promise to reach a big step towards our understanding of pulsar emission mechanisms.

\section*{Acknowledgements}

We are grateful to the referee for helpful comments and suggestions. This work has been supported by the CEFIPRA grant IFC/F5904-B/2018 and ANR-20-CE31-0010. We acknowledge the High Performance Computing center of the University of Strasbourg for supporting this work by providing scientific support and access to computing resources. We thank Lucas Guillemot for sending us some radio polarization data and David Smith for stimulating discussions. DM acknowledges the support of the
Department of Atomic Energy, Government of India, under project no. 12-R\&D-TFR-5.02-0700.

%




\bibliographystyle{mn2e}
\bibliography{/home/petri/zotero/Ma_bibliotheque}

\begin{thebibliography}{38}
\expandafter\ifx\csname natexlab\endcsname\relax\def\natexlab#1{#1}\fi

\bibitem[{Abdo {et~al}\mbox{.}(2010)Abdo, Ackermann, Ajello, Atwood, Axelsson,
  Baldini, Ballet, Barbiellini, Baring, Bastieri, Baughman, Bechtol,
  Bellazzini, Berenji, Blandford, Bloom, {E. Bonamente}, Borgland, Bregeon,
  Brez, Brigida, Bruel, Burnett, Buson, Caliandro, Cameron, Camilo, Caraveo,
  Casandjian, Cecchi, Çelik, Charles, {A. Chekhtman}, Cheung, Chiang, Ciprini,
  Claus, Cognard, Cohen-Tanugi, Cominsky, {J. Conrad}, Corbet, Cutini, Hartog,
  Dermer, Angelis, Luca, Palma, Digel, Dormody, Silva, Drell, Dubois, Dumora,
  Espinoza, Farnier, {C. Favuzzi}, Fegan, Ferrara, Focke, Fortin, Frailis,
  Freire, Fukazawa, {S. Funk}, Fusco, Gargano, Gasparrini, Gehrels, Germani,
  Giavitto, Giebels, Giglietto, {P. Giommi}, Giordano, Glanzman, Godfrey,
  Gotthelf, Grenier, Grondin, Grove, {L. Guillemot}, Guiriec, Gwon, Hanabata,
  Harding, Hayashida, Hays, Hughes, Jackson, Jóhannesson, Johnson, Johnson,
  Johnson, Johnson, Johnston, {T. Kamae}, Kanbach, Kaspi, Katagiri, Kataoka,
  Kawai, Kerr, Knödlseder, Kocian, {M. Kramer}, Kuss, Lande, Latronico,
  Lemoine-Goumard, Livingstone, Longo, Loparco, {B. Lott}, Lovellette, Lubrano,
  Lyne, Madejski, Makeev, Manchester, Marelli, Mazziotta, McConville, McEnery,
  McGlynn, Meurer, Michelson, Mineo, {W. Mitthumsiri}, Mizuno, Moiseev, Monte,
  Monzani, Morselli, Moskalenko, Murgia, {T. Nakamori}, Nolan, Norris, Noutsos,
  Nuss, Ohsugi, Omodei, Orlando, Ormes, {M. Ozaki}, Paneque, Panetta, Parent,
  Pelassa, Pepe, Pesce-Rollins, Piron, Porter, Rainò, Rando, Ransom, Ray,
  Razzano, Rea, Reimer, Reimer, Reposeur, {S. Ritz}, Rodriguez, Romani, Roth,
  Ryde, Sadrozinski, Sanchez, Sander, Parkinson, Scargle, Schalk, Sellerholm,
  Sgrò, Siskind, Smith, Smith, Spandre, Spinelli, Stappers, Starck, Striani,
  Strickman, Strong, Suson, Tajima, Takahashi, Takahashi, Tanaka, Thayer,
  Thayer, Theureau, Thompson, Thorsett, Tibaldo, Tibolla, Torres, Tosti,
  Tramacere, Uchiyama, Usher, Etten, Vasileiou, Venter, Vilchez, Vitale, Waite,
  Wang, Wang, {K. Watters}, Weltevrede, Winer, Wood, Ylinen, \&
  Ziegler}]{abdo_first_2010}
Abdo A.~A. {et~al.}, 2010, ApJS, 187, 460

\bibitem[{Abdo {et~al}\mbox{.}(2013)Abdo, Ajello, Allafort, Baldini, Ballet,
  Barbiellini, Baring, Bastieri, {A. Belfiore}, Bellazzini, Bhattacharyya,
  Bissaldi, Bloom, Bonamente, Bottacini, Brandt, Bregeon, Brigida, Bruel,
  Buehler, Burgay, Burnett, Busetto, Buson, Caliandro, Cameron, Camilo,
  Caraveo, Casandjian, Cecchi, Çelik, Charles, {S. Chaty}, Chaves, Chekhtman,
  Chen, Chiang, Chiaro, Ciprini, Claus, Cognard, {J. Cohen-Tanugi}, Cominsky,
  Conrad, Cutini, D'Ammando, Angelis, DeCesar, Luca, Hartog, Palma, Dermer,
  Desvignes, Digel, Venere, Drell, {A. Drlica-Wagner}, Dubois, Dumora,
  Espinoza, Falletti, Favuzzi, Ferrara, Focke, {A. Franckowiak}, Freire, Funk,
  Fusco, Gargano, Gasparrini, Germani, Giglietto, {P. Giommi}, Giordano,
  Giroletti, Glanzman, Godfrey, Gotthelf, Grenier, Grondin, Grove, Guillemot,
  Guiriec, Hadasch, Hanabata, Harding, Hayashida, Hays, {J. Hessels}, Hewitt,
  Hill, Horan, Hou, Hughes, Jackson, Janssen, Jogler, {G. Jóhannesson},
  Johnson, Johnson, Johnson, Johnson, Johnston, Kamae, {J. Kataoka}, Keith,
  Kerr, Knödlseder, Kramer, Kuss, Lande, Larsson, Latronico, {M.
  Lemoine-Goumard}, Longo, Loparco, Lovellette, Lubrano, Lyne, Manchester, {M.
  Marelli}, Massaro, Mayer, Mazziotta, McEnery, McLaughlin, Mehault, Michelson,
  Mignani, Mitthumsiri, Mizuno, Moiseev, Monzani, Morselli, Moskalenko, Murgia,
  Nakamori, Nemmen, Nuss, Ohno, Ohsugi, Orienti, Orlando, Ormes, Paneque,
  Panetta, Parent, Perkins, Pesce-Rollins, Pierbattista, Piron, {G. Pivato},
  Pletsch, Porter, Possenti, Rainò, Rando, Ransom, Ray, Razzano, {N. Rea},
  Reimer, Reimer, Renault, Reposeur, Ritz, Romani, Roth, Rousseau, Roy, {J.
  Ruan}, Sartori, Parkinson, Scargle, Schulz, Sgrò, Shannon, Siskind, Smith,
  Spandre, Spinelli, Stappers, Strong, Suson, Takahashi, Thayer, Thayer,
  Theureau, Thompson, Thorsett, Tibaldo, Tibolla, Tinivella, Torres, {G.
  Tosti}, Troja, Uchiyama, Usher, Vandenbroucke, Vasileiou, Venter, Vianello,
  {V. Vitale}, Wang, Weltevrede, Winer, Wolff, Wood, Wood, Wood, \&
  Yang}]{abdo_second_2013}
Abdo A.~A. {et~al.}, 2013, ApJS, 208, 17

\bibitem[{Benli {et~al}\mbox{.}(2021)Benli, Pétri, \&
  Mitra}]{benli_constraining_2021}
Benli O., Pétri J., Mitra D., 2021, A\&A, 647, A101, publisher: EDP Sciences

\bibitem[{Blaskiewicz {et~al}\mbox{.}(1991)Blaskiewicz, Cordes, \&
  Wasserman}]{blaskiewicz_relativistic_1991}
Blaskiewicz M., Cordes J.~M., Wasserman I., 1991, ApJ, 370, 643

\bibitem[{Bogovalov(1999)}]{bogovalov_physics_1999}
Bogovalov S.~V., 1999, A\&A, 349, 1017

\bibitem[{Cao {et~al}\mbox{.}(2016{\natexlab{a}})Cao, Zhang, \&
  Sun}]{cao_oblique_2016}
Cao G., Zhang L., Sun S., 2016{\natexlab{a}}, MNRAS, 461, 1068

\bibitem[{Cao {et~al}\mbox{.}(2016{\natexlab{b}})Cao, Zhang, \&
  Sun}]{cao_spectral_2016}
Cao G., Zhang L., Sun S., 2016{\natexlab{b}}, MNRAS, 455, 4267

\bibitem[{Cerutti {et~al}\mbox{.}(2015)Cerutti, Philippov, Parfrey, \&
  Spitkovsky}]{cerutti_particle_2015}
Cerutti B., Philippov A., Parfrey K., Spitkovsky A., 2015, MNRAS, 448, 606

\bibitem[{Contopoulos {et~al}\mbox{.}(1999)Contopoulos, Kazanas, \&
  Fendt}]{contopoulos_axisymmetric_1999}
Contopoulos I., Kazanas D., Fendt C., 1999, ApJ, 511, 351

\bibitem[{Deutsch(1955)}]{deutsch_electromagnetic_1955}
Deutsch A.~J., 1955, Annales d'Astrophysique, 18, 1

\bibitem[{Everett \& Weisberg(2001)}]{everett_emission_2001}
Everett J.~E., Weisberg J.~M., 2001, ApJ, 553, 341

\bibitem[{Johnston \& Kerr(2018)}]{johnston_polarimetry_2018}
Johnston S., Kerr M., 2018, MNRAS, 474, 4629, publisher: Oxford Academic

\bibitem[{Kalapotharakos {et~al}\mbox{.}(2012)Kalapotharakos, Contopoulos, \&
  Kazanas}]{kalapotharakos_extended_2012}
Kalapotharakos C., Contopoulos I., Kazanas D., 2012, MNRAS, 420, 2793

\bibitem[{Kalapotharakos {et~al}\mbox{.}(2017)Kalapotharakos, Harding, Kazanas,
  \& Brambilla}]{kalapotharakos_fermi_2017}
Kalapotharakos C., Harding A.~K., Kazanas D., Brambilla G., 2017, ApJ, 842, 80

\bibitem[{Komissarov(2006)}]{komissarov_simulations_2006}
Komissarov S.~S., 2006, MNRAS, 367, 19

\bibitem[{Kramer \& Johnston(2008)}]{kramer_high-precision_2008}
Kramer M., Johnston S., 2008, Monthly Notices of the Royal Astronomical
  Society, 390, 87

\bibitem[{Li {et~al}\mbox{.}(2012)Li, Spitkovsky, \&
  Tchekhovskoy}]{li_resistive_2012}
Li J., Spitkovsky A., Tchekhovskoy A., 2012, ApJ, 746, 60

\bibitem[{Michel(1973)}]{michel_rotating_1973}
Michel F.~C., 1973, The Astrophysical Journal Letters, 180, L133

\bibitem[{Mitra(2017)}]{mitra_nature_2017}
Mitra D., 2017, J Astrophys Astron, 38, 52

\bibitem[{Mitra \& Li(2004)}]{mitra_comparing_2004}
Mitra D., Li X.~H., 2004, A\&A, 421, 215

\bibitem[{Mitra \& Rankin(2011)}]{mitra_toward_2011}
Mitra D., Rankin J.~M., 2011, ApJ, 727, 92

\bibitem[{Mitra {et~al}\mbox{.}(2007)Mitra, Rankin, \&
  Gupta}]{mitra_absolute_2007}
Mitra D., Rankin J.~M., Gupta Y., 2007, MNRAS, 379, 932

\bibitem[{Mitra \& Seiradakis(2004)}]{mitra_effect_2004}
Mitra D., Seiradakis J.~H., 2004, in {arXiv}:astro-ph/0401335, arXiv:
  astro-ph/0401335

\bibitem[{Parfrey {et~al}\mbox{.}(2012)Parfrey, Beloborodov, \&
  Hui}]{parfrey_introducing_2012}
Parfrey K., Beloborodov A.~M., Hui L., 2012, MNRAS, 423, 1416

\bibitem[{Phillips(1992)}]{phillips_radio_1992}
Phillips J.~A., 1992, ApJ, 385, 282

\bibitem[{Pierbattista {et~al}\mbox{.}(2016)Pierbattista, Harding, Gonthier, \&
  Grenier}]{pierbattista_young_2016}
Pierbattista M., Harding A.~K., Gonthier P.~L., Grenier I.~A., 2016, A\&A, 588,
  A137

\bibitem[{Pierbattista {et~al}\mbox{.}(2015)Pierbattista, Harding, Grenier,
  Johnson, Caraveo, Kerr, \& Gonthier}]{pierbattista_light-curve_2015}
Pierbattista M., Harding A.~K., Grenier I.~A., Johnson T.~J., Caraveo P.~A.,
  Kerr M., Gonthier P.~L., 2015, A\&A, 575, A3

\bibitem[{Pétri(2011)}]{petri_unified_2011}
Pétri J., 2011, MNRAS, 412, 1870

\bibitem[{Pétri(2012)}]{petri_pulsar_2012}
Pétri J., 2012, MNRAS, 424, 605

\bibitem[{Pétri(2018)}]{petri_general-relativistic_2018}
Pétri J., 2018, MNRAS, 477, 1035

\bibitem[{Radhakrishnan \& Cooke(1969)}]{radhakrishnan_magnetic_1969}
Radhakrishnan V., Cooke D.~J., 1969, Ap. Lett., 3, 225

\bibitem[{Rookyard {et~al}\mbox{.}(2015)Rookyard, Weltevrede, \&
  Johnston}]{rookyard_investigating_2015}
Rookyard S.~C., Weltevrede P., Johnston S., 2015, {\textbackslash}mnras, 446,
  3356

\bibitem[{Seyffert {et~al}\mbox{.}(2011)Seyffert, Venter, de~Jager, \&
  Harding}]{seyffert_geometric_2011}
Seyffert A.~S., Venter C., de~Jager O.~C., Harding A.~K., 2011, arXiv e-prints,
  1105, arXiv:1105.4094

\bibitem[{Spitkovsky(2006)}]{spitkovsky_time-dependent_2006}
Spitkovsky A., 2006, ApJ, 648, L51

\bibitem[{Theureau {et~al}\mbox{.}(2011)Theureau, Parent, Cognard, Desvignes,
  Smith, Casandjian, Cheung, Craig, Donato, Foster, Guillemot, Harding,
  Lestrade, Ray, Romani, Thompson, Tian, \& Watters}]{theureau_psrs_2011}
Theureau G. {et~al.}, 2011, A\&A, 525, A94

\bibitem[{Watters {et~al}\mbox{.}(2009)Watters, Romani, Weltevrede, \&
  Johnston}]{watters_atlas_2009}
Watters K.~P., Romani R.~W., Weltevrede P., Johnston S., 2009, ApJ, 695, 1289

\bibitem[{Weltevrede \& Johnston(2008)}]{weltevrede_profile_2008}
Weltevrede P., Johnston S., 2008, MNRAS, 391, 1210

\bibitem[{Weltevrede \& Wright(2009)}]{weltevrede_mapping_2009}
Weltevrede P., Wright G., 2009, MNRAS, 395, 2117

\end{thebibliography}




%


\end{document}